\documentclass[preprint,numberedappendix]{emulateapj}
\usepackage{graphics,graphicx}
\usepackage{subfigure}
\usepackage{float}
\usepackage{xspace}
\usepackage{amsmath,amsthm,amssymb}
\usepackage{tabularx,booktabs}
\usepackage{enumerate}
\usepackage[referable]{threeparttablex}
\usepackage{color}
\usepackage{hyperref}
\hypersetup{
    bookmarks=true,         
    unicode=false,          
    pdftoolbar=true,        
    pdfmenubar=true,        
    pdffitwindow=false,     
    pdfstartview={FitH},    
    pdftitle={My title},    
    pdfauthor={Author},     
    pdfsubject={Subject},   
    pdfcreator={Creator},   
    pdfproducer={Producer}, 
    pdfkeywords={keyword1} {key2} {key3}, 
    pdfnewwindow=true,      
    colorlinks=true,       
    linkcolor=red,          
    citecolor=blue,        
    filecolor=magenta,      
    urlcolor=cyan           
}
\def\red#1 {\textcolor{red}{#1}\ }   
\def\blue#1 {\textcolor{blue}{#1}\ }   
\newcommand{\gadget}{{\sc GADGET-3}\ }
\newcommand{\gadgettwo}{{\sc GADGET-2}\ }

\def\gs{\mathrel{\raise0.35ex\hbox{$\scriptstyle >$}\kern-0.6em \lower0.40ex\hbox{{$\scriptstyle \sim$}}}}
\def\ls{\mathrel{\raise0.35ex\hbox{$\scriptstyle <$}\kern-0.6em \lower0.40ex\hbox{{$\scriptstyle \sim$}}}}

\def\oi{\mbox{\rm{O}{\sc i}}\xspace}

\def\h2{\mbox{{\sc H}$_2$}\xspace}
\def\hi{\mbox{\rm{H}\textsc{i}}\xspace}
\def\hii{\mbox{\rm{H}\textsc{ii}}\xspace}
\def\cii{\mbox{\rm{[C}\textsc{ii}]}\xspace}

\def\ci{\mbox{\rm{C}\textsc{i}}\xspace}

\def\Mh2{\mbox{$M_{\rm{H}_2}$}\xspace}
\def\mH{\mbox{$m_{\rm{H}}$}\xspace}
\def\mgmc{\mbox{$m_{\rm{GMC}}$}\xspace}
\def\Mstar{\mbox{$M_{\rm{\ast}}$}\xspace}
\def\Mgas{\mbox{$M_{\rm{gas}}$}\xspace}
\def\mmol{\mbox{$m_{\rm{mol}}$}\xspace}

\def\Mneu{\mbox{$M_{\rm{neutral}}$}\xspace}
\def\mneu{\mbox{$m_{\rm{neutral}}$}\xspace}
\def\msph{\mbox{$m_{\rm{SPH}}$}\xspace}
\def\gassd{\mbox{$\Sigma_{\rm{gas}}$}\xspace}
\def\starssd{\mbox{$\Sigma_{\rm{*}}$}\xspace}

\def\nH{\mbox{$n_{\rm{H}}$}\xspace}
\def\nhi{\mbox{$n_{\rm{H}\textsc{i}}$}\xspace}
\def\ne{\mbox{$n_{\rm{e}}$}\xspace}
\def\nav{\mbox{$\langle n_{\rm H} \rangle$}\xspace}
\def\xe{\mbox{$x_{\rm{e}}$}\xspace}
\def\next{\mbox{$n_{\rm{H,ext}}$}\xspace}
\def\nh2{\mbox{$n_{\rm{H}_2}$}\xspace}

\def\ncii{\mbox{$n_{\rm{C}\textsc{ii}}$}\xspace}

\def\NH2{\mbox{$N_{\rm{H}_2}$}\xspace}
\def\nhi{\mbox{$n_{\rm{H}\textsc{i}}$}\xspace}

\def\Xc{\mbox{$X_{\rm{C}}$}\xspace}

\def\Tk{\mbox{$T_{\rm{k}}$}\xspace}

\def\Hpe{\mbox{$\Gamma_{\rm{PE}}$}\xspace}
\def\Hcrhi{\mbox{$\Gamma_{\rm{CR,H}\textsc{i}}$}\xspace}
\def\Hcrh2{\mbox{$\Gamma_{\rm{CR,H}_2}$}\xspace}

\def\Ch2{\mbox{$\Lambda_{\rm{H}_2}$}\xspace}

\def\Ccii{\mbox{$\Lambda_{\rm{C}\rm{\textsc{ii}}}$}\xspace}

\def\Coi{\mbox{$\Lambda_{\rm{O}\rm{\textsc{i}}}$}\xspace}

\def\fmol{\mbox{$f'_{\rm{mol}}$}\xspace}

\def\g0{\mbox{$G_0$}\xspace}
\def\d0{\mbox{$D_0$}\xspace}

\def\cri{\mbox{$\zeta_{\rm{CR}}$}\xspace}

\def\Pe{\mbox{$P_{\rm{ext}}$}\xspace}
\def\Z{\mbox{$Z'$}\xspace}
\def\Zn{\mbox{$Z'$}}

\def\sv{\mbox{$\sigma_{v}$}\xspace}

\def\SFRsd{\mbox{$\Sigma_{\rm{SFR}}$}\xspace}

\def\rh2{\mbox{$R_{\rm{H}_2}$}\xspace}
\def\rrh2{\mbox{$r_{\rm{H}_2}$}\xspace}
\def\rrci{\mbox{$r_{\rm{{\sc C\textsc{i}}}}$}\xspace}
\def\rci{\mbox{$R_{\rm{C}\textsc{i}}$}\xspace}

\def\rcl{\mbox{$R_{\rm{GMC}}$}\xspace}

\def\CIIsd{\mbox{$\Sigma_{\rm{[C\textsc{ii}]}}$}\xspace}
\def\Lcii{\mbox{$L_{\rm{[C}\rm{\textsc{ii}]}}$}\xspace}

\def\lsun{\mbox{${\rm L_\odot}$}\xspace}
\def\cms{\mbox{cm$^{2}$}\xspace} 
\def\cmpc{\mbox{cm$^{-3}$}\xspace} 
\def\kms{\mbox{km\,s$^{-1}$}\xspace}
\def\ps{\mbox{s$^{-1}$}\xspace}

\def\msun{\mbox{$\rm{M}_\odot$}\xspace}
\def\sfru{\mbox{$\rm{M}_\odot$~yr$^{-1}$}\xspace}

\def\sigame{\texttt{S\'IGAME}\xspace}

\newcommand{\e}  [1]{\ensuremath{\times10^{#1}}}

\begin{document}

\shorttitle{Modeling \cii emission from Galaxies}
\shortauthors{Olsen et al.}

\title{SImulator of GAlaxy Millimeter/submillimeter Emission (S\'IGAME):  the \cii$-$SFR relationship of massive z=2 main sequence galaxies}

\author{Karen P. Olsen\altaffilmark{1}}
\author{Thomas R. Greve\altaffilmark{2}}
\author{Desika Narayanan\altaffilmark{3}}
\author{Robert Thompson\altaffilmark{4}}
\author{Sune Toft\altaffilmark{1}}
\author{Christian Brinch\altaffilmark{5,6}}

\altaffiltext{1}{Dark Cosmology Centre, Niels Bohr Institute, University of Copenhagen, Juliane Maries Vej 30, DK-2100 Copenhagen, Denmark; karen@dark-cosmology.dk}
\altaffiltext{2}{Department of Physics and Astronomy, University College London, Gower Street, London WC1E 6BT, UK}
\altaffiltext{3}{Department of Physics and Astronomy, Haverford College, 370 W Lancaster Ave., Haverford, PA 19041, USA}
\altaffiltext{4}{University of the Western Cape, 7535 Bellville, Cape Town, South Africa}
\altaffiltext{5}{Centre for Star and Planet Formation (Starplan) and Niels Bohr Institute, University of Copenhagen, 
Juliane Maries Vej 30, DK-2100 Copenhagen, Denmark}
\altaffiltext{6}{DeIC, Technical University of Denmark, Building 309, DK-2800 Kgs. Lyngby, Denmark}

\email{karen@dark-cosmology.dk}

\begin{abstract}
We present \sigame simulations of the \cii$157.7\,{\rm \mu m}$ fine structure
line emission from cosmological smoothed particle hydrodynamics (SPH)
simulations of seven main sequence galaxies at $z=2$. Using sub-grid physics
prescriptions the gas in our simulations is modeled as a multi-phased
interstellar medium (ISM) comprised of molecular gas residing in giant
molecular clouds, an atomic gas phase associated with photo-dissociation regions
(PDRs) at the cloud surfaces, and a diffuse, ionized gas phase.  Adopting
logotropic cloud density profiles and accounting for heating by the local FUV
radiation field and cosmic rays by scaling both with local star formation rate
(SFR) volume density, we calculate the \cii emission using a photon escape 
probability formalism. The \cii
emission peaks in the central $\ls 1\,{\rm kpc}$ of our galaxies as do the SFR radial profiles, 
with most \cii ($\gs 70\%$) originating in the molecular gas phase, whereas further out 
($\gs 2\,{\rm kpc}$), the atomic/PDR gas dominates ($\gs 90\%$) the \cii emission, 
no longer tracing on-going star formation.
Throughout, the ionized gas contribution is negligible ($\ls3\%$).  
The \cii luminosity vs. SFR (\cii-SFR) relationship, integrated as well as
spatially resolved (on scales of $1\,{\rm kpc}$),
delineated by our simulated galaxies is in good agreement 
with the corresponding relations observed
locally and at high redshifts. In our simulations, the molecular gas dominates 
the \cii budget at ${\rm SFR}\gs 20\,{\rm \msun\,yr^{-1}}$ ($\SFRsd \gs 0.5\,{\rm \msun\,yr^{-1}\,kpc^{-2}}$), 
while atomic/PDR gas takes over at lower SFRs, suggesting a 
picture in which \cii predominantly traces the molecular
gas in high-density/pressure regions where star formation is on-going, 
and otherwise reveals the atomic/PDR gas phase.
\end{abstract}


\section{Introduction}

Single ionized carbon (C{\sc ii}) can be found throughout the interstellar
medium (ISM) of galaxies where gas is exposed to UV radiation with energies
above the ionization potential of neutral carbon ($11.3\,{\rm eV}$ cf.\
$13.6\,{\rm eV}$ for hydrogen).  C{\sc ii} is found both in regions of ionized
and neutral gas where, depending on the gas phase, its fine structure line \cii
$^2P_{3/2}-^2P_{1/2}$ ($\lambda_{\rm rest} = 157.714\,{\rm \mu m}$) is
collisionally excited by electrons, \hi or \h2. The $^2P_{3/2}$ upper level lies
$91\,{\rm K}$ ($= h\nu/k_{\rm B}$) above the $^2P_{1/2}$ ground state and, over
a large temperature range ($\sim 20-8000\,{\rm K}$), the critical density of
\cii is only $\sim 5-44$, $\sim 1600-3800$ and $\sim 3300-7600\,{\rm cm^{-3}}$
for collisions with $e^-$, \hi and \h2, respectively \citep{goldsmith12}.  
As a consequence of these characteristics, 
\cii is observed to be one of the strongest cooling lines of the ISM, with a line
luminosity equivalent to $\sim 0.1-1\%$ of the far-infrared (FIR) luminosity of
galaxies \citep[e.g.,][]{stacey91,brauher08,casey14}.

Due to high atmospheric opacity at FIR wavelengths, observations of
\cii\ in the local Universe must be done at high altitudes or in
space.  Indeed, the very first detections of \cii\ towards Galactic
objects \citep{russell80,stacey83,kurtz83} and other galaxies
\citep{crawford85,stacey91,madden92} were done with airborne
observatories such as the NASA Lear Jet and the Kuiper Airborne
Observatory.  The Infrared Space Observatory (ISO)
allowed for the first systematic \cii\ surveys of local galaxies
\citep[e.g.,][]{malhotra97,luhman98,luhman03}.  Detections of \cii\ at
high redshifts ($z > 1$) have also become feasible in recent years, with ground-based
facilities
\citep[e.g.,][]{maiolino05,maiolino09,hailey-dunsheath10,stacey10} and
the {\it Herschel Space Observatory} \citep[e.g.,][]{gullberg15}.  The
Atacama Large Millimeter Array (ALMA), owing to its tremendous
collecting area and high angular resolution, is now resolving \cii\ in
high-$z$ galaxies \citep{debreuck14,wang13} and pushing \cii\ observations of
high-$z$ galaxies to much lower luminosity than before
\citep{ouchi13,maiolino15,capak15}.

In spite of the observational successes, the interpretation of the \cii\ line as
a diagnostic of the ISM and the star formation conditions in galaxies is
complicated by the fact that the \cii\ emission can originate from different
phases of the ISM. In our Galaxy, about $30\,$\% of the total \cii emission is
found to come from dense photo-dominated regions (PDRs), $25\,$\% from cold \hi
gas, $25\,$\% from CO-dark H$_2$ gas, and $20\,\%$ from ionized gas
\citep{pineda14}. We expect these percentages to be different in other galaxies
where high levels of star formation and/or accretion onto the supermassive black
hole can boost the energy injected into the ISM and change the carbon ionization
balance. Models predict a reduction in the \cii\ emission from the extreme X-ray
dominated regions (XDRs) associated with active galactic nuclei (AGNs)
\citep{meijerink07,herrera-camus15}. In regions of intense FUV-fields ($\gs
10\times$ the local Galactic field), the CO emitting surfaces of clouds will
shrink due to photo-dissociation of CO, resulting in a thicker envelope layer of
(self-shielding) H$_2$ where carbon exist only as C{\sc i} or C{\sc ii}
\citep[e.g.,][]{wolfire10}.  Also, extreme cosmic ray (CR) energy densities
(i.e., $\gs 1000\times$ that of the Galactic average, as is found in local
starbursts galaxies; \citealt{acero09}), might lead to the efficient destruction
of CO (in reactions with He$^+$ which forms via cosmic ray ionization), and
leave behind C{\sc ii} deep in the FUV-shielded regions of the molecular medium
\citep[e.g.,][]{bisbas15}.

The sensitivity of \cii to the presence of FUV radiation led to the line being
suggested as a tracer of recent star formation. Observations seem to bear out
this notion, with normal star-forming galaxies ($L_{\rm IR} \le 10^{11}\,\lsun$)
in the local universe exhibiting a nearly linear relation between \cii\
luminosity ($L_{\rm [CII]}$) and IR luminosity ($L_{\rm IR}$) over several dex
\citep[e.g.,][]{malhotra97,malhotra01}.  Similar ${\rm \cii-SFR}$ calibrations
for local star-forming galaxies based on alternative star formation rate (SFR) tracers such as
H$\alpha$ and the UV have also been established
\citep[e.g.,][]{boselli02,delooze11}.  Recent {\it Herschel} studies of local
spirals and dwarf galaxies have found that \cii\ remains a robust tracer of star
formation over scales ranging from $\sim 20$\,pc to $\sim 1\,{\rm kpc}$
\citep[e.g.,][hereafter referred to as D14, H15, and K15,
respectively]{delooze14,herrera-camus15,kapala15}.  \citet{pineda14} found that
our Galaxy matches these resolved extragalactic ${\rm \cii-SFR}$ relations, both
in terms of the slope and overall normalization, only when the \cii\ emission of
all the gas phases in our Galaxy are combined.

The scatter in the ${\rm \cii-SFR}$ relation is non-negligible -- about
$0.2-0.3\,{\rm dex}$ in the surface density relations and $\sim 0.3\,{\rm dex}$
in the luminosity relations (at $z\ls0.3$) -- and must be characterized as best as possible in
order to optimize \cii\ as a star formation indicator.  While some of the
scatter can be ascribed to imperfect star formation tracers, it does tend to
increase systematically in regions of low metallicity, warm dust temperatures,
and large filling factors of ionized, diffuse gas (D14, H15).  Understanding
this behavior is of importance for studies of the low-metallicity, UV-intense
environments that we expect to encounter in normal galaxies at $z\gs 6$, where
in some cases the \cii\ line is strongly detected \citep{capak15}, in accordance
with the notion that it should be one of the brightest available diagnostics of
the ISM at those epochs, while in other instances the absence of \cii\ down to
faint flux levels places the sources $\gs 10$ times below the local ${\rm
\cii-SFR}$ relation \citep{walter12,kanekar13,ouchi13,schaerer15,maiolino15}.

With the \cii\ line increasingly being used as a tracer of gas and star
formation at high redshifts, efforts have also recently been made to simulate
the \cii\ emission from galaxies. Various sub-grid `gas physics' approaches
have been applied to both semi-analytical \citep{popping14b,munoz14} and
hydrodynamical \citep{nagamine06,vallini13,vallini15} simulations of galaxies.
The simulations by \cite{nagamine06} focus on the \cii detectability of lyman
break galaxies (LBGs) at $z=3$, while \cite{vallini13} apply their simulations to
observed upper limits on the \cii emission from the $z=6.6$ Ly$\alpha$ emitter
(LAE) Himiko \citep{ouchi13} finding that its metallicity must be subsolar.
Both set of simulations consider a two-phase ISM consisting of a cold and a
warm neutral in pressure equilibrium, and both find that the \cii emission is
dominated by the cold neutral medium (CNM). In an update of their 2013
simulation, that includes the \cii contribution from PDRs and a non-uniform
metallicity distribution in the gas phase, \cite{vallini15} finds that most of
the \cii emission from their $z=6.6$ models originates in the PDRs with only
$\sim 10\%$ coming from the CNM. 

In this paper we present an adapted version of our code SImulator of GAlaxy
Millimeter/submillimeter Emission \citep[\sigame;][submitted]{olsen15} that is
capable of incorporating \cii\ emission into smoothed particle hydrodynamics
(SPH) simulations of galaxies. We consider a multi-phased ISM consisting of
molecular clouds, whose surface layers are stratified by FUV-radiation from
localized star formation, embedded within a neutral medium of atomic gas.  In
addition, we include the diffuse ionized gas intrinsic to the SPH simulations
as a third ISM phase. The temperatures of the molecular and atomic gas are
calculated from thermal balance equations sensitive to the local FUV-radiation
and CR ionization rate.  We apply \sigame to GADGET-3 cosmological SPH
simulations of seven star-forming galaxies on the main-sequence (MS) at $z=2$ in
order to simulate the \cii\ emission from normal star-forming galaxies at this
epoch, examine the relative contributions to the emission from the molecular,
atomic and ionized ISM phases, and the relationship to the star formation
activity in the galaxies.  Throughout, we adopt a flat cosmology with
$\Omega_{\rm M} = 0.27, \Omega_{\Lambda} = 0.73$, and $h = 0.71$
\citep{spergel2003}. 

\begin{figure*}[t]
\begin{center}
\includegraphics[width=1.6\columnwidth]{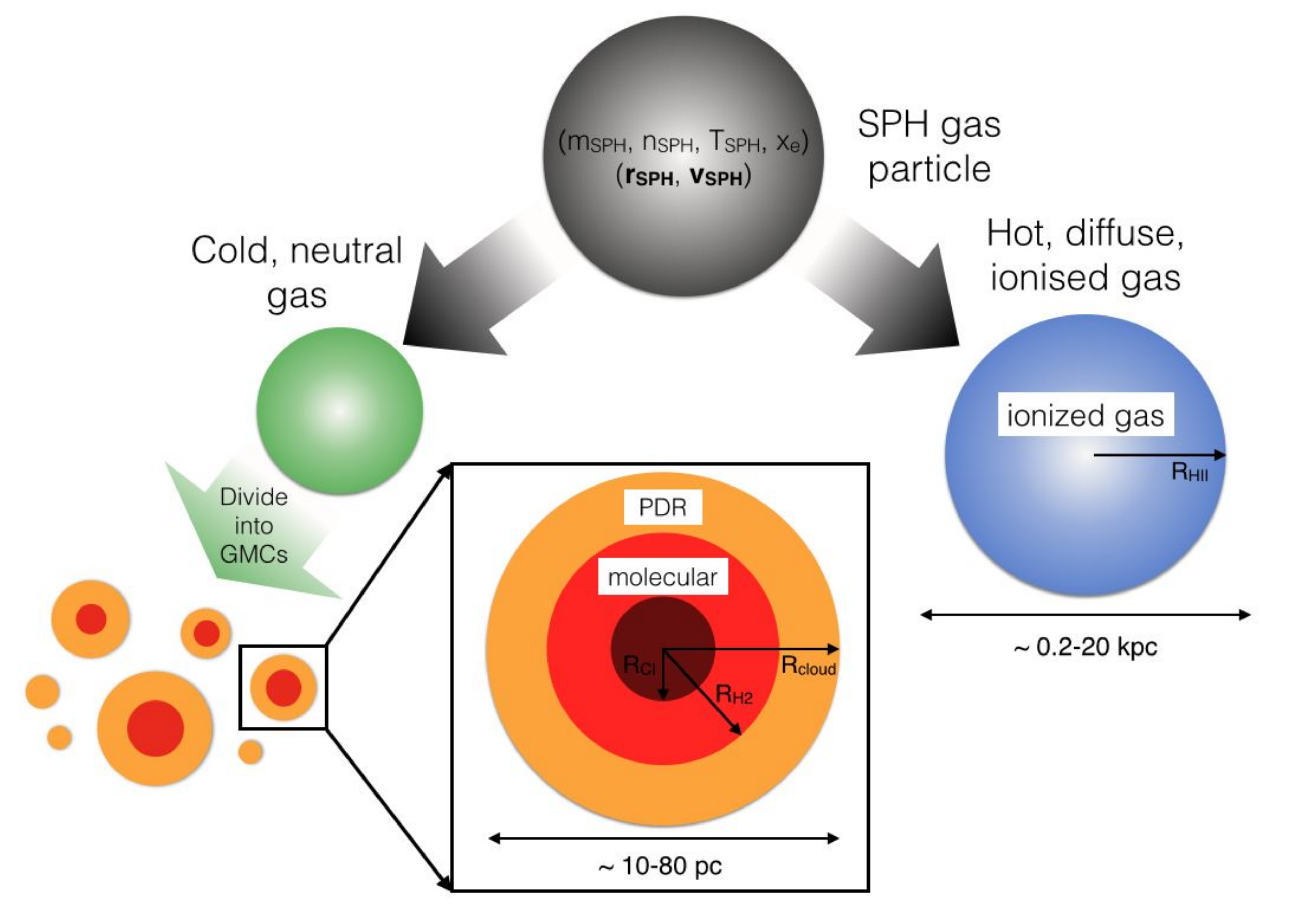}
\end{center}
\caption{Schematic illustrating the sub-grid procedures applied to the SPH
simulation in post-processing. Each SPH particle is a hybrid of neutral and
ionized gas. The neutral gas associated with each SPH gas particle is divided
into GMCs with masses and sizes following the Galactic mass-spectrum and
mass-size relation for GMCs.  Each GMC has an onion-layer structure, set by the
stratification of the impinging FUV-field, which consists of outer layer of PDR/atomic
gas of H{\sc i} and C{\sc ii}, and an inner molecular region where
carbon is found in its single ionized state, and in neutral form further in
(Section \ref{split1}).  The ionized gas associated with each SPH particle is
assumed to reside in spherical clouds with radii and temperatures given by the
SPH smoothing length and gas temperature (Section \ref{split2}).}
\label{figure:cartoon}
\end{figure*}

\section{Methodology overview}\label{section:methodology}
\sigame is applied at the post-processing stage of a SPH simulation and takes as
its input the following quantities associated with each SPH particle: the
position ($[x,y,z]$), velocity ($[v_x,v_y,v_z]$), smoothing length ($h$), gas
mass (\msph), hydrogen density (\nH), gas kinetic temperature (\Tk), electron
fraction ($\xe = n_{e}/n_{\rm H}$), SFR, metallicity ($Z$)
as well as the relative abundances of carbon ([C/H]) and oxygen ([O/H]). The key
steps involved in the post-processing are illustrated in
Fig.\,\ref{figure:cartoon} and briefly listed below (with details given in
subsequent sections):
\begin{enumerate}
\item The SPH gas is separated into its neutral and ionized constituents as
dictated by the electron fraction provided by the \gadget simulations.

\item The neutral gas is divided into giant molecular clouds (GMCs) according to
the observed mass function of GMCs in the Milky Way (MW) and nearby quiescent
galaxies. The GMCs are modeled as logotropic spheres with their sizes and
internal velocity dispersions derived according to pressure-normalized scaling
relations.

\item Each GMC is assumed to consist of three spherically symmetric regions: (1)
a FUV-shielded molecular core region where all carbon is locked up in \ci and
CO; (2) an outer molecular region where both \ci\ and C{\sc ii} can exist; and
(3) a largely neutral atomic layer of \hi, \hii ~and C{\sc ii}.  The last region
mimics the FUV-stratified PDRs observed at the surfaces of molecular clouds
\citep{hollenbach99}. 
This layer can contain both atomic and ionized gas, but 
we shall refer to it simply as the PDR gas.
The relative extent of these regions within each cloud,
and thus the densities at which they occur, ultimately depends on the strength
of the impinging FUV-field and CR ionization rate. The latter are set to scale
with the local SFR volume density and, by requiring thermal balance with the
cooling from line emission (from C{\sc ii}, \oi, and \h2), determine the
temperatures of the molecular and atomic gas phases.

\item The remaining ionized gas of the SPH simulation is divided into \hii\ clouds 
of radius equal to the smoothing lengths, temperature equal to that of the SPH 
simulation and constant density. 

\item The \cii\ emission from the molecular, PDR, and diffuse ionized gas
is calculated separately and summed to arrive at the total \cii\ emission from
the galaxy. In doing so it is assumed that there is no radiative coupling
between the clouds in the galaxy.
\end{enumerate}
The SPH simulations used in this paper, and the galaxies extracted from them,
are described in the following section.

%
\begin{table*}
\centering
\caption{Global properties of the seven simulated galaxies used for this work at $z=2$.} 
\begin{tabular}{lccccccc}
\hline
                                		& G1        & G2        & G3       	&	G4		&	G5		&	G6     &	G7		  \\	
\toprule                                                                                                       
\Mstar [10$^{10}$\,\msun]			& 0.36		&	0.78    &	0.95	& 	1.80	&	4.03	&	5.52   &	6.57	 \\
\Mgas [10$^{10}$\,\msun]			& 0.42		&   0.68	&	1.26	&  	1.43	&	2.59	&	2.16   &	1.75	 \\
\Mneu [10$^{10}$\,\msun]			& 0.09		&   0.13	&	0.57	&  	0.20	&	0.29	&	0.29   &	0.39	 \\
$M_{\rm ionized}$ [10$^{10}$\,\msun]	        & 0.33		&   0.55	&	0.69	&  	1.23	&	1.30	&	1.87   &	1.36	 \\
SFR [\sfru]				        & 4.9		&   10.0 	&	8.8	&	25.1	&	19.9	&	59.0   &	37.5	 \\
\SFRsd [\sfru kpc$^{-2}$]	    		& 0.016		&   0.032	&	0.028	&	0.080	&	0.063	&	0.188  &	0.119	 \\	
\Z	    					& 0.43		&   0.85	&	0.64	&	1.00	&	1.00	&	1.67   &	1.72	 \\	
\hline                                                       	                    
\hline                                                                           
\label{table:prop}
\end{tabular}

All quantities have been calculated at $z=2$ using a fixed cut-out radius of
$R_{\rm cut}=10\,{\rm kpc}$, which is the radius at which the accumulative stellar
mass function of each galaxy flattens. \Mgas is the total gas mass, and \Mneu
and $M_{\rm ionized}$ the gas masses in neutral and ionized form, respectively
(see Section \ref{modelISM}). The metallicity ($\Z=Z/Z_{\odot}$) is the mean of
all SPH gas particles within $R_{\rm cut}$. \label{table:global-properties}
\end{table*}

\section{SPH Simulations}\label{sph}
Our simulations are evolved with an updated version of the public \gadget
cosmological SPH code (\citealt{Springel05} and S. Huang et al. 2015 in preparation).  It
includes cooling processes using the primordial abundances as described in
\citet{Katz96}, with additional cooling from metal lines assuming
photo-ionization equilibrium from \citet{wiersma09}.  We use the more recent
`pressure-entropy' formulation of SPH which resolves mixing issues when
compared with standard `density-entropy' SPH algorithms \citep[see][for further
details]{Saitoh13,Hopkins13DISPH}.  Our code additionally implements the
time-step limiter of \cite{Saitoh09}, \cite{Durier12} which improves the accuracy of
the time integration scheme in situations where there are sudden changes to a
particle's internal energy.  To prevent artificial fragmentation
\citep{Schaye08,Robertson08}, we prohibit gas particles from cooling below
their effective Jeans temperature which ensures that we are always resolving at
least one Jeans mass within a particle's smoothing length.  This is very
similar to adding pressure to the ISM as in
\cite{Springel03}, \cite{Schaye08}, except instead of directly pressurizing the gas we
prevent it from cooling and fragmenting below the Jeans scale.

We stochastically form stars within the simulation from molecular gas following
a \citet{Schmidt59} law with an efficiency of 1\% per local free-fall time
\citep{Krumholz07,Lada10}.  The molecular content of each gas particle is
calculated via the equilibrium analytic model of
\citet{Krumholz08,Krumholz09,mckee10}.  This model allows us to regulate star
formation by the local abundance of H$_2$ rather than the total gas density,
which confines star formation to the densest peaks of the ISM.
Further implementation details can be found in \citet{thompson14}.  Galactic
outflows are implemented using the hybrid energy/momentum-driven wind (ezw)
model fully described in \citet{Dave13,Ford15}.  We also account for metal
enrichment from Type II supernovae (SNe), Type Ia SNe, and AGB stars as
described in \citet{Oppenheimer08}.

\subsection{SPH simulations of $z=2$ MS galaxies} \label{tcase}
We use the cosmological zoom-in simulations presented in \citet{thompson15}, and
briefly summarized here.  Initial conditions were generated using the {\small
MUSIC} code \citep{MUSIC} assuming cosmological parameters consistent with
constraints from the {\it Planck} \citep{Planck14} results, namely
$\Omega_{m}=0.3, \Omega_{\Lambda}=0.7,H_0=70,\sigma_8=0.8,n_s=0.96$. Six target
halos were selected at $z=2$ from a low-resolution $N$-body simulation consisting
of $256^3$ dark-matter particles in a $(16h^{-1}{\rm Mpc})^3$ volume with an
effective co-moving spatial resolution of $\epsilon=1.25\,h^{-1}$ kpc. Each
target halo is populated with higher resolution particles at $z=249$, with the
size of each high resolution region chosen to be $2.5$ times the maximum radius
of the original low-resolution halo. The majority of halos in our sample are
initialized with a single additional level of refinement
($\epsilon=0.625\,h^{-1}$kpc), while the two smallest halos are initialized with
two additional levels of refinement ($\epsilon=0.3125\,h^{-1}$kpc).

The six halos produce seven star-forming galaxies at $z=2$ that are free from
all low-resolution particles within the virial radius of their parent halo.
Their stellar masses (\Mstar) range from $3.6\times10^9$ to $6.6\times
10^{10}\,\msun$ and their SFRs from $5$ to $60$\,\sfru (Table
\ref{table:global-properties}). We hereafter label the galaxies G1, ..., G7 in
order of increasing \Mstar. Other relevant global properties directly inferred
from the SPH simulations, such as total gas mass (\Mgas), neutral and ionized
gas masses (\Mneu and $M_{\rm ionized}$, respectively), average SFR surface
density ($\Sigma_{\rm SFR}$), and average metallicity (\Z), can also be found in
Table \ref{table:global-properties}.
\begin{figure}[t]
\begin{center}
\includegraphics[width=\columnwidth]{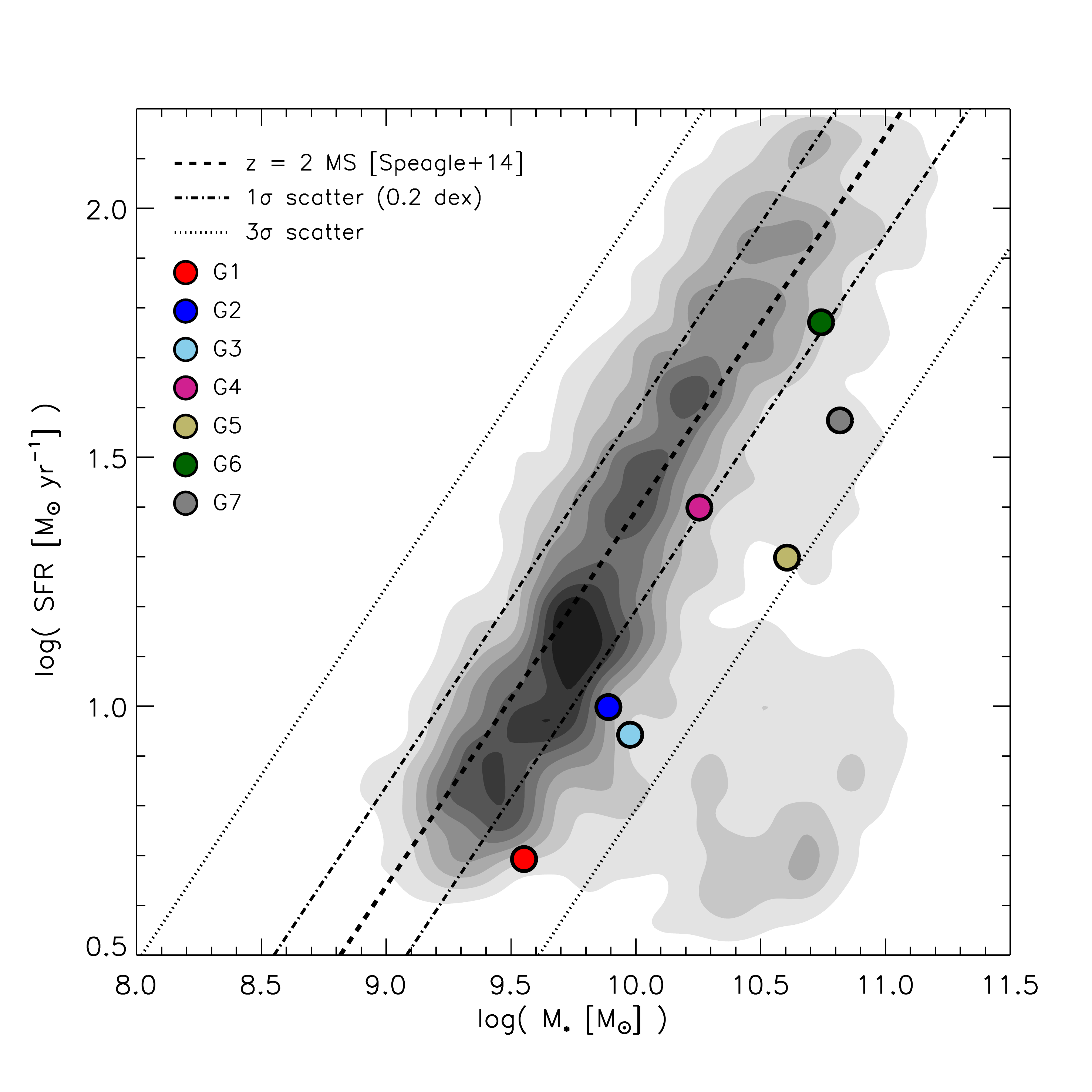}
\end{center}
\caption{The SFR--M$_{\ast}$ relation at $z\simeq 2$ as determined by
\citet{speagle14} (dashed line) with the location of our seven simulated
galaxies highlighted (filled circles). Dotted-dashed and dotted lines indicate the
$1\sigma$ and $3\sigma$ scatter around the relation of \cite{speagle14}. For
comparison we also show the locus defined by $3754$ $1.4<z<2.5$ galaxies from
the NEWFIRM Medium-Band Survey (gray filled contours), with masses and SFRs
calculated using a Kroupa IMF \citep{whitaker11}.}  
\label{M_SFR}
\end{figure}

Fig.\,\ref{M_SFR} shows the locations of G1, ..., G7 in the SFR$-$\Mstar
diagram.  The galaxies are consistent with observational determinations of the
$z\sim2$ MS of star-forming galaxies
\citep{whitaker11,speagle14}.

\section{Modeling the ISM} \label{modelISM}
\begin{figure}[htbp] 
\hspace*{-0.5cm}
\includegraphics[width=1.11\columnwidth]{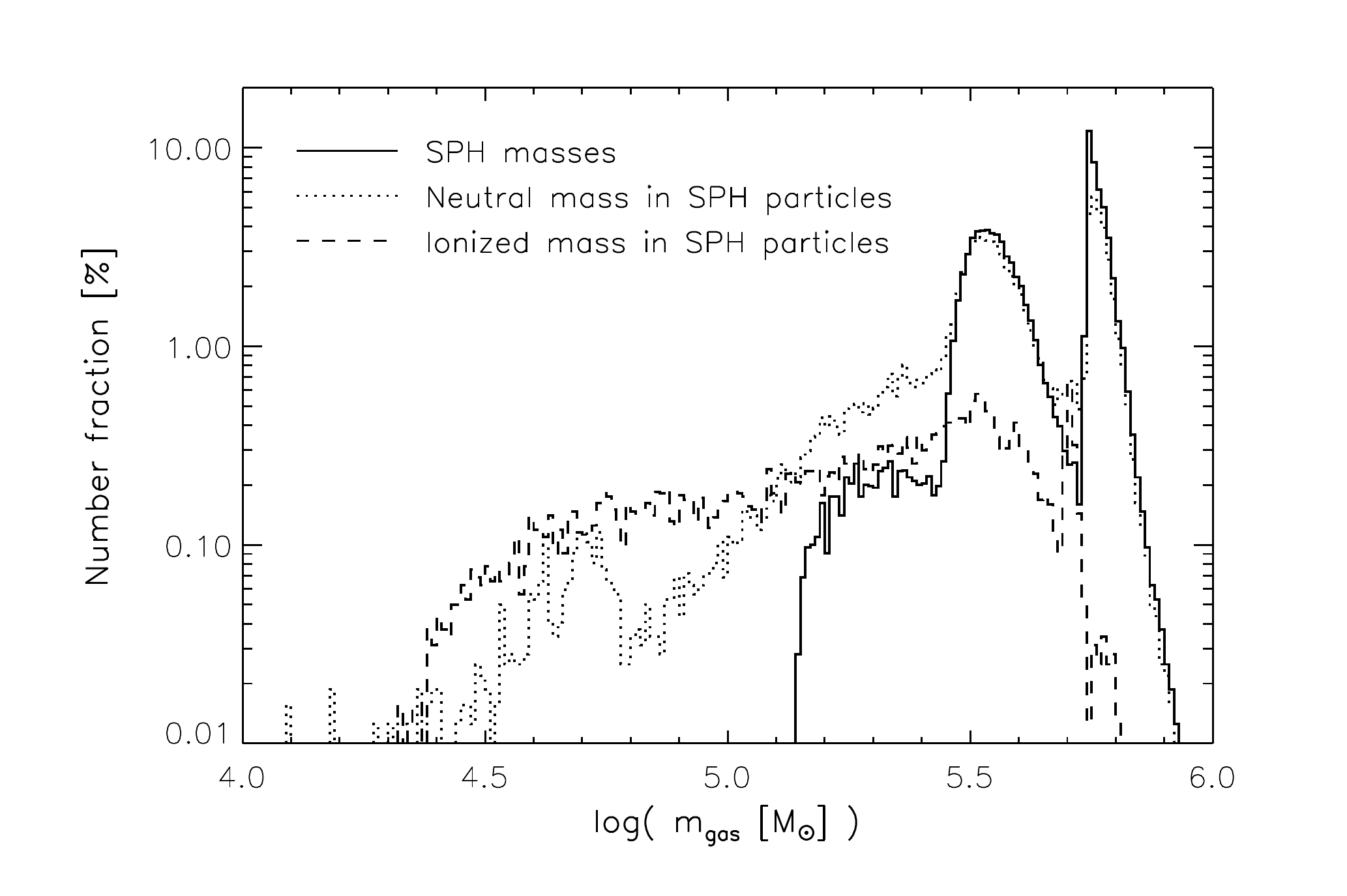}
\caption{The distribution of SPH gas particle masses (solid histogram) in G4.
The distribution peaks at two characteristic masses ($\sim 10^{5.5}\,\msun$ and
$\sim 10^{5.8}\,\msun$), where the lower mass peak represents gas particles left
over from the first generation of stars in the simulation. Splitting the gas
into its neutral and ionized components according to eqs.\ \ref{eq:mwcnm} and
\ref{eq:mhii} results in the mass distribution given by the dotted and dashed
histograms, respectively.}
\label{m_sph}
\end{figure}

As illustrated in Fig.\,\ref{figure:cartoon}, the first step in modeling the
ISM is to split each SPH particle into an ionized and a neutral gas component.
This is done using the electron fraction, $\xe$, associated with each SPH
particle, i.e.,: 
\begin{align}
	m_{\rm neutral}			&=	(1-\xe) \msph	\label{eq:mwcnm} \\
	m_{\rm ionized}			&=	\xe \msph		\label{eq:mhii}.
\end{align}
The electron fraction from \gadget gives the density of electrons relative to
that of hydrogen, $\ne/\nH$, and can therefore reach values of $\sim1.16$ in the
case where helium is also ionized. As a result we re-normalized the distribution
of \xe ~values to a maximal \xe ~of 1 so as to not exceed the total gas mass in
the simulation. Fig.\ \ref{m_sph} shows the distribution of SPH gas particles
masses in G4 -- chosen for its position near the center of the stellar and gas
mass ranges of G1, ..., G7 -- along with the mass distributions of the neutral
and ionized gas components obtained from eqs.\ \ref{eq:mwcnm} and \ref{eq:mhii}.
The ionized gas is seen to have a relatively flat distribution spanning the mass
range $\sim 10^{4.3} - 10^{5.8}\,\msun$. The neutral gas, however, peaks at two
characteristic masses ($\sim 10^{5.5}\,\msun$ and $\sim 10^{5.8}\,\msun$), where
the lower mass peak represents gas particles left over from the first generation
of stars in the simulation.

\subsection{GMCs}\label{split1}
\subsubsection{Masses, sizes and velocity dispersion}\label{split11}
The neutral gas mass, \mneu, associated with a given SPH particle is divided
into GMCs by randomly sampling the GMC mass spectrum as observed in the
Galactic disk and Local Group galaxies: $\frac{dN}{dm_{\rm GMC}} \propto m_{\rm
GMC}^{-\beta}$ with $\beta=1.8$ \citep{blitz07}.  Similar to \citet{narayanan08,narayanan08a}, 
a lower and upper cut in mass of $10^4\,\msun$
and $10^6\,\msun$, respectively, are enforced in order to ensure the GMC masses
stay within the range observed by \cite{blitz07}. Up to 40 GMCs are created per
SPH particle but typically most ($>90\%$) of the SPH particles are split into
four GMCs or less.  While \mneu~never exceeds the upper GMC mass limit
($10^6\,\msun$), there are instances where \mneu~is below the lower GMC mass
limit ($10^4\,\msun$).  In those cases we simply discard the gas, i.e., remove it
from any further sub-grid processing. For the highest resolution simulations in
our sample (G1 and G2), the discarded neutral gas amounts to $\sim6\,\%$ of the
total neutral gas mass, and $\ls 0.02\,\%$ in the remaining five galaxies.  We
shall therefore assume that it does not affect our results significantly.

The GMCs are randomly distributed within $0.2\times$ the smoothing length of the
original SPH particle, but with the radial displacement scaling inversely with
GMC mass in order to retain the original gas mass distribution as closely as
possible. To preserve the overall gas kinematics as best possible, all GMCs
associated with a given SPH particle are given the same velocity as that of the
SPH particle.

GMC sizes are obtained from the pressure-normalized scaling relations for
virialized molecular clouds which relate cloud radius (\rcl) with mass ($m_{\rm
GMC}$) and external pressure (\Pe):
\begin{align}
 	\frac{\rcl}{\text{pc}} =\left( \frac{\Pe/k_{\text{B}}}{10^4\,\cmpc\,\text{K}} \right)^{-1/4} \left (\frac{\mgmc}{290\,\text{\msun}}\right )^{1/2}.	
	\label{Pe_size}
\end{align}
We assume  $\Pe = P_{\rm tot}/(1+\alpha_0+\beta_0)$ for relative cosmic and
magnetic pressure contributions of $\alpha_0=0.4$ and $\beta_0=0.25$
\citep{elmegreen89b}. For the total pressure ($P_{\rm tot}$) we adopt the
external hydrostatic pressure at mid-plane for a rotating disk of gas and stars,
i.e.,:
\begin{equation}
	P_{\rm tot}\approx\frac{\pi}{2}\rm{G}\gassd \left[ \gassd+\left( \frac{\sigma_{\rm gas,\perp}}{\sigma_{\rm *,\perp}} \right) \starssd\right], 
\end{equation}
where \gassd\ and \starssd\ are the local surface densities of gas and stars,
respectively, and $\sigma_{\rm gas,\perp}$ and $\sigma_{\rm *,\perp}$ their
local velocity dispersions measured perpendicular to the mid-plane (see e.g.,
\cite{elmegreen89b,swinbank11}). For each SPH particle, all of these quantities
are calculated directly from the simulation output (using a radius of $R=1$\,kpc
from each SPH particle), and it is assumed that the resulting $\Pe$ is the
external pressure experienced by all of the GMCs generated by the SPH particle.
We find that GMCs in our simulated galaxies are subjected to a wide range of
external pressures ($\Pe/k_{\rm B} \sim10^2-10^7\,{\rm cm^{-3}\,K}$). For
comparison, the range of pressures experienced by clouds in our Galaxy and in
Local Group galaxies is $\Pe/k_{\rm B} \sim10^3-10^7\,{\rm cm^{-3}\,K}$ with an
average of $P_{\rm ext}/k_{\rm B} \sim10^4$\,\cmpc\,K in Galactic clouds
\citep{elmegreen89b,blitz07}. This results in GMC sizes in our simulations
ranging from $\rcl=1 - 300\,{\rm pc}$. 

The internal velocity dispersion ($\sigma_v$) of the GMCs is inferred from the virial
theorem, which provides us with a pressure-normalized $\sv-R_{\rm GMC}$
relation:
\begin{equation}
	\sigma_v = 1.2\,\kms \left( \frac{\Pe/k_{\rm B}}{10^4\,\cmpc\,{\rm K}}\right)^{1/4}\left ( \frac{\rcl}{{\rm pc}}\right )^{1/2},
	\label{sigma_v_Pe}
\end{equation}
where the normalization of $1.2\,{\rm km\,s^{-1}}$ comes from studies of Galactic GMCs
\citep[][]{larson81,elmegreen89b,swinbank11}.

\subsubsection{GMC density and temperature structure} \label{split12}
\begin{figure}[htbp] 
\hspace*{-0.5cm}
\includegraphics[width=1.1\columnwidth]{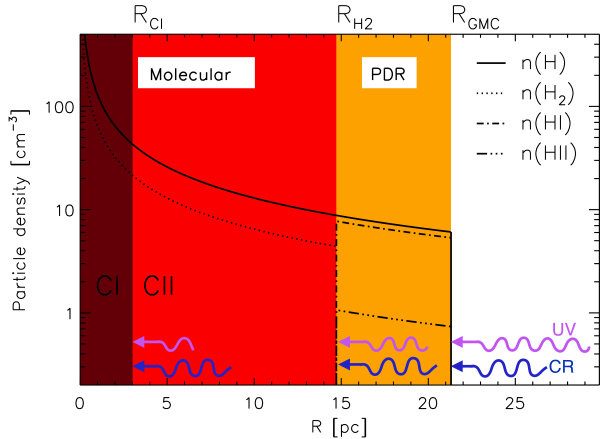}
\caption{Example H number density profile (solid line) for a GMC of mass $m_{\rm
GMC}=1.3\times 10^{4}$\,\msun ~and radius $R_{\rm GMC}=21$\,pc. Also shown are
the density profiles of H$_2$ ($=0.5 n_{\rm H}$), H{\sc i} and H{\sc ii}. Note,
the transition from molecular to atomic H is assumed to happen instantaneously
at $R_{\rm H_2}$. The total C abundance follows that of H but scaled with the
[C/H] abundance (as provided by the `parent' SPH particle, see Section
\ref{section:methodology}).  C{\sc ii} can exist throughout the cloud except
for the very inner region ($R < R_{\rm CI}$; indicated in brown), with the C{\sc
i} to C{\sc ii} transition happening instantaneously at $R_{\rm CI}$. \cii emission
from the layer $R_{\rm CI}\le R \le R_{\rm H_2}$ (indicated in red) is referred
to as `molecular' emission, while \cii emission from $R_{\rm H_2}< R \le R_{\rm
GMC}$ (indicated in orange) is referred to as `PDR' emission.  The
relative thickness of these layers is set by the impinging FUV radiation field
and cosmic rays (illustrated as purple and blue arrows, respectively), with the former
undergoing attenuation further into the cloud.}
\label{onion}
\end{figure}

We assume a truncated logotropic profile for the total hydrogen number density
of the GMCs, i.e.,: 
\begin{align}
	\nH(R)			=	\next\left( \frac{\rcl}{R} \right), 	
	\label{eq:nH}
\end{align}
where $\nH(R>\rcl)=0$. For such a density profile it can be shown that the
external density, \next, is 2/3 of the average density:
\begin{align}
	\next			=	2/3\nav=2/3\frac{\mgmc}{4/3\pi\mH \rcl^3}.
	\label{eq:next}
\end{align}
While the total hydrogen density follows a logotropic profile, the transition from
H$_2$$\longrightarrow$H{\sc i}/H{\sc ii} is assumed to be sharp. Similarly for
the transition from C{\sc i}$\longrightarrow$C{\sc ii}. This is illustrated in
Fig.\,\ref{onion}, which shows an example density profile of a GMC from our simulations.
From the center of the GMC and out to \rh2, hydrogen is in molecular form.
Beyond \rh2, hydrogen is found as \hi and \hii out to \rcl. 

The size of the molecular region, when adopting the logotropic density profile,
is related to the total molecular gas mass fraction (\fmol) of each
GMC:
\begin{align}
	\fmol	&=	\frac{\mmol}{\mgmc}
			=	\frac{\int_0^{R_{\rm H_2}} \rho_{\rm ext}\frac{R_{\rm GMC}}{R} 4\pi R^2 dR}
			{\int_0^{R_{\rm GMC}}\rho_{\rm ext}\frac{R_{\rm GMC}}{R} 4\pi R^2 dR}\\
			&=	\frac{\int_0^{R_{\rm H_2}} R \,dR}{\int_0^{R_{\rm GMC}} R \,dR}
			=	\left( \frac{R_{\rm H_2}}{R_{\rm GMC}} \right)^2\\
	&\Rightarrow \rrh2	=	\sqrt{\fmol}, \label{eq:fmol}
\end{align}
where \rrh2 is the fractional radius, $\rrh2=\rh2/\rcl$.
We find \fmol by assuming $\hi\leftrightarrow\h2$ equilibrium and using
the analytical steady-state approach of \cite{pelupessy06} for inferring
\fmol for a logotropic cloud
subjected to a radially incident FUV radiation field \citep[see also][]{olsen15}.  
In this framework, the value of \fmol\ (and thereby $\rrh2$)
depends on:
\begin{enumerate}
\item The cloud boundary pressure (\Pe), which is calculated as
explained in Section\,\ref{split11}.

\item The metallicity (\Z) of the GMC, which is inherited from
its parent SPH particle and assumed constant throughout the cloud.

\item The kinetic temperature of the gas at the GMC surface
($T_k(R_{\rm GMC})$). This quantity is calculated in an iterative process
together with \fmol\ by solving the following thermal balance equation:
\begin{equation}
 	\Hpe+\Hcrhi = \Ccii+\Coi \label{Tk_rcl}, 
\end{equation}
where \Hpe is the heating rate associated with the photoelectric ejection of
electrons from dust grains by the FUV field and \Hcrhi is the heating rate by
cosmic rays in atomic gas. The main cooling agents are assumed to be due to \cii
and [O{\sc i}]($63\,{\rm \mu m}$ and $145\,{\rm \mu m}$) line emission (i.e.,
\Ccii and \Coi, respectively). \Hpe, \Hcrhi, and \Ccii all depend on the
electron fraction at $R_{\rm GMC}$, which is determined by the degree of H{\sc i}
ionization by the local FUV radiation field and CR ionization rate 
(see below for how these quantities are derived). For analytical expressions for
the heating rates, we refer to \cite{olsen15}. For $\Lambda_{\rm
O\textsc{i}}$ we use the expressions given in \citet{rollig06}. The
calculation of \Ccii at the GMC surface is detailed in Section \ref{cii_em} and
Appendix \ref{apD}.

\item The strength of the local FUV radiation field (\g0) and the
CR ionization rate (\cri) impinging on the GMCs.  These quantities do not come
out from the simulation directly, and instead they are calculated by scaling the
Galactic FUV field ($G_{\rm 0,MW}$) and CR ionization rate ($\zeta_{\rm CR,MW}$)
with the local SFR volume density in the simulations, i.e., $G_{\rm 0} \propto
G_{\rm 0,MW} \left ({\rm SFRD}_{\rm local}/{\rm SFRD}_{\rm MW}\right )$ and
$\zeta_{\rm CR} \propto \zeta_{\rm CR,MW} \left ({\rm SFRD}_{\rm local}/{\rm
SFRD}_{\rm MW}\right )$, where ${\rm SFRD_{\rm local}}$ is estimated for each
SPH particle as the volume averaged SFR within a $2\,{\rm kpc}$ radius.  We have
adopted Milky Way values of $G_{\rm 0,MW} = 0.6\,{\rm Habing}$ \citep{seon11}
and $\zeta_{\rm CR, MW} = 3\times 10^{-17}\,{\rm s^{-1}}$ \citep{webber98}.  For
${\rm SFRD}_{\rm MW}$ we adopt $0.0024\,{\rm \msun\,yr^{-1}\,kpc^{-3}}$,
inferred from the average Galactic SFR ($0.3\,{\rm \msun\,yr^{-1}}$) within a
disk $10\,{\rm kpc}$ in radius and $0.2\,{\rm kpc}$ in height
\citep{heiderman10,bovy12}.

\item The electron fraction at the cloud boundary. This fraction is
not the previously introduced \xe, which was inherent to the SPH simulations and
used to split the SPH gas into a neutral and ionized gas phase. Instead, it is
the electron fraction given by the degree of ionization of H{\sc i} caused by
the \g0 and \cri impinging on the cloud.  This fraction (and thus the H{\sc
i}:H{\sc ii} ratio) is calculated with \texttt{CLOUDY} v13.03 \citep{ferland13}
given the hydrogen density and temperature at the cloud boundary and assuming an
unattenuated \g0 and \cri at $R_{\rm GMC}$.

\end{enumerate}

\bigskip

The \cii emitting region in each of our GMCs is defined as the layer between the
surface of the cloud and the depth at which the abundances of C and C$^+$ are
equal. Hence, if the latter occurs at a radius $\rci$ from the cloud center, the
thickness of the layer is $\rcl - \rci$ (Fig.\ \ref{onion}). At radii $< \rci$,
all carbon atoms are for simplicity assumed to be in neutral form. In order to
determine the fractional radius \rrci ($=R_{\rm CI}/R_{\rm GMC}$) we follow the
work of \cite{rollig06} \citep[but see also][]{pelupessy09}, who considers the
following dominant reaction channels for the formation and destruction of C$^+$:
\begin{align}
    \rm{C} +\gamma             &\rightarrow    \rm{C}^+ + {\it e^-}  \label{CIIform1} \\
    \rm{C}^+ +{\it e^-}           &\rightarrow    \rm{C} + \gamma  \label{CIIform2} \\
    \rm{C}^+ +\rm{H}_2      &\rightarrow    \rm{CH}^+_2 + \gamma.   \label{CIIform3}
\end{align}
In this case, \rrci can be found by solving the following equation:
\begin{align}
    5.13\times10^{-10}&{\rm s^{-1}}\g0\int_{1}^{\infty}\frac{e^{-\mu \xi_{\rm FUV}A_{\rm V}(\rrci)}}{\mu^2}d\mu \label{eq:R06} \\
    =&~n_{\rm H}(r_{\rm C{\textsc i}})[a_{\rm C}\Xc+0.5k_{\rm C}], \nonumber
\end{align}
where the left-hand side is the C$^+$ formation rate due to photo-ionization by
the attenuated FUV field at \rrci (eq.\ \ref{CIIform1}), and the right-hand side
is the destruction rate of C$^+$ due to recombination and radiative association
(eqs.\ \ref{CIIform2} and \ref{CIIform3}). The constants $a_{\rm
C}=3\e{-11}$\,\cmpc\ps ~and $k_{\rm C}=8\e{-16}$\,\cmpc\ps ~are the
recombination and radiative association rate coefficients.  Note, we have
accounted for an isotropic FUV field since $\mu = \cos \theta$, where $\theta$
is the angle between the Poynting vector and the normal direction.  $A_{\rm
V}(\rrci)$ is the visual extinction corresponding to the $\rcl - \rci$ layer,
and is given by $A_{\rm V}(\rrci) = 0.724\sigma_{\rm dust}\Z \langle n_{\rm
H}\rangle\rcl \ln\left( \rrci^{-1} \right)$, where $\sigma_{\rm dust} =
4.9\e{-22}\,\cms$ is the FUV dust absorption cross section
\citep{pelupessy09,mezger82}. $\xi_{\rm FUV}$ accounts for the difference in
opacity between visual and FUV light and is set to $3.02$.  \Xc~is the carbon
abundance relative to H and is calculated by adopting the carbon mass fractions
of the parent SPH particle (self-consistently calculated as part of the overall
SPH simulation) and assuming it to be constant throughout the GMC.  
%
%
\begin{figure}[htbp] 
\hspace*{-0.5cm}
\includegraphics[width=1.11\columnwidth]{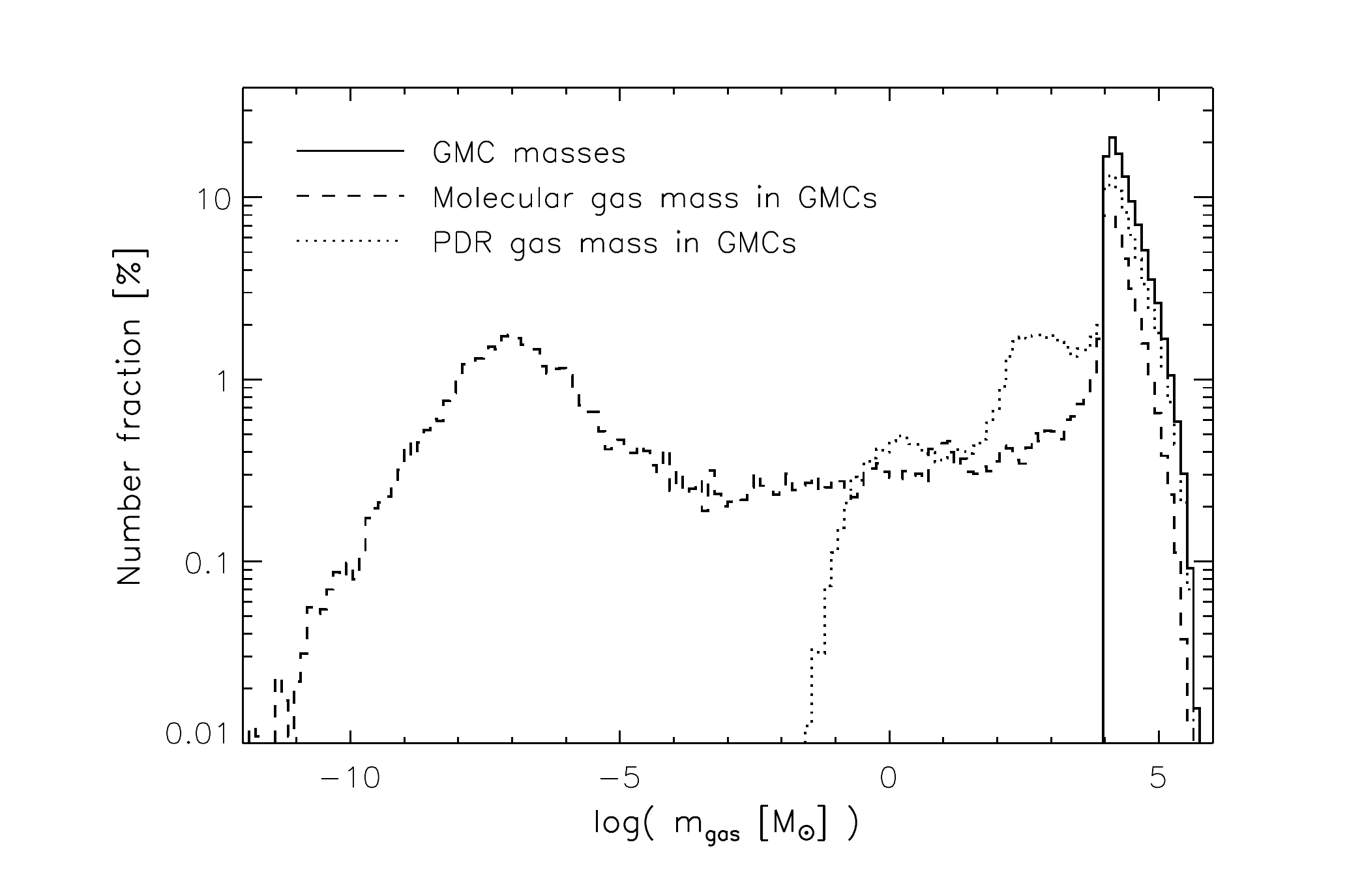}
\hspace*{-0.5cm}
\includegraphics[width=1.11\columnwidth]{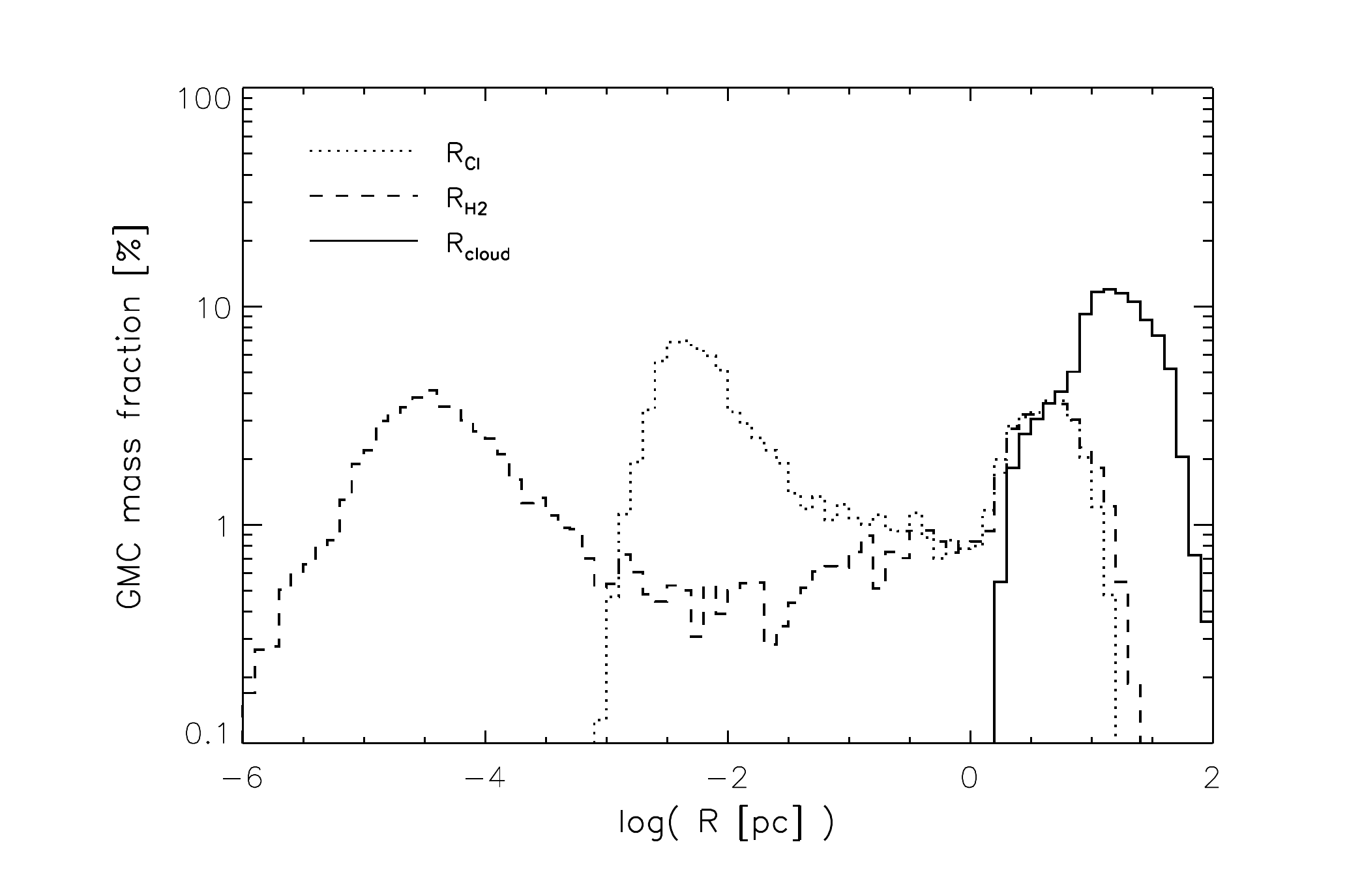}
\hspace*{-0.5cm}
\includegraphics[width=1.11\columnwidth]{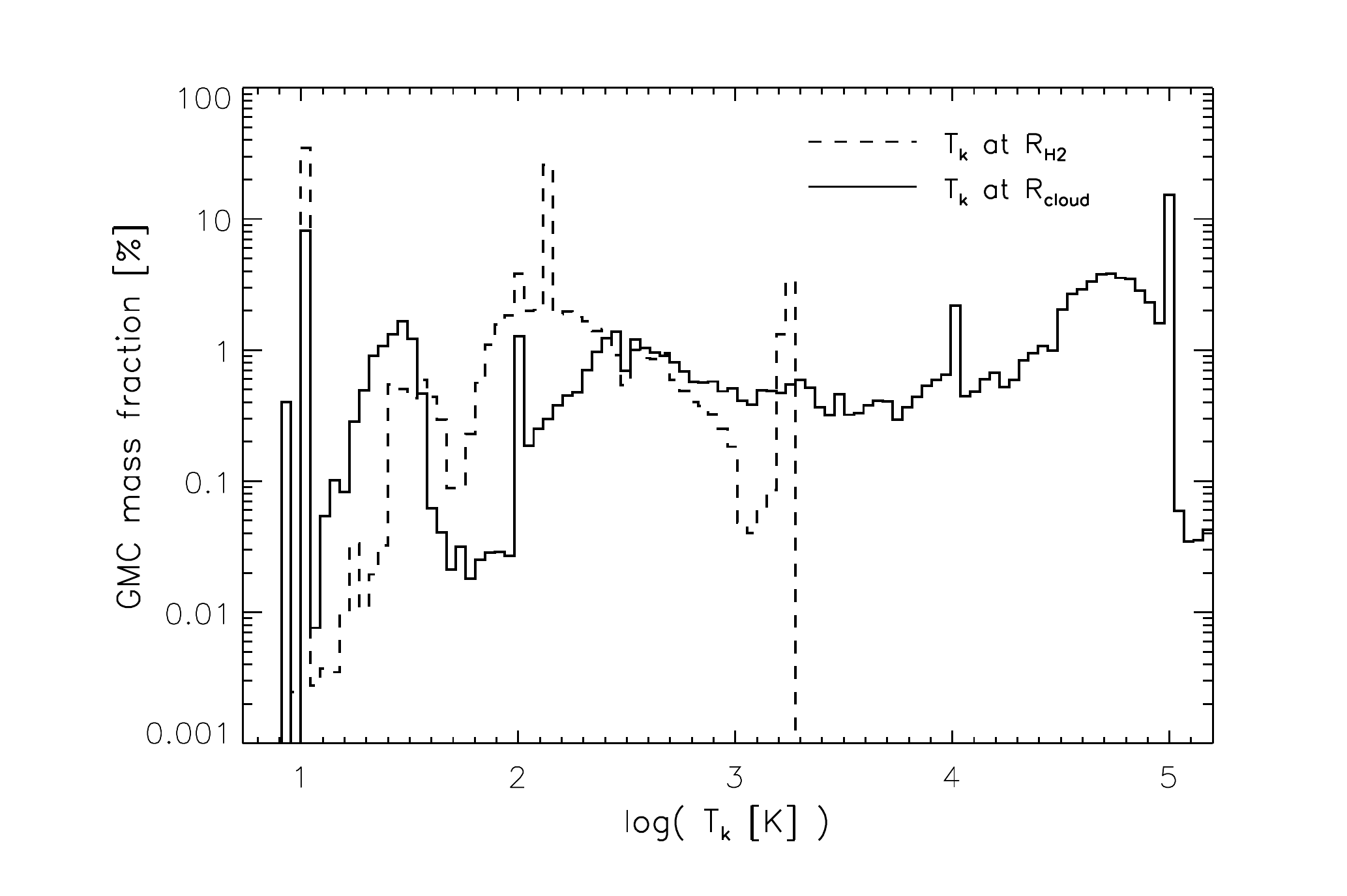}\\
\caption{{\bf Top:} the number distribution of GMC masses (solid histogram)
together with the number distributions of the molecular (dashed histogram) and
PDR (dotted histogram) gas component.  {\bf Middle:} GMC-mass-weighted
distributions of \rci (dotted), \rh2 (dashed) and \rcl for GMCs in G4. {\bf Bottom:}
GMC-mass-weighted distributions of the temperature at \rh2 (dashed) and 
at the cloud surfaces (solid) for GMCs in G4.}
\label{figure:R-T-distributions-GMCs}
\end{figure}

\bigskip

The temperature of the \cii-emitting molecular gas (i.e., from the gas layer
between \rci and \rh2) is assumed to be constant and equal to the temperature at
\rh2. The latter is given by the thermal balance:
\begin{equation}
   	\Hpe+\Hcrh2 =\Ch2+\Ccii+\Coi,
	\label{Tk_rH2}
\end{equation}
where \Hcrh2 ~is the CR heating in molecular gas, and \Ch2 is the
cooling due to the S(0) and S(1) rotational lines of H$_2$ (\citet{papa14}; see
also \citealt{olsen15}). \Hcrh2 depends on \xe, which is calculated with
\texttt{CLOUDY} for the local CR ionization rate and the attenuated FUV field at
\rrh2, i.e.\ $\g0 e^{-\xi_{\rm FUV}A_{\rm V}(\rrh2)}$. $A_{\rm V}(\rrh2)$
corresponds the extinction through the outer atomic transition layer of each
GMC. 

The temperature of the PDR gas (i.e., from the gas between \rh2 and \rcl)
is assumed to be constant and equal to the temperature at \rcl.

\bigskip

For each GMC we solve in an iterative manner simultaneously for \rh2 (eq.\
\ref{eq:fmol}) and \rci (eq.\ \ref{eq:R06}), as well as for the gas temperature
at \rcl and at \rh2 (eqs.\ \ref{Tk_rcl} and \ref{Tk_rH2}, respectively)

The resulting distributions of \rh2 and \rci for the GMC population in G4 are
shown in Fig.\,\ref{figure:R-T-distributions-GMCs} (middle panel), along with
the distribution of $R_{\rm GMC}$ (obtained from eq.\ \ref{Pe_size}).  In GMCs
in general, we expect $R_{\rm C\textsc{i}} < R_{\rm H_2}$, due to efficient
H$_2$ selfshielding.  However, as Fig.\ \ref{figure:R-T-distributions-GMCs}
shows some of the GMCs in our simulations have very small \rh2 (due to them
having virtually zero molecular gas fractions), and in those cases \rci can be
equal to or even exceed \rh2. The latter implies that the \cii emission is only
coming from the PDR phase, with no contribution from the molecular gas.

The distributions of the kinetic temperature of the gas at \rh2 and \rcl for the
GMC population in G4 are also shown in Fig.\,\ref{figure:R-T-distributions-GMCs}
(bottom panel). The temperatures at the cloud surfaces range from $\sim 8\,{\rm
K}$ to $\sim10^{5.5}\,{\rm K}$, and the temperatures at \rh2 lies between $\sim
8$ and $1800\,{\rm K}$.  

With \rh2 and \rci determined for each GMC we can calculate the gas masses
associated with the molecular and PDR gas phase, respectively. The
resulting mass distributions are shown in the top panel of
Fig.\,\ref{figure:R-T-distributions-GMCs}, along with the distribution of total
GMC masses ($m_{\rm GMC}$) as determined by the adopted GMC mass spectrum
(Section \ref{split11}). We see that most of the molecular and PDR gas
masses follow the total GMC mass spectrum. There is, however, a fraction of GMCs
with extremely small molecular gas masses (corresponding to $\fmol\sim 0$).  On
the other hand, there are some GMCs with small PDR gas masses, i.e.,
clouds that are so shielded from FUV radiation that they are almost entirely
molecular.

\subsection{The ionized gas} \label{split2}
\begin{figure}[htbp] 
\hspace*{-0.5cm}
\includegraphics[width=1.11\columnwidth]{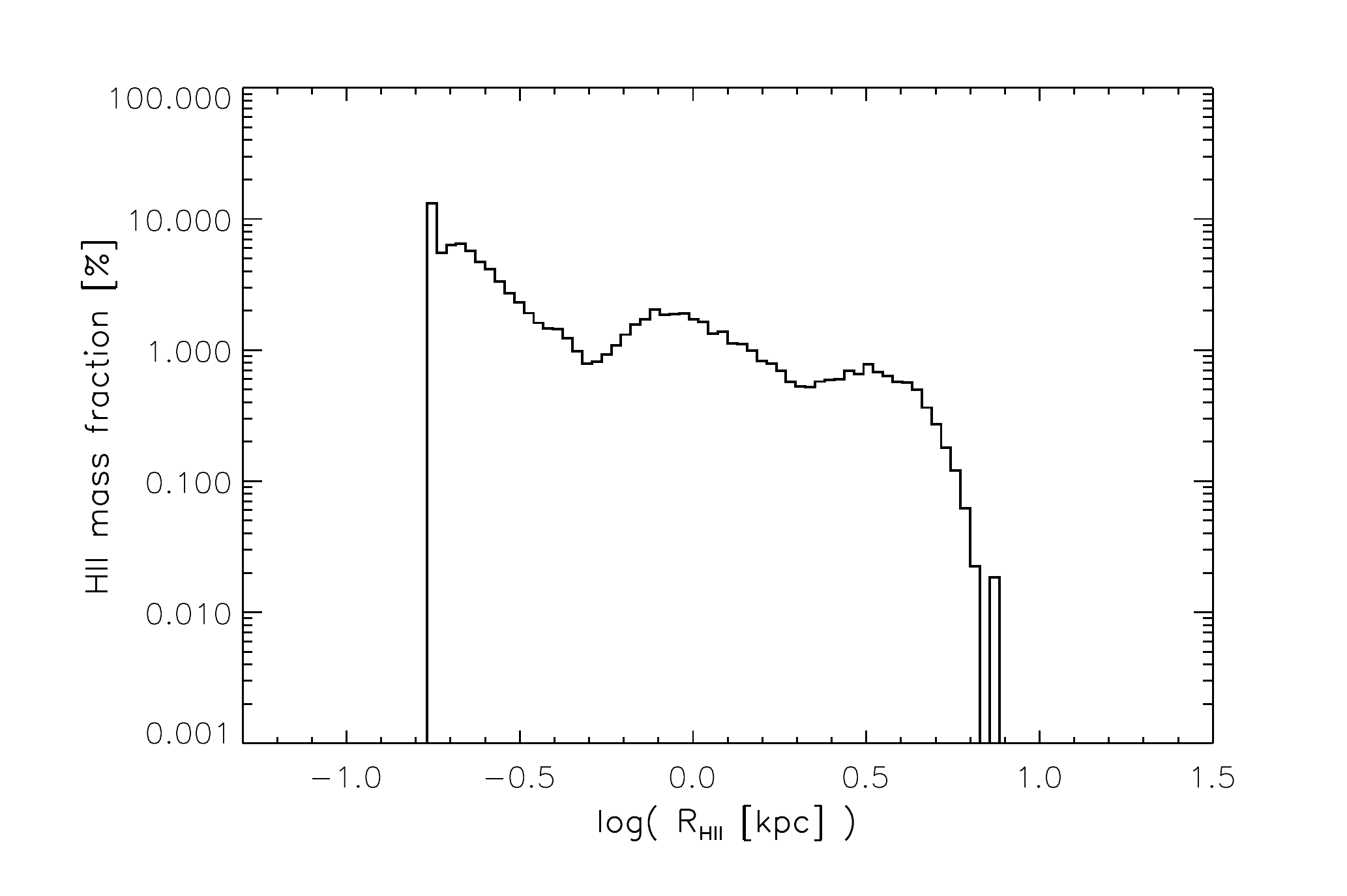}
\hspace*{-0.5cm}
\includegraphics[width=1.11\columnwidth]{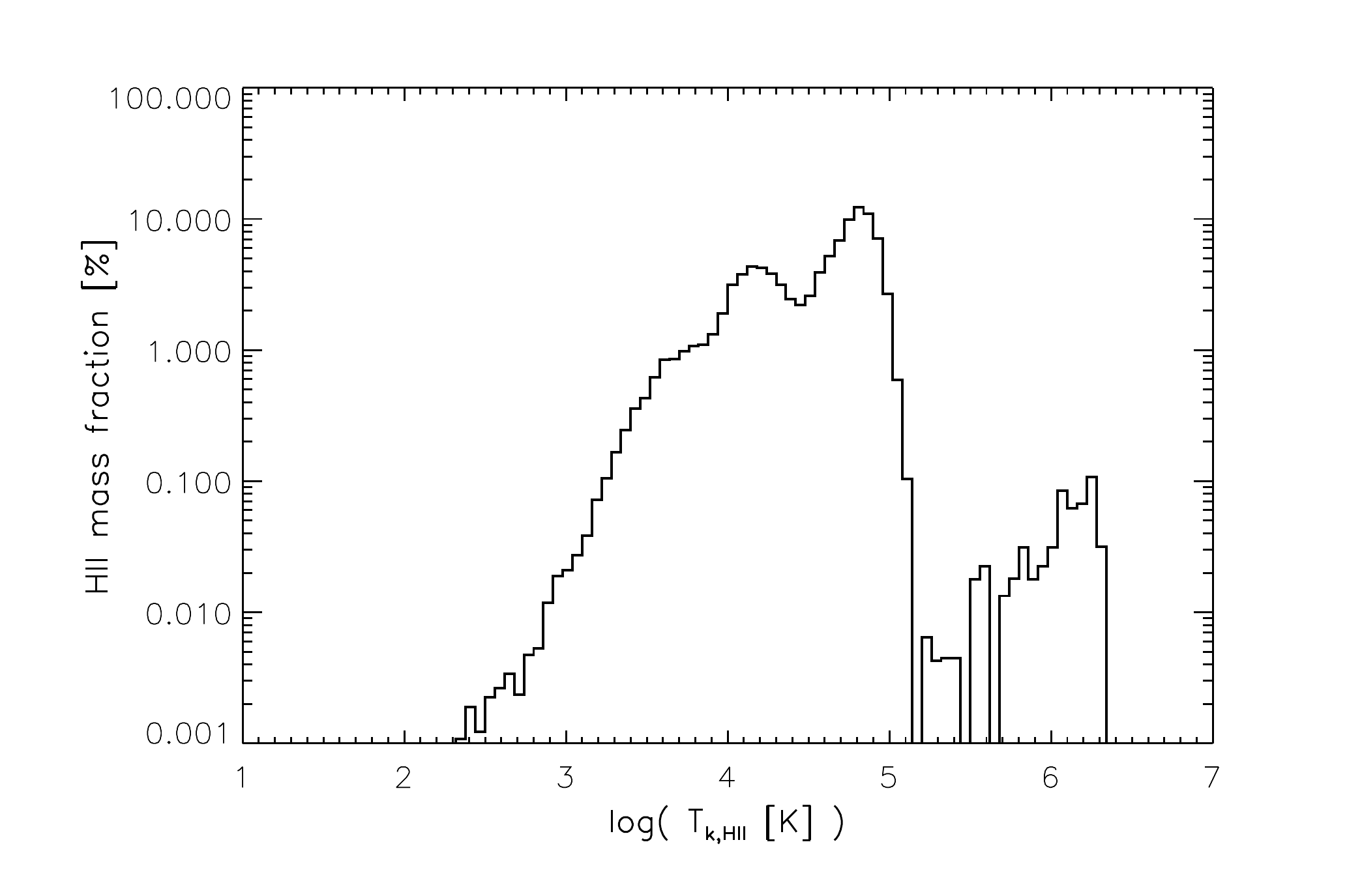}
\caption{Mass-weighted histograms of the size (top) and temperature (bottom)
distributions of the ionized clouds in G4.}
\label{figure:R-T-distributions-HII}
\end{figure}

The ionized gas in our simulations (see eq.\ \ref{eq:mhii}), is assumed to be
distributed in spherical clouds of uniform densities and with radii ($R_{\rm
H\textsc{ii}}$) equal to the smoothing lengths of the original SPH particles.
These ionized regions in our simulations are furthermore assumed to be
isothermal with the temperatures equal to that of the SPH gas.  Fig.\
\ref{figure:R-T-distributions-HII} shows the size (top) and temperature (bottom)
distribution for the ionized clouds in G4. Cloud sizes range from $\sim
0.1$ to $\sim 10\,{\rm kpc}$, with more than $60\,\%$ of the ionized gas
mass residing in clouds of size $\ls 1000\,{\rm pc}$. For comparison, the range
of observed sizes of \hii clouds in nearby galaxies is $10-1000\,{\rm pc}$
\citep{oey97,hodge99}.  The temperatures range from $\sim 10^2\,{\rm K}$ to
$\sim 10^6\,{\rm K}$ with the bulk of the ionized gas having temperatures $\sim
10^{4-5}\,{\rm K}$.

\section{The \cii\ line emission}\label{cii_em}
The \cii\ luminosity of a region of gas is the volume-integral of the effective \cii\
cooling rate per volume, i.e.,:
\begin{align}
	\Lcii			=	\int_{\rm \Delta V} \Ccii \,dV.
	\label{eq:Lcii}
\end{align}
Since we have adopted spherical symmetry in our sub-grid treatment of the clouds
(both neutral and ionized), we have: 
\begin{align}
	\Lcii 	&= 4\pi\int_{R_1}^{R_2} \Ccii R^2 dR,
	\label{eq:CII-integral}
\end{align}
where $R_1 = R_{\rm C\textsc{i}}$ and $R_2 = R_{\rm H_2}$ for the molecular
phase; $R_1 = R_{\rm H_2}$ and $R_2 = R_{\rm GMC}$ for the PDR region, and $R_1
= 0$ and $R_2 = R_{\rm H\textsc{ii}}$ for the ionized gas.

The effective \cii cooling rate is: 
\begin{align}
	\Ccii			=	A_{\rm ul}\beta f_{\rm u}\ncii h \nu,
	\label{eq:ccii}
\end{align}
where $A_{\rm ul}$ ($=$2.3\e{-6}$\,\ps$)\footnote{Einstein coefficient for
spontaneous emission taken from the LAMDA database:
\url{http://www.strw.leidenuniv.nl/~moldata/}, \cite{schoier05}} is the Einstein
coefficient for spontaneous decay. We ignore the effects of any background
radiation field from dust and the cosmic microwave background (CMB). $\beta$ is the \cii photon escape
probability for a spherical geometry (i.e., $\beta = (1 - exp(-\tau))/\tau$,
where $\tau$ is the \cii optical depth). $f_u$ is the fraction of singly ionized
carbon in the upper $^2P_{3/2}$ level and is determined by radiative processes
and collisional (de)excitation (see Appendix \ref{apD}).  The latter can occur
via collisions with $e^-$, H{\sc i}, or H$_2$, depending on the state of the
gas. In our simulations the collisional partner is H$_2$ in the molecular phase,
H{\sc i} and $e^-$ in the PDR regions, and $e^-$ in the ionized gas.  Analytical
expressions for the corresponding collision rate coefficients as a function of
temperature are given in Appendix \ref{apD}.  \ncii\ is the number density
of singly ionized carbon and is given by $\ncii = X_{\rm C} f_{\rm C\textsc{ii}}
n_{\rm H}$, where $f_{\rm C\textsc{ii}}$ is the fraction of carbon atoms in the
singly ionized state.  For the latter we used tabulated fractions from
\texttt{CLOUDY} v13.03 over a wide range in temperature, hydrogen density, FUV
field strength, and CR ionization rate.

We calculate the integral in eq.\ \ref{eq:CII-integral} numerically by splitting
the $R_2 - R_1$ region up into 100 radial bins. In each bin, $n_{\rm H}$ is set
to be constant and -- in the case of the molecular and PDR regions -- given by
the logotropic density profile at the radius of the given bin
(Fig.\,\ref{onion}). For the ionized clouds, $n_{\rm H}$ is constant throughout
(Section \ref{split2}). For the PDR regions (i.e., from $R_{\rm H_2}$ to $R_{\rm
GMC}$) we assume that the temperature, electron fraction and \g0 are kept fixed
to the outer boundary value at $R_{\rm GMC}$ (i.e., no attenuation of the FUV
field). This implies that the \cii luminosity from the PDR gas is an upper
limit. Similar for the \cii emission from the molecular region (i.e., from
$R_{\rm C\textsc{i}}$ to $R_{\rm H_2}$), where we assume that the temperature
throughout this region is fixed to its values at $R_{\rm H_2}$. Also, throughout
this region we adopt the attenuated FUV field at $R_{\rm H_2}$.

\section{Results and discussion}
Having divided the ISM in our galaxies into molecular, atomic and ionized gas
phases, and having devised a methodology for calculating their \cii\ emission,
we are now in a position to quantify the relative contributions from the
aforementioned gas phases to the total \cii emission, and examine their
relationship to the on-going star formation.
\begin{figure*}[t!] 
  \begin{center}
  \includegraphics[width=0.98\textwidth]{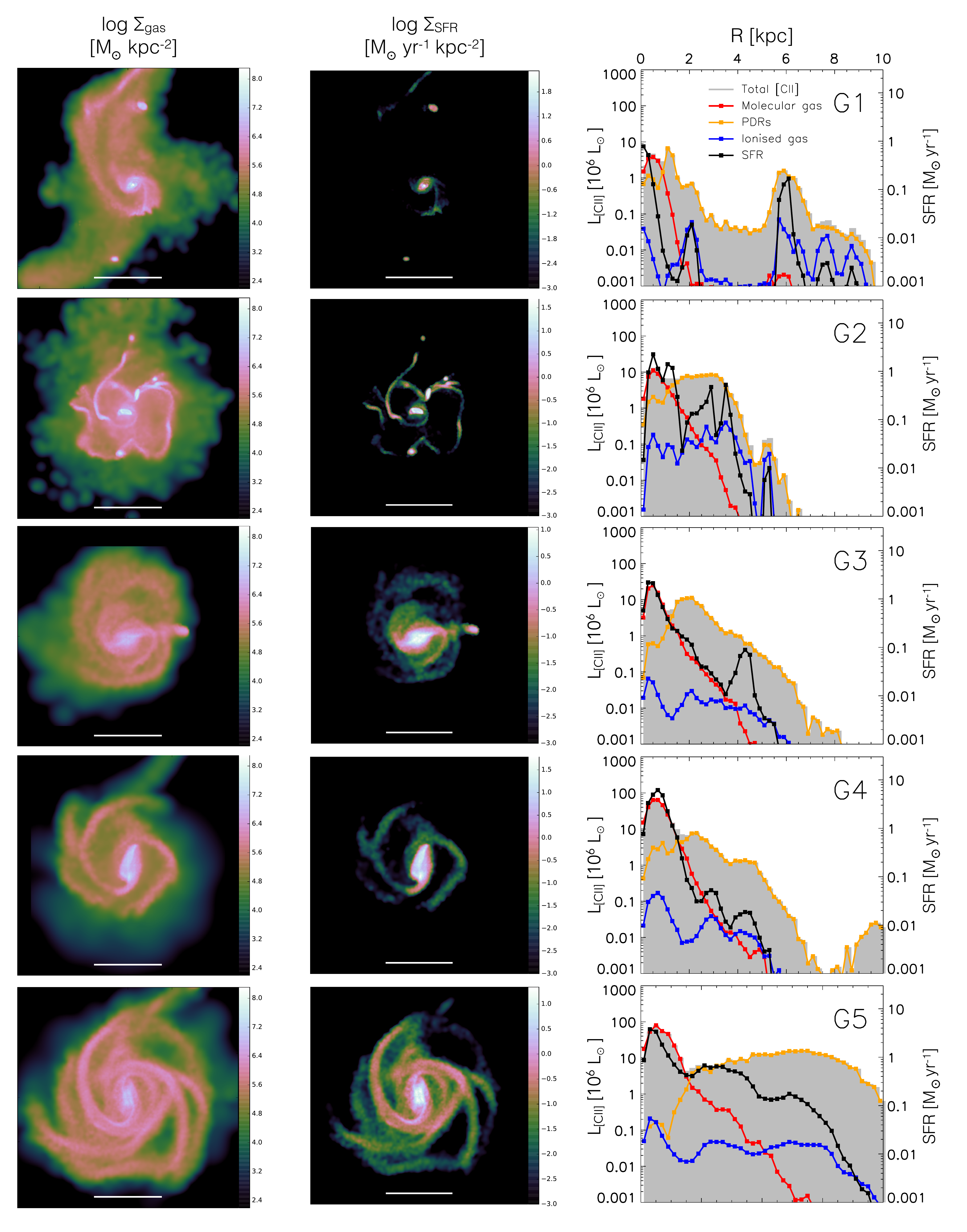}
  \caption{Gas and SFR surface density maps (left and middle columns,
  respectively) of our SPH simulated galaxies viewed face-on (G1, ..., G7 from
  top to bottom; see also \citet{thompson15}).  The horizontal white bars
  correspond to a physical scale of $5\,{\rm kpc}$. The right-hand column shows
  the radial profiles of the total \cii luminosity (gray histogram) and the
  contributions from molecular gas (red curve), PDR (orange curve) and
  ionized gas (blue curve). The SFR radial profiles are also shown (black
  curve).  The radial profiles were determined by summing up the \cii luminosity
  and SFR within concentric rings with fixed width of $0.2\,{\rm kpc}$.}
 \label{fig:maps1}
 \end{center}
\end{figure*}
\setcounter{figure}{6}
\begin{figure*}[t!]
\begin{center}  
  \includegraphics[width=0.98\textwidth]{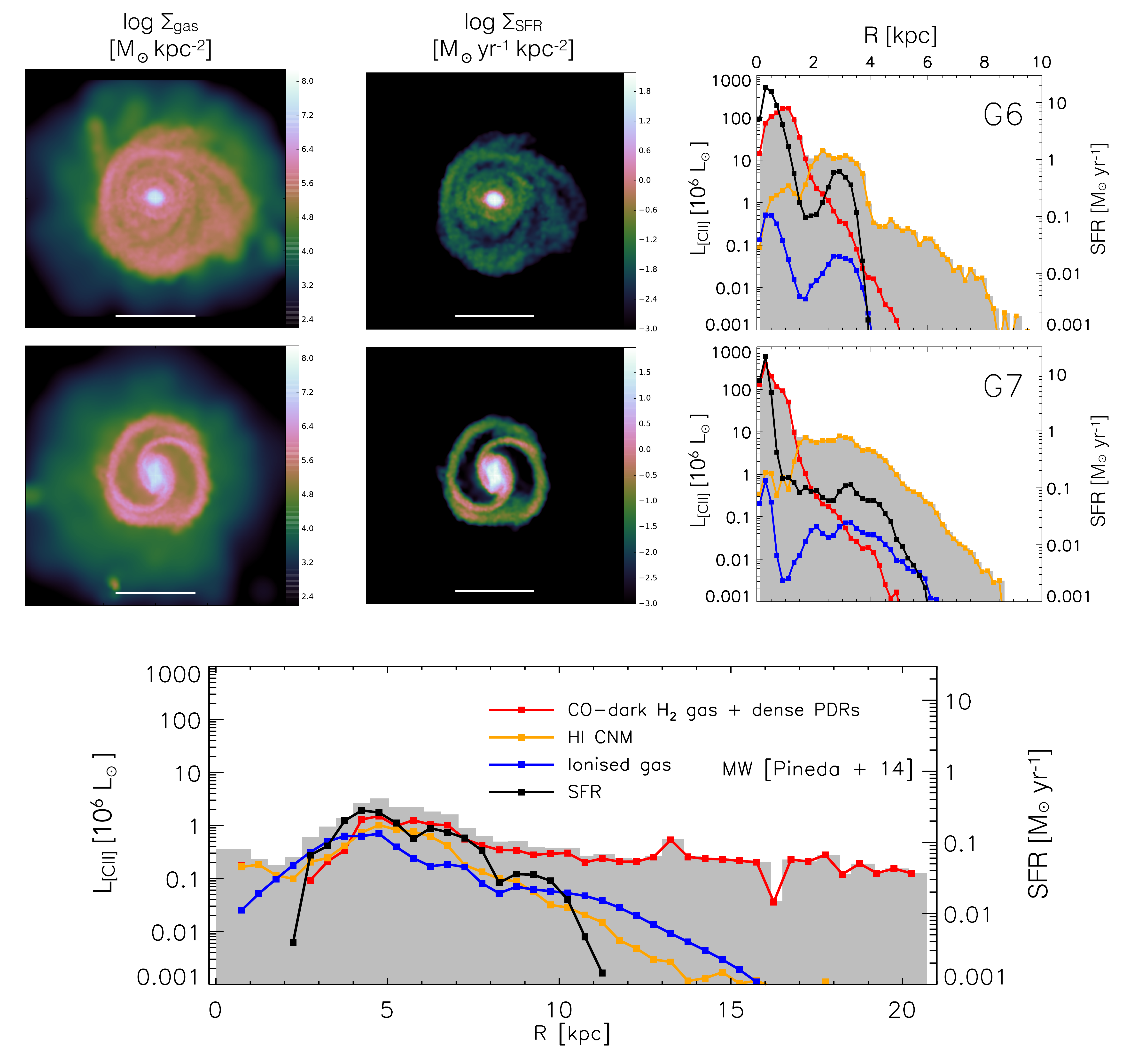}
  \caption{Continued. In the bottom panel we show for reference the radial \Lcii
  profile of the Galaxy (gray histogram) along with the contributions from
  CO-dark H$_2$ gas and dense PDRs (red curve), cold H\,{\sc i} (orange curve),
  and ionized gas (blue curve) \citep{pineda14}. Also shown is the radial SFR
  profile (black curve) inferred from the $1.4\,$GHz intensity distribution.
  Fixed radial bin widths of $0.5\,{\rm kpc}$ have been adopted (see
  \citet{pineda14} for details).}
  \label{fig:maps2}
\end{center}
\end{figure*}

\subsection{Radial \cii luminosity profiles}\label{subsection:CII-profiles}
First, however, to get a sense of the distribution of gas and star formation in
our simulated galaxies, we show in Fig.\ \ref{fig:maps1} surface density maps of
the total SPH gas (left column) and star formation rate (middle column) when
viewed face-on. The maps reveal spiral galaxy morphologies, albeit with some
variety: some (G1, G2 and G3) show perturbed spiral arms due to on-going mergers
with satellite galaxies; others (G4, G5, G6 and G7) have seemingly undisturbed,
grand-design spiral arms; a central bar-like structure is also seen in some (G2,
G4, G5 and G7). Overall, the star formation is seen to be much more centrally
concentrated than the SPH gas. This is especially true for G1 and G2, which have
very centrally peaked star formation. The radial SFR and \cii luminosity
profiles of G1, ..., G7 -- derived by summing up the SFR and the \cii luminosity
within concentric rings (of fixed width: $0.2\,{\rm kpc}$) -- are also shown in
Fig.\,\ref{fig:maps1}.  We have inferred the radial \cii luminosity distribution
for the full ISM as well as for the individual gas phases.

The radial SFR profiles of our model galaxies typically peak at $R\sim0.5\,{\rm
kpc}$ (in G1 at $R\ls 0.5\,{\rm kpc}$) and then tail off with radius.  In some
cases, local peaks in the star formation activity occur at galactocentric
distances $\gs 2\,{\rm kpc}$, corresponding to the locations of either satellite
galaxies (G2 and G3) or spiral arms (G5 and G7). 

The total \cii luminosity profiles (gray histograms) also peak at
$R\sim0.5\,{\rm kpc}$, and within the central $R\ls 1.5\,{\rm kpc}$ there is in
general a good correspondence between the total \cii emission and the star
formation activity. This correspondence is driven by the molecular gas phase
which dominates the \cii emission in the central regions. The molecular \cii
emission is seen to correlate strongly with the star formation out to radii
$\sim 5\,{\rm kpc}$ and beyond (e.g., G3 and G4). There are cases, however,
where localized enhancements in the SFR are not matched by
increased \cii emission from the molecular gas (e.g.\ G2, and G3).  These SFR
enhancements are reflected in the \cii emission profile of the ionized gas,
which at galactocentric distances $\gs 2\,{\rm kpc}$ follow the SFR closely
(despite contributing only a small fraction to the total \cii emission
budget, see below). In contrast, the \cii emission from the PDR gas,
which dominates the total \cii luminosity at $2.0\,{\rm kpc} \ls R \ls 10\,{\rm
kpc}$, does not appear to be a sensitive tracer of the SFR. At $R\ls 1.5\,{\rm
kpc}$ where the SFR peaks, the \cii emission from this phase is seen to drop. At
larger radii, the PDR \cii emission declines but at a more gradual rate
than the star formation.

\bigskip

In the bottom right panel of Fig.\,\ref{fig:maps1} we show the Galactic SFR and
\cii luminosity radial profiles from \cite{pineda14}, who observed the \cii
emission from (1) CO-dark H$_2$ gas and dense PDRs, (2) cold neutral \hi gas, and
(3) hot ionized gas in our own Galaxy. In order to facilitate an approximate
comparison with our simulations we identify these three Galactic ISM phases with
the molecular, atomic, and ionized gas in our simulations. It is important to
keep in mind, however, that the ISM in our simulations has a higher pressure, is
kinematically more violent, and is more actively forming stars than is the 
case in our Galaxy. 

The SFR and \cii profiles in our Galaxy peak at larger radii ($R\sim4-5\,$kpc)
than in our simulated galaxies. This is not surprising given the low content of
star formation and gas in the Galactic bulge, and the fact that the bulk of star
formation in our Galaxy takes place in the disk. In contrast, gas is still being
funneled toward the central regions of our simulated galaxies where it is converted into stars.
Thus the SFR level in our simulated galaxies is much higher (by $\gs 10\times$)
and more centrally concentrated than in our Galaxy, where stars form at a rate
of $\ls 0.1-1\,{\rm \msun\,yr^{-1}}$ across the disk. Remarkably,
significant levels of \cii emission extend out to $R\sim 20\,{\rm kpc}$ in our
Galaxy, well beyond the point where star formation has ceased. Here, the
emission is completely dominated by CO-dark H$_2$ gas and dense PDR regions.
This is similar to the picture seen in our simulated galaxies where the \cii
emission at large radii is dominated by a neutral gas phase -- designated PDR
gas in our simulations, dubbed CO-dark H$_2$ $+$ dense PDR gas in
\cite{pineda14} -- that is largely uncoupled from star formation.  The radial
\cii profile of the ionized gas in the Galaxy roughly follows the SFR profile,
as is the case in our simulations. 

\bigskip

\begin{figure}[h] 
\includegraphics[width=1.0\columnwidth]{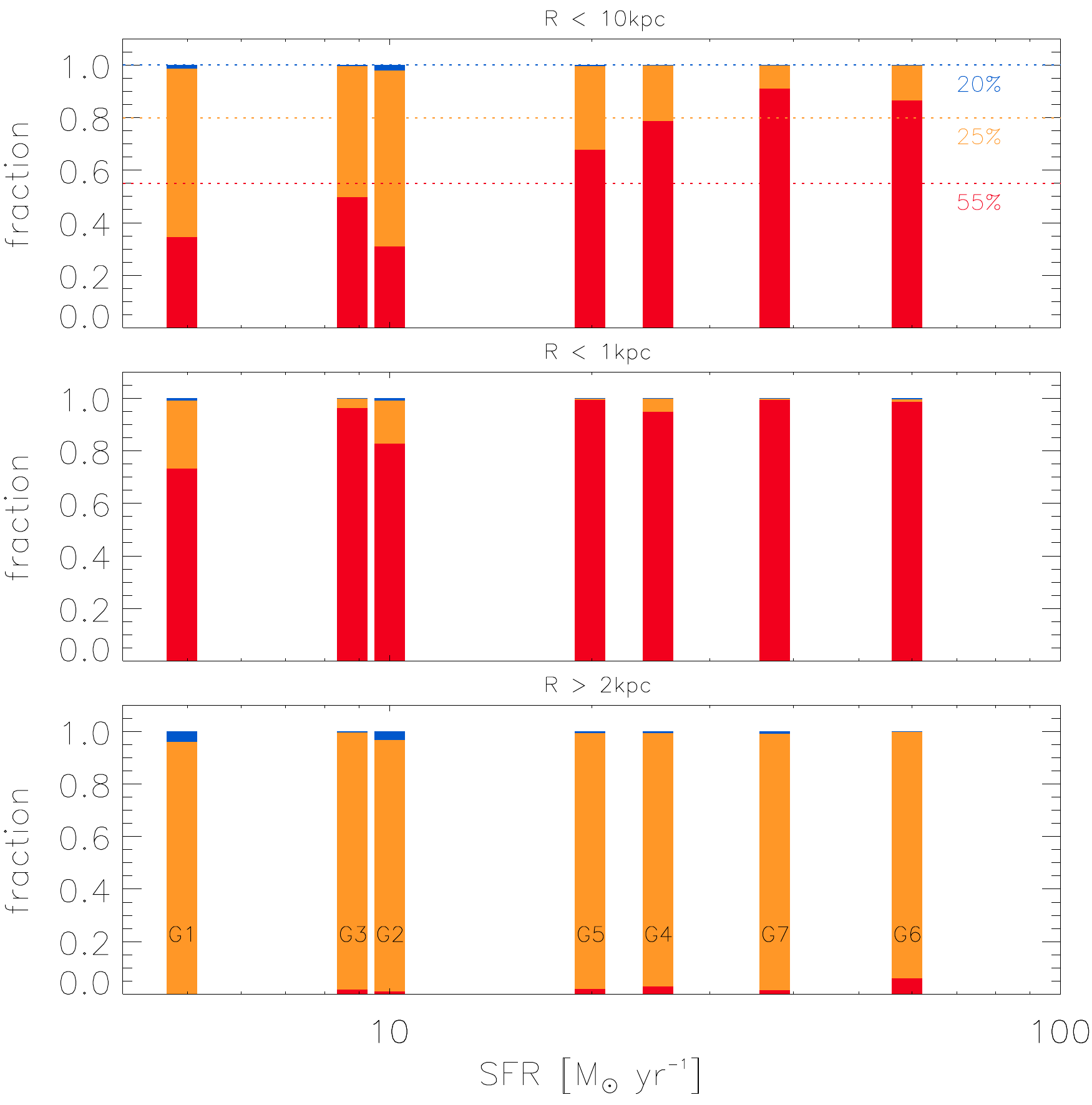}
\caption{Contributions to the total \cii luminosity from the molecular (red),
PDR (yellow), and ionized (blue) gas phases for each of our simulated galaxies,
ranked according to their total SFRs. The three panels show the relative
contributions within $R < 10\,{\rm kpc}$ (top), for $R < 1\,{\rm kpc}$ (middle),
and for $R > 2\,{\rm kpc}$ (bottom). The horizontal dashed lines and the
percentages given indicate the relative contributions to the total \cii
luminosity of our own Galaxy from CO-dark H$_2$ + dense PDR gas (red; 55\%),
cold atomic gas (yellow; 25\%), and ionized gas (blue; 20\%) \citep{pineda14}.
The total SFR of the Galaxy is $1.9\,{\rm \msun\,yr^{-1}}$ \citep{chomiuk11}. }
\label{figure:CII-percentages}
\end{figure}

Fig.\,\ref{figure:CII-percentages} displays the fractional \cii luminosity from
the different ISM phases: top panel for the entire disk ($R<10\,{\rm kpc}$),
middle panel for the central region ($R\le 1\,{\rm kpc}$), and bottom panel for
the outer disk ($R>2\,{\rm kpc}$).  Within $R\le 10\,{\rm kpc}$, the molecular
gas can constitute from $\sim31\%$ (G2) to $\sim 91\,\%$ (G7) of the total \cii
luminosity; for the PDR gas the range is $\sim 9\%$ (G7) to $\sim 67\%$ (G2).
Fig.\ \ref{figure:CII-percentages} shows that the contribution from the
molecular gas to the total \cii emission increases with the overall SFR of the
galaxy. A reverse trend is seen for the PDR gas.  As expected, the total
\cii emission from the central regions ($R \le 1\,{\rm kpc}$) is dominated by
the molecular gas ($\gs 70\%$) while further our ($R > 2\,{\rm kpc}$) the
PDR gas phase dominates ($\gs 90\%$)  (see Fig.\
\ref{figure:CII-percentages}, bottom two panels).  For reference, we note that
in our own Galaxy about $55\,\%$ of the total Galactic \cii luminosity (within
$R\ls 20\,{\rm kpc}$) is from molecular gas and dense PDRs, $25\,\%$ from cold
H{\sc i}, and $20\,\%$ from the ionized gas \citep{pineda14}. Thus, the ionized
phase is a more important contributor to the overall \cii budget in our Galaxy
than in the simulated galaxies presented here where the contribution from the
ionized gas is $\ls 3\%$ in all cases. 

\subsection{The integrated $\Lcii-{\rm SFR}$ relation}\label{subsection:integrated-CII-SFR-relation}
\begin{figure*}[htbp] 
\includegraphics[width=1.0\textwidth]{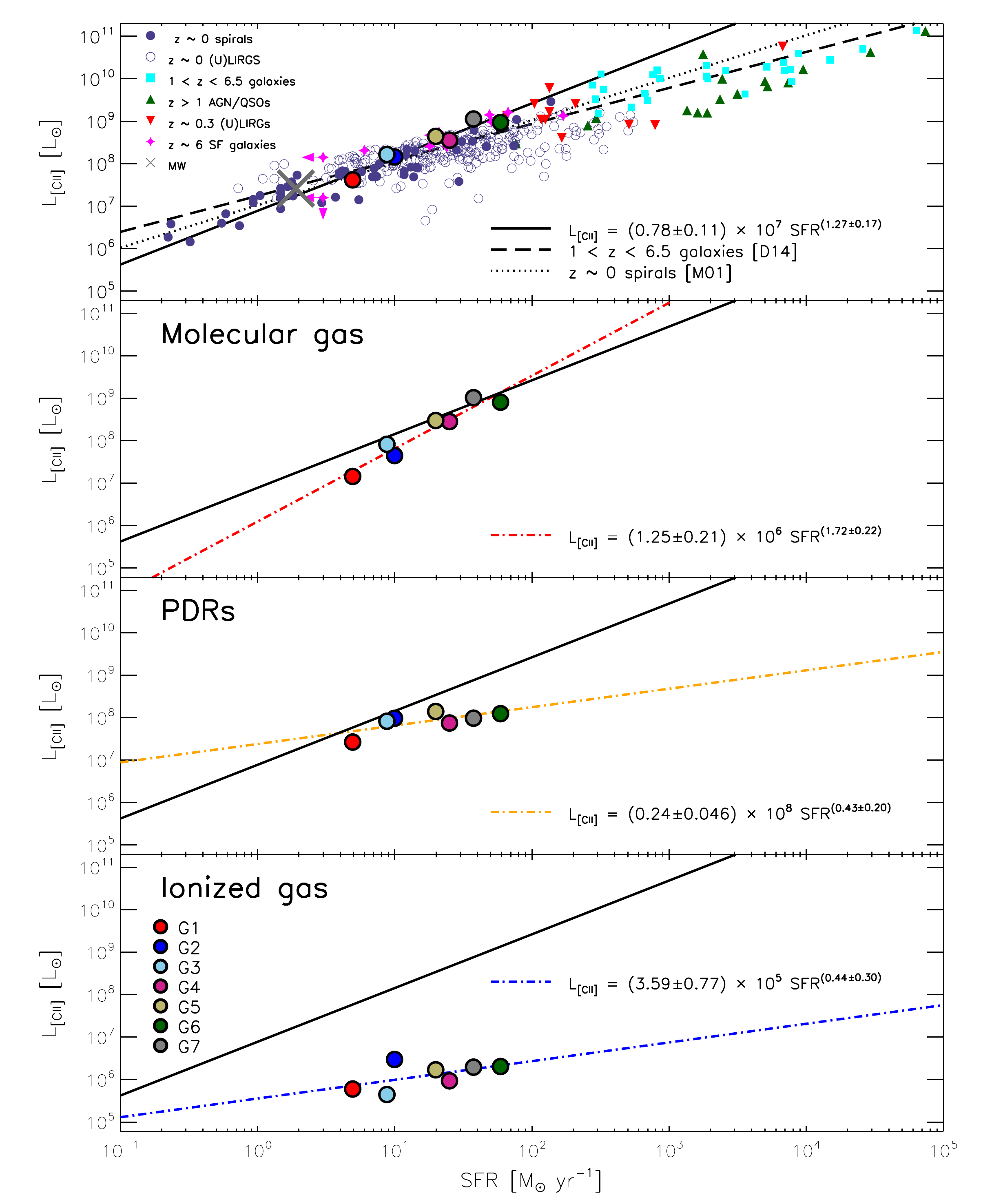}
\caption{\Lcii vs.\ SFR for our simulated galaxies (big filled circles).  The
same color-coding as in Fig.\,\ref{M_SFR} is used and listed again for
convenience in the bottom panel. From top to bottom panel we show the \cii
luminosity from the full ISM (gray), the molecular gas (red), PDRs (orange), and
ionized gas (blue).  For comparison, we show individual \cii observations of 54
$z\sim 0$ spirals \citep[purple filled circles;][]{malhotra01}, with the best
fit to this sample given by the dotted line (see main text for details).  Also
shown are  240 $z\sim 0$ (U)LIRGs \citep[purple open
circles;][]{diaz-santos13,farrah13}, and 12 $z\sim0.3$ (U)LIRGs \citep[red
triangles;][]{magdis14}.  The high-$z$ comparison samples include 25 $1<z<6$
star-forming galaxies (cyan squares; D14) and the dashed line indicate the best fit
to this sample.  Also shown are 16 $z\sim4-7$ quasars \citep[green
triangles;][]{iono06,walter09,wagg10,gallerani12,wang13,venemans12,carilli13,willott13},
and 10 normal star-forming $z\sim5-6$ galaxies (magenta stars) observed by
\cite{capak15}. The Galaxy is shown for reference with a gray cross \citep{pineda14}.}
\label{cii_sfr}
\end{figure*}

Fig.\,\ref{cii_sfr} shows the integrated ${\rm \Lcii-SFR}$ relations for our
simulated galaxies: top panel for the full ISM and, in separate panels below,
for each of the three ISM phases considered in our simulations.

When considering the entire ISM a tight correlation between \Lcii and SFR
emerges, which is well fit in log-log space by a straight line with slope
$1.27\pm 0.17$ (solid line in the top panel). This relation is largely set by
the molecular and PDR gas phases. The molecular gas phase, itself
exhibiting a strong correlation between \cii\ and SFR with slope $1.72\pm
0.22$, drives the slope of the total correlation. The PDR gas on the
other hand, shows a weaker \cii$-$SFR correlation with a slope of $0.43\pm
0.20$ but contributes significantly to the normalization of the total ${\rm
\cii-SFR}$ relation, especially at the low SFR end (see Fig.\ \ref{cii_sfr}).
The \cii emission from the ionized gas also shows a weak dependency on SFR
(slope $0.44\pm 0.30$) but does not contribute significantly to the total \cii\
emission, its normalization factor being $\gs 10\times$ below that of the
molecular and PDR gas (Fig.\ \ref{cii_sfr}, bottom panel). 

\smallskip

In the top panel of Fig.\ \ref{cii_sfr} we compare the ${\rm \Lcii-SFR}$
relation obtained from our simulated galaxies with samples of \cii\-detected
galaxies in the redshift range $z\sim0-6.5$ compiled from the literature.  Our
simulated galaxies are seen to match the observed ${\rm \Lcii-SFR}$ relation
both in terms of the slope of the relation and its overall normalization. A
power-law fit to the simulated galaxies (shown as solid line in Fig.\
\ref{cii_sfr}) yields a near-linear slope ($1.27\pm 0.17$).  Normal star-forming
galaxies at $z\sim 0$, with similar levels of SFR ($\sim 0.2 - 100\,{\rm
\msun\,yr^{-1}}$; \citealt{malhotra01}) as our simulated galaxies, are
consistent with a linear correlation given by $L_{\rm [C\,\textsc{ii}]} =
1.05\times 10^7{\rm SFR}$ (dotted line in Fig.\ \ref{cii_sfr}) with a scatter
of $0.3\,{\rm dex}$ \citep{magdis14}\footnote{This expression is inferred
from a power-law fit by \citet{magdis14} to the \cii and FIR ($42.5-122.5\,{\rm
\mu m}$) luminosities (in units of \lsun) of the \cite{malhotra01} sample:
$L_{\rm [C\,\textsc{ii}]} = 10^{-2.51\pm 0.39} L_{\rm FIR}$, where we have made
use of the conversion ${\rm SFR} = L_{\rm FIR} / 3.4\times 10^9$.}. 
The scatter of our simulated galaxies around their best-fit relation is
$0.14\,{\rm dex}$, i.e., significantly lower. We attribute this to the fact
that our simulated galaxies constitute a fairly homogeneous (and small) sample
spanning a rather small range in SFR, $L_{\rm [C\textsc{ii}]}$, and \Z, unlike
the observed samples with which we are comparing.

\bigskip

A direct comparison with \cii-detected galaxies at high redshifts is
complicated by the fact that the latter typically have significantly larger SFRs
than our model galaxies. Furthermore, high-$z$ samples are often heterogeneous
\citep[e.g., significant AGN contribution, cf.][]{gullberg15}.  One exception,
however, is the recent \cii-detected sample of star-forming galaxies at $z\simeq 5-6$
presented by \citet{capak15} (shown as magenta stars in Fig.\ \ref{cii_sfr}),
which span the same range in SFR as our simulations. These galaxies are seen to
be in excellent agreement with the ${\rm \Lcii-SFR}$ relation defined by our
simulations, both in terms of slope and normalization.  

Extrapolating our best-fit relation to SFRs $\gs
300\,{\rm \msun\,yr^{-1}}$ in order to compare with other high-$z$
samples, the relation is seen to overshoot the data. A power-law fit
to the $z > 1$ star-forming galaxies with SFR\,$\gtrsim300$\,\sfru
compiled by D14 yields: $L_{\rm [C\,\textsc{ii}]} = 1.7\times 10^7{\rm
  SFR}^{0.85}$ (D14; shown as the dashed line in Fig.\ \ref{cii_sfr}),
i.e., formally, a shallower relation than that of our simulated
galaxies and that of the $z\sim 0$ sample (albeit less so).  Finally,
we stress that the \cii-detected galaxies at $z > 1$ with SFRs $\gs
300\,{\rm \msun\,yr^{-1}}$ likely derive from rather complex
environments \citep[e.g.][]{narayanan15}, which may not correspond to
the relatively quiescent MS star-forming galaxies modeled
here.

\subsection{The resolved $\CIIsd-\SFRsd$ relation}
\begin{figure*}[htbp] 
\includegraphics[width=1\textwidth]{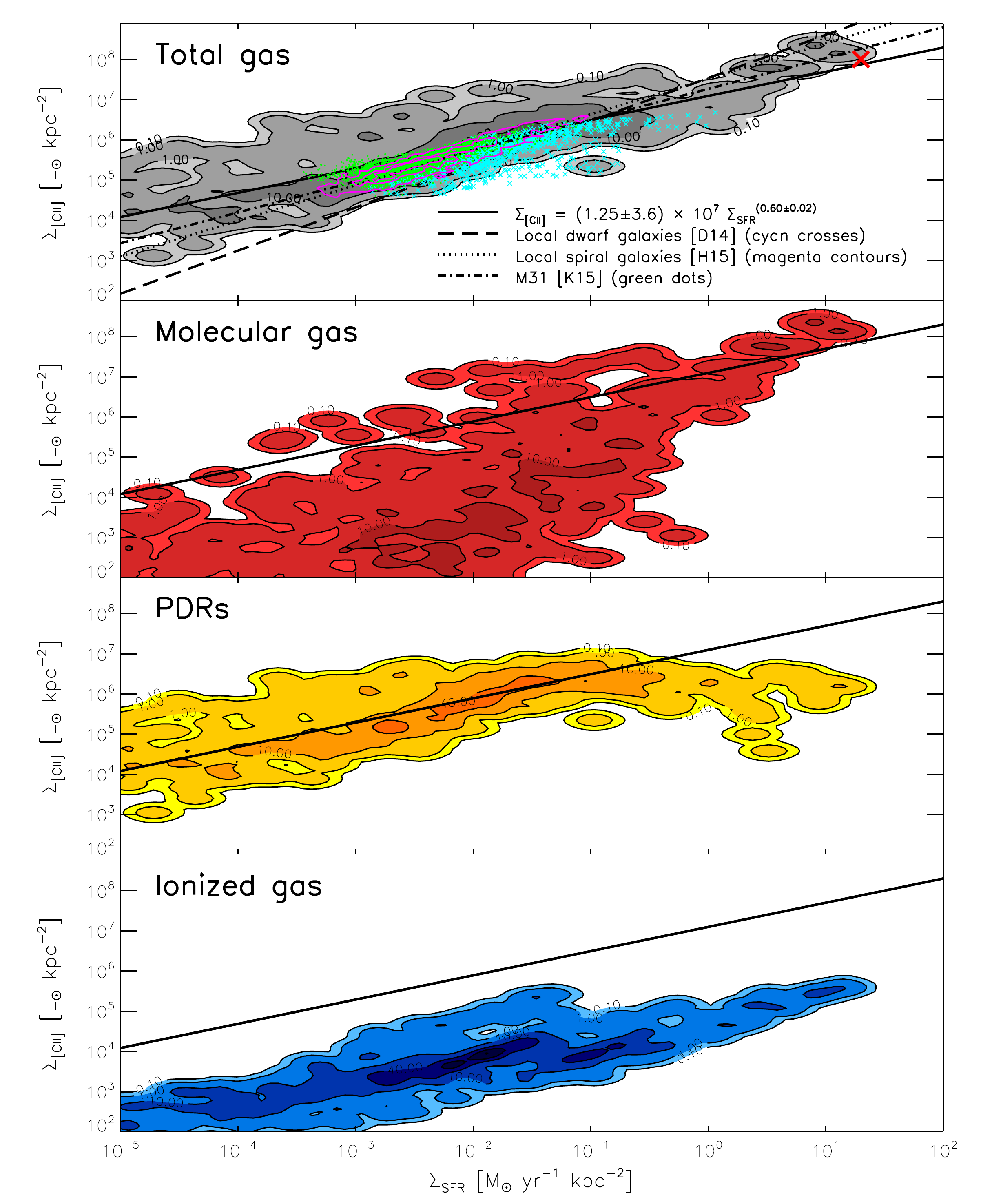} 
\caption{\cii\ luminosity surface density (\CIIsd) vs.\ SFR surface density
(\SFRsd) for the full ISM in our simulated galaxies (top panel), and for each of
the three ISM phases (bottom three panels). \CIIsd and \SFRsd are determined
over $1\,{\rm kpc}\times 1\,{\rm kpc}$ regions within all 7 galaxies. The filled
colored contours indicate the number of such regions with this combination of
\SFRsd and \CIIsd as a percentage (at 0.1\%, 1\%, 10\%, 40\% and 70\%) of the maximum
number of regions. The solid line shown in all panels is the best-fit power-law
to the total gas \CIIsd vs.\ \SFRsd (see legend). For comparison we show
observed \CIIsd vs.\ \SFRsd for individual $20\,{\rm pc}$ regions in M31
(green dots; K15), $1\,{\rm kpc}$ regions in nearby (mostly spiral) galaxies
(magenta contours, 
indicating regions containing 25\%, 45\%, and 95\% of the total
number of data points; H15), and $1\,{\rm kpc}$ regions in local dwarf galaxies
(cyan crosses; D14).  
The best-fit power-laws to these three data-sets are shown
as dashed-dotted, dotted, and dashed lines, respectively. In the case of M31 the
power-law is fitted to $50\,{\rm pc}$ regions (see K15). We also show the
galaxy-averaged \CIIsd and \SFRsd of ALESS\,73.1 (red cross), a $z=4.76$
submillimeter-selected galaxy which was marginally resolved in \cii and in the
FIR continuum with ALMA, revealing a source size of $R\sim 2\,{\rm kpc}$,
$L_{\rm [CII]} = 5.15\times 10^9\,{\rm \lsun}$, and ${\rm SFR}\sim 1000\,{\rm
\msun\,yr^{-1}}$ \citep{debreuck14}.
}
\label{cii_sfr_res}
\end{figure*}

In Fig.\,\ref{cii_sfr_res} we show the combined ${\rm \CIIsd-\SFRsd}$
relation of all seven simulated galaxies in their face-on
configuration. The relation is shown for the entire ISM (top panel)
and for each of the separate gas phases (bottom three panels). Surface
densities were determined within 1\,kpc $\times$ 1\,kpc
regions. Contours reflect the number of regions at a given
(\SFRsd,\CIIsd)-combination and are given as percentages of the peak
number of regions.

Given the variations in the local star formation conditions within our model
galaxies, we can explore the relationship between \cii\ and star formation over
a much wider range of star formation intensities than is possible with the
integrated quantities. A ${\rm \CIIsd - \SFRsd}$ correlation spanning more than
five decades in \SFRsd\ is seen for the entire ISM as well as for the individual
ISM phases. Similar to the integrated \cii\ vs.\ SFR relations in the previous
section, the molecular and PDR phases dominate the resolved \cii\ emission
budget at all SFR surface densities, and with the molecular
relation being steepest and exhibiting the largest degree of scatter.  The
resolved \cii emission from the ionized phase, while clearly correlated with
\SFRsd, is largely negligible at all star formation densities.

We compare the ${\rm \CIIsd-\SFRsd}$ relation for our simulated galaxies
to that of three resolved surveys of nearby galaxies: (1) the $\sim
50\,$pc-scale relation derived from five $3'\times 3'$ fields toward
M\,31 (K15; shown as green points in Fig.\ \ref{cii_sfr_res}), (2) the
kpc-scale relation obtained for 48 local dwarf galaxies covering a wide
range in metallicities ($Z/Z_{\rm \odot} = 0.02-1$, D14; cyan crosses),
and (3) the kpc-scale relations for local, mostly spiral, galaxies (H15;
magenta contours). The data from these surveys have been converted to
the Chabrier IMF assumed by our simulations by multiplying \SFRsd with a
factor 0.94 when a Kroupa IMF was adopted (D14; K15) or 0.92 when a
truncated Salpeter IMF was used (H15), following \cite{calzetti07} and
\cite{speagle14}.  Regarding the observations by H15, we adopt the raw
\CIIsd measurements rather than the IR-color-corrected ones. From Fig.\
\ref{cii_sfr_res} we see that over the range in \SFRsd
($\sim0.001-1\,\sfru$\,kpc$^{-2}$) spanned by these three surveys, the
${\rm \CIIsd - \SFRsd}$ relation defined by our simulations is in
excellent agreement with the observations. Also, the majority of our
simulated $1\,{\rm kpc}\times 1\,{\rm kpc}$ regions fall within the
observed \SFRsd and \CIIsd ranges. A small fraction (a few procent) of
regions in our simulations exhibit an excess in \CIIsd for a given
\SFRsd relative to the observed relations, but the majority coincide
with the observations.  We note that the observed relations exhibit
significant scatter ($\sim 0.2-0.3\,{\rm dex}$; D14, H15, K15) as well
as small systematic offsets relative to each other (in particular in the
case of D14).  The ${\rm \CIIsd-\SFRsd}$ relations observed in the five
fields in M\,31 by K15 yield best-fit power-law slopes in the range
$\sim 0.67-1.03$, with an average of $0.77$ (shown as dashed-dotted line
in Fig.\ \ref{cii_sfr_res}). Similar power-law fits to the samples of
D14 and H15 yield slopes of $\sim 1.07$ and $\sim 0.88$, respectively
(shown as dashed and dotted lines in Fig.\ \ref{cii_sfr_res}). In
comparison, a power-law fit to our simulations -- across the full
\SFRsd-range -- results in a slope of $\sim 0.60$, i.e., on the low side
of the observed range.  However, if instead we fit only to simulated
regions within the $\SFRsd =0.001-0.1\,\sfru$\,kpc$^{-2}$ range, thereby
matching the range with the most observations, we find a slope of $\sim
0.75$, i.e., well within the observed range.

Beyond the above \SFRsd-range a comparison with the observations has to rely on
extrapolations of the simple power-law fits to the observed ${\rm \CIIsd -
\SFRsd}$ relations.  At low \SFRsd\ ($\ls 0.001\,{\rm \sfru\,kpc^{-2}}$), the
simulations broadly follow the extrapolations of the observed relations, except
for D14 where the systematic offset noted at higher \SFRsd is compounded at the
lower \SFRsd values owing to the relatively steep slope of the fit to the D14
data.  At these low \SFRsd levels a larger (but still minor, overall) fraction
of the simulated regions display excess \cii\ emission levels ($\gs10\times$)
compared to the survey-based power-law fits.  This excess \cii\ emission is
driven by the PDR gas in our simulations (second panel in
Fig.\,\ref{cii_sfr_res}), which dominates the \cii emission at these \SFRsd
levels.  Interestingly, in at least two of the five fields in M\,31 studied by
K15, a similar \cii-excess relative to the power-law fit is observed at
$\SFRsd\ls 0.001\,$\sfru\,kpc$^{-2}$ (see Fig.\ 7 in K15). K15 argues that the
excess is due to a contribution from diffuse, ionized gas (\hii), although they
cannot discount the possibility that the larger dispersion is at least partly
due to being close to the sensitivity limit of their survey at such low \CIIsd.
Our simulations, however, clearly show no significant contribution from the
diffuse \hii gas phase at low \SFRsd.

At high \SFRsd ($\gs 1\,{\rm \msun\,yr^{-1}\,kpc^{-2}}$), the scatter in the
${\rm \CIIsd - \SFRsd}$ relation from our simulations decreases significantly,
and is seen to broadly match the extrapolated locally observed relations.
Furthermore, the simulation contours are seen to agree with the few galaxies
observed to date to have \SFRsd$\gs1$\,\sfru\,kpc$^{-2}$: a few sources from the
D14 sample (there are two D14 galaxies with \SFRsd$\gs 1$\,\sfru\,kpc$^{-2}$)
and ALESS\,73.1, a $z=4.76$ sub-millimeter selected galaxy, marginally resolved
in \cii and with a disk-averaged \SFRsd of $\sim 80\,{\rm
\msun\,yr^{-1}\,kpc^{-2}}$ (\citealt{debreuck14}; indicated by a red cross in
Fig.\ \ref{cii_sfr_res}).

The aforementioned studies of local galaxies typically measure both obscured
(e.g.\ from $24\,\mu$m) and un-obscured (from either FUV or H$\alpha$) SFRs in
order to estimate the total \SFRsd and, while intrinsic uncertainties are
inherent in the empirical $24\,\mu$m/FUV/H$\alpha\rightarrow\,$SFR calibrations,
the \SFRsd values should be directly comparable to those from our simulations.
Nonetheless, some caution is called for when comparing our simulations to
resolved observations, primarily regarding the determination of \SFRsd on
$\lesssim\,$kpc-scales.  In particular, as pointed out by D14, the emission from
old stellar populations in diffuse regions with no ongoing star formation can
lead to overestimates of the SFR when using the above empirical SFR
calibrations. Both K15 and H15 account for this by subtracting the expected
cirrus emission from old stars in $24\,\mu$m (using the method presented in
\citealt{leroy12}), although on scales $\ll 1\,{\rm kpc}$ this correction for
$24\,\mu$m cirrus emission becomes particularly challenging as it was calibrated
on $\gs 1\,$kpc scales.  K15 further points to the problem arising from the
possibility of photons leaking between neighboring stellar populations, meaning
that average estimates of the SFR on scales $<50$\,pc might not represent the
true underlying SFR.

\subsection{Physical underpinnings of the ${\rm \cii-SFR}$ relationship}
\begin{figure}[htbp] 
\includegraphics[width=\columnwidth]{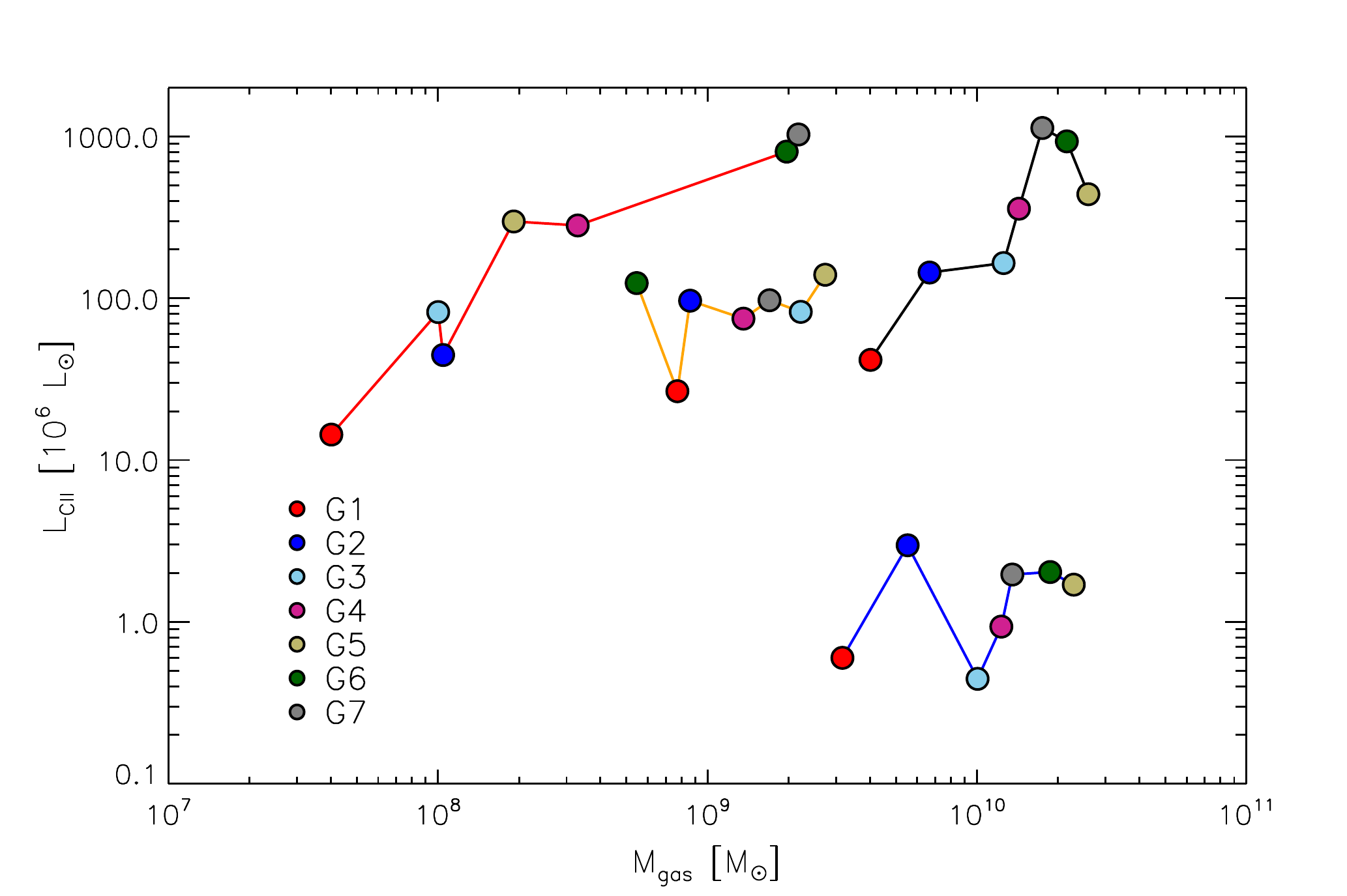}
\caption{\cii luminosity versus gas mass of the total ISM (black line), the
molecular (red line), the PDR (orange line), and the \hii (blue line) gas
phase for each of our seven simulated galaxies. The \cii\ emission from the
molecular gas is seen to increase with the amount of molecular gas available,
whereas this is not the case for gas associated with PDR and \hii regions.}
\label{m_cii}
\end{figure}

The SFRs of galaxies at both low and high redshifts is
observed to strongly correlate with their molecular gas content (typically
traced by CO or dust emission)
\citep[e.g.][]{kennicutt98,daddi10a,genzel10,narayanan12a}.  This ${\rm
SFR}-{\rm H_2}$ dependency is incorporated into our SPH simulations (see
Section \ref{sph}), and it is therefore natural to ask whether the integrated
${\rm \cii-SFR}$ relation examined in Section
\ref{subsection:integrated-CII-SFR-relation} might simply be a case of galaxies
with higher SFRs having larger (molecular) gas masses with
higher associated \cii luminosities. To investigate this, we show in
Fig.\,\ref{m_cii} \cii luminosity vs.\ gas mass for the full ISM in our
simulated galaxies and for the individual gas phases separately.
For the molecular gas phase we see a significant luminosity$-$mass scaling (red
curve), which would explain the strong $L_{\rm [C\textsc{ii}]}-{\rm SFR}$
relation for this phase and, at least in part, the $L_{\rm [C\textsc{ii}]}-{\rm
SFR}$ relation for the full ISM. The PDR gas and the ionized gas
do not exhibit similar luminosity$-$mass scaling relations. Since, on average,
significantly more mass resides in each of these two phases than in the
molecular phase, this has the effect of weakening the correlation between $L_{\rm
[C\textsc{ii}]}$ and SFR when considering the full ISM.

\begin{figure}[htbp] 
\begin{center} 
\includegraphics[width=1\columnwidth]{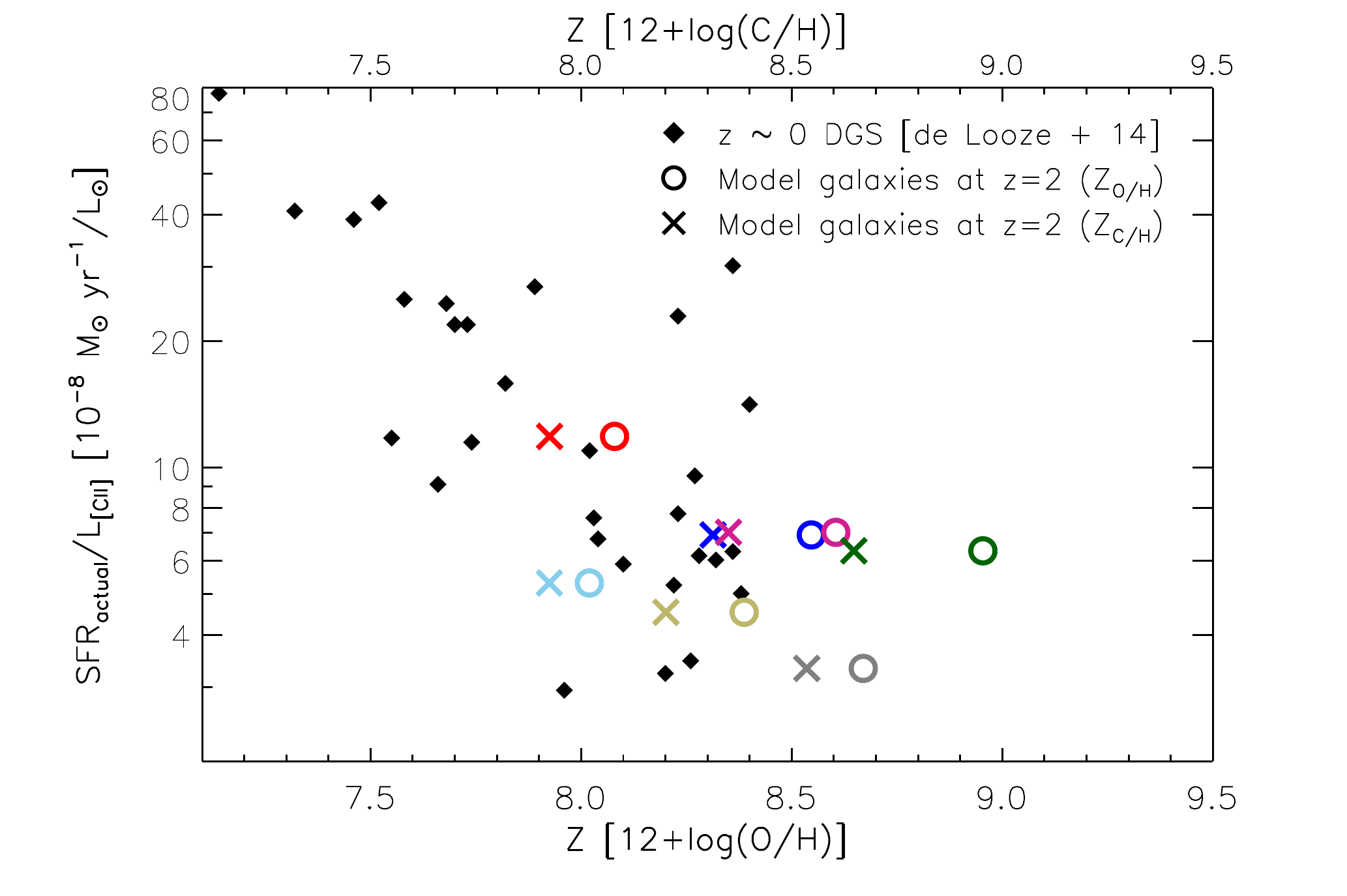}
\end{center}
\caption{${\rm SFR}/\Lcii$ as a function of metallicity for our simulated
galaxies.  The metallicity is parametrized both relative to the oxygen (circles)
and the carbon (crosses) abundance. The galaxies are color coded in the usual
manner.  For comparison, we show ${\rm SFR}/\Lcii$ vs.\ $Z_{\rm [O/H]}$ for the
local dwarf galaxy sample of D14, where SFRs are from combined UV and $24\,\mu$m
measurements.}
\label{cii_zm}
\end{figure}

The abundances of metals in the ISM is expected to affect the \cii\ emission.
In their study of low-metallicity dwarf galaxies, D14 compared total (IR$+$UV)
SFRs to SFRs derived using the best-fit $\Lcii-{\rm SFR}$ relation to their
entire sample.  They found an increase in the SFR$_{\rm{IR+UV}}$/SFR$_{{\rm
C\textsc{ii}}}$ fraction, corresponding to weaker \cii emission, toward lower
metallicities.  In Fig.\ \ref{cii_zm} we show ${\rm SFR/\Lcii}$ as a function of
$Z_{\rm [O/H]}$ ($=12+\log {\rm [O/H]}$) for our simulated galaxies as well as
for the dwarf galaxy sample of D14.  As D14 points out, $Z_{\rm [O/H]}$ does not
take into account potential deficits of carbon relative to the [O/H] abundance,
potentially obscuring any potential correlation between \cii luminosity and
actual carbon abundance. In Fig.\ \ref{cii_zm} we therefore also plot ${\rm
SFR/\Lcii}$ as a function of  $Z_{\rm [C/H]}$ ($=12+\log {\rm [C/H]}$) for our
simulated galaxies. Arguably, our simulated galaxies exhibit a weak trend in
${\rm SFR/\Lcii}$ with metallicity (in particular for [C/H]). Certainly, there
is a very good overall agreement with the findings of D14. The lack of a strong
trend within our simulation sample is not suprising, however, since our galaxies
span a limited range in metallicity ($Z_{\rm [O/H]} = 8.03-8.96$) and, as we saw
in Section \ref{subsection:integrated-CII-SFR-relation}, form a tight $L_{\rm
CII}-{\rm SFR}$ relation.  

\begin{figure}[htbp] 
\includegraphics[width=\columnwidth]{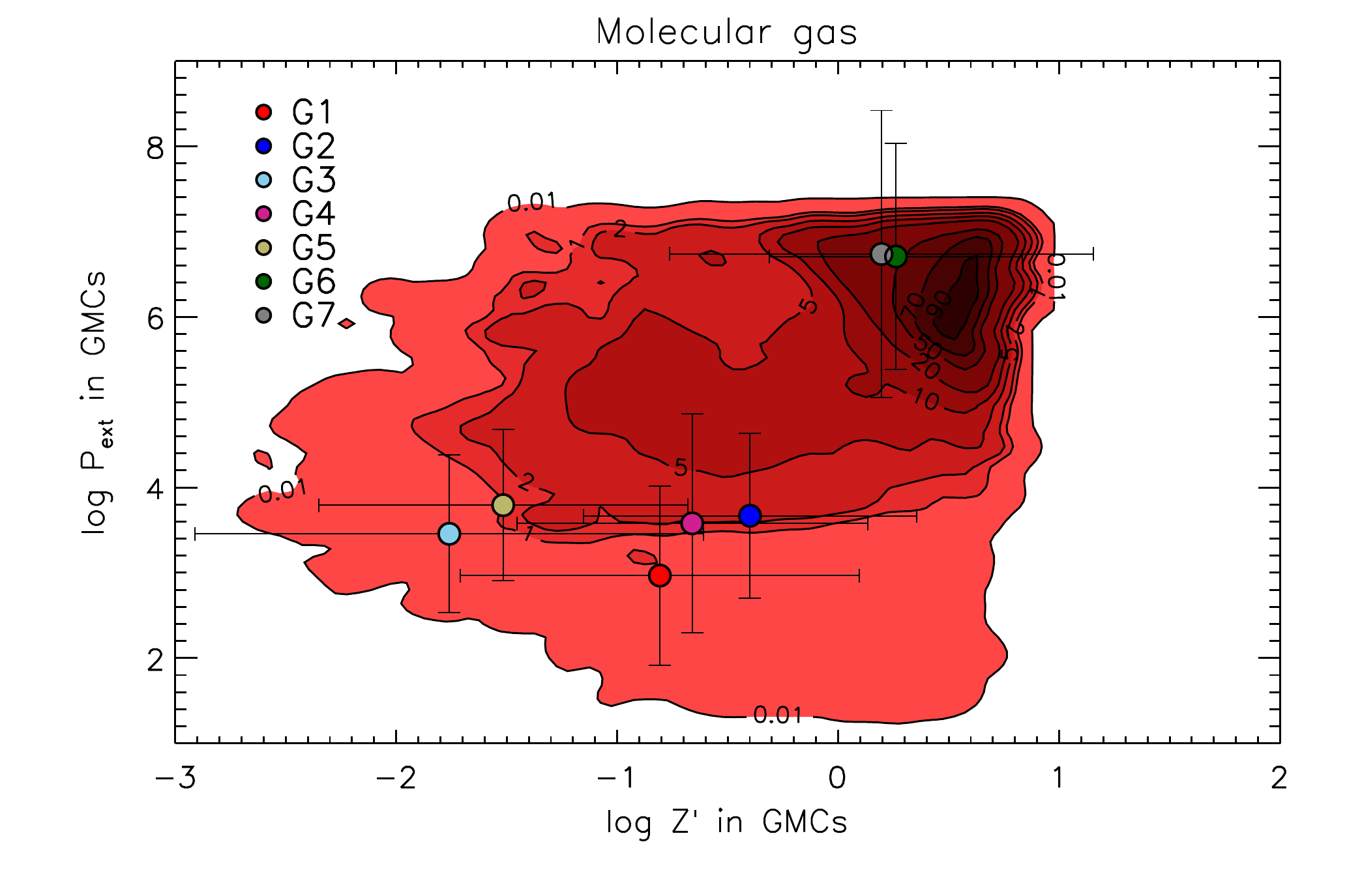}
\includegraphics[width=\columnwidth]{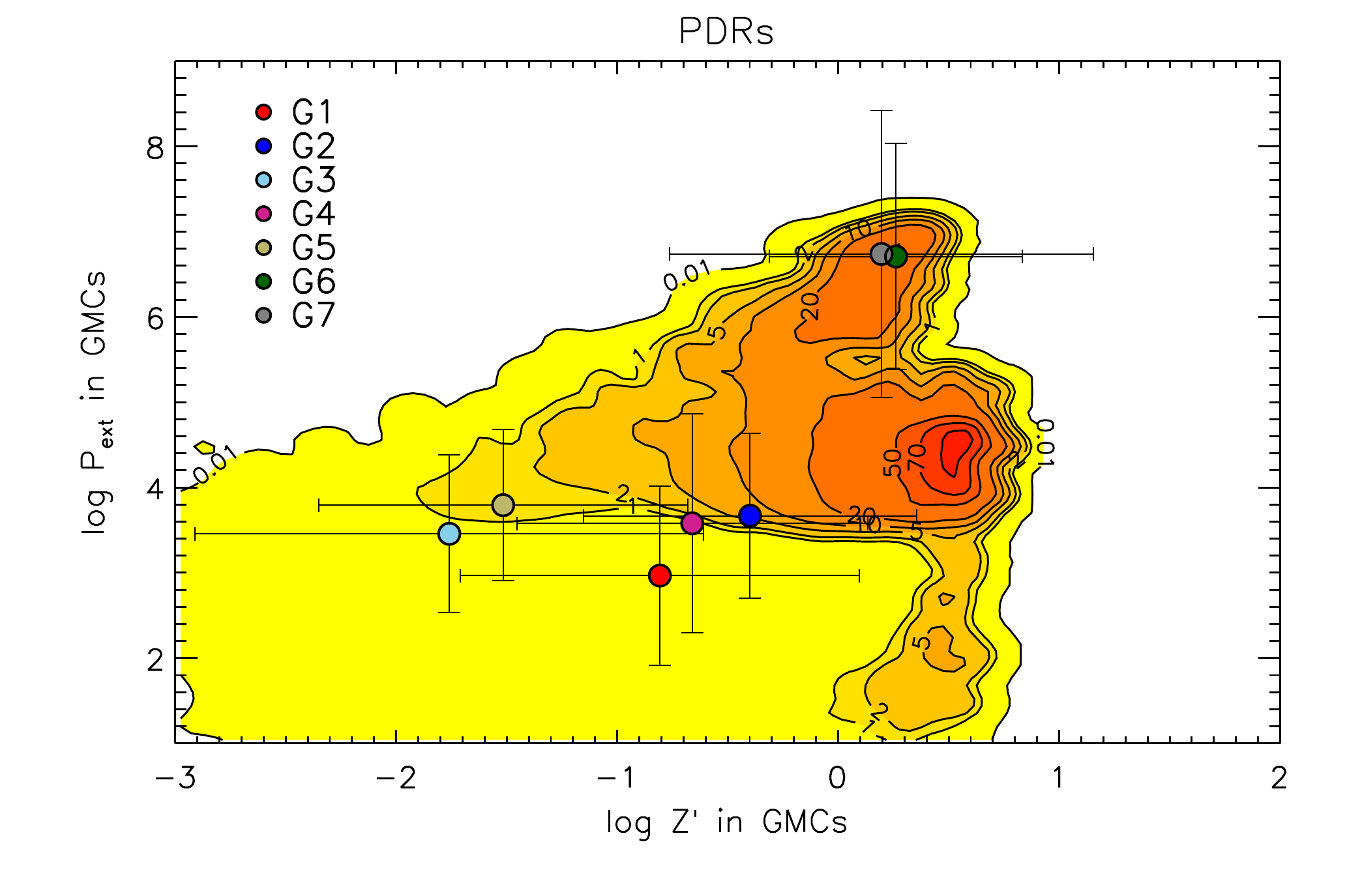}
\includegraphics[width=\columnwidth]{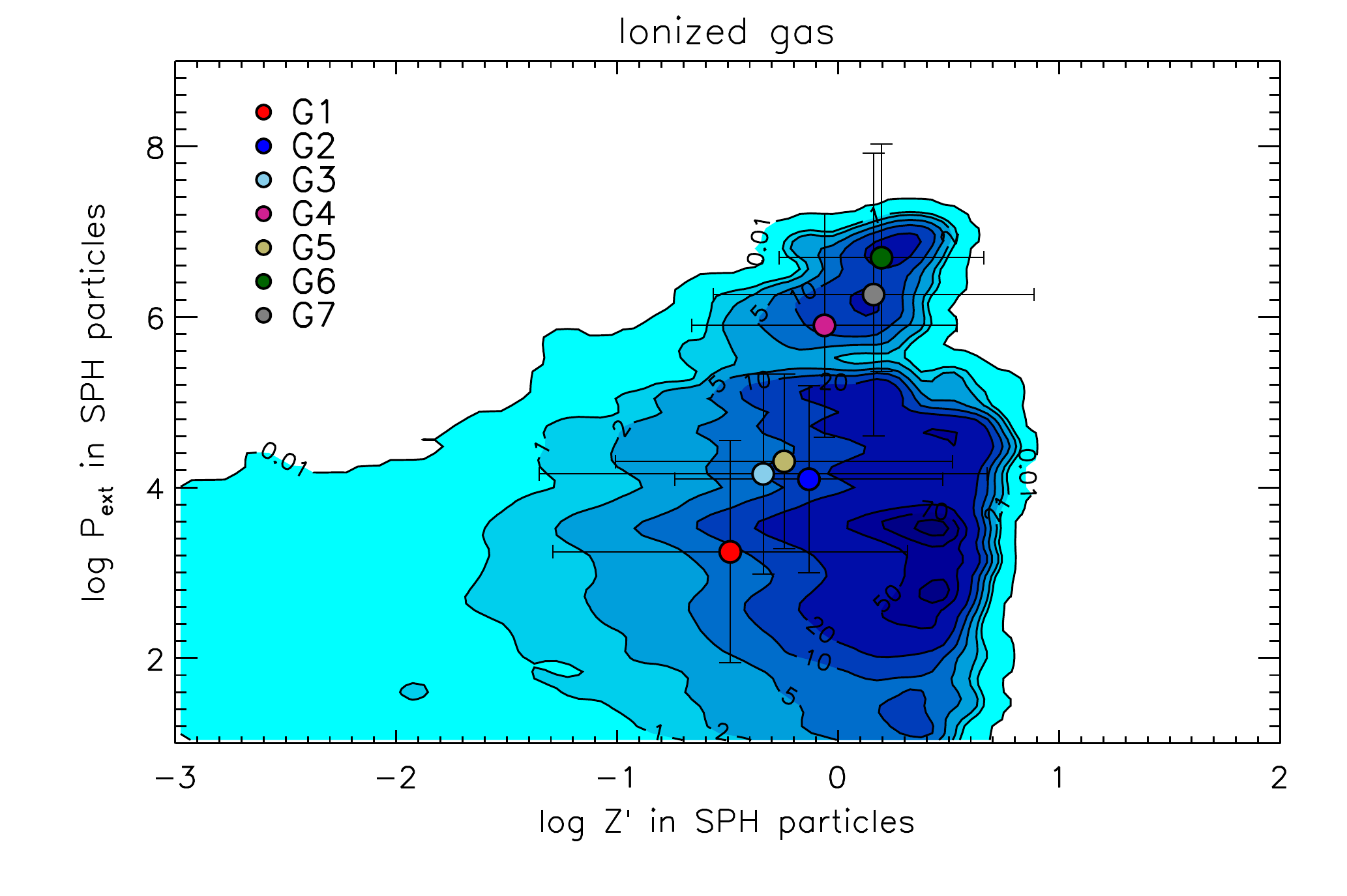}
\caption{Contours of (median) $\epsilon_{\rm [C\textsc{ii}]}$ (i.e.\ \cii
luminosity per gas mass) as a function of \Zn\ and $\Pe/k_{\rm B}$ for the
molecular (top), PDR (middle), and \hii\ (bottom) gas phases in our model
galaxies. In each panel the contours indicate 0.01\%, 1\%, 2\%, 5\%, 10\%, 20\%, 50\%, 70\%, and
90\% percentages of the maximum efficiency. For reference, the maximum (median)
$\epsilon_{\rm [C\textsc{ii}]}$ for the molecular, PDR, and ionized gas phases
are: \blue{$3.7$, $6.6$, and $0.02\,{\rm \lsun/\msun}$}, respectively.  Median values
of \Zn\ and $\Pe/k_{\rm B}$ and for the GMC and ionized cloud population in each
model galaxies are indicated as colored circles (error bars are the $1\sigma$
dispersion of the distributions).}
\label{Pe_Z_grid}
\end{figure}

To better understand how the metallicity and also the ISM pressure affects the
\cii emission, we define a `\cii\ emission efficiency' ($\epsilon_{\rm
[C\textsc{ii}]}$) of a given gas phase as the \cii luminosity of the gas phase
divided by its mass, and examine how it depends on \Zn\ and $\Pe/k_{\rm B}$.
Specifically, we create a grid of (\Zn, $\Pe/k_{\rm B}$) values, and in each
grid point the median $\epsilon_{\rm [C\textsc{ii}]}$ of all clouds in our
simulated galaxies is calculated.  The resulting $\epsilon_{\rm [C\textsc{ii}]}$
contours as a function of \Zn\ and $\Pe/k_{\rm B}$ are shown in Fig.\
\ref{Pe_Z_grid} for the molecular (top), PDR (middle), and \hii\ (bottom)
phases. For comparison, we also show the median \Zn- and $\Pe/k_{\rm B}$-values
for each of our galaxies.

The molecular phase in our simulations is seen to radiate most efficiently in
\cii at relatively high metallicities ($\log \Z\sim0.5$) and cloud external
pressures ($\Pe/k_{\rm B} \sim10^6\,{\rm cm^{-3}\, K}$).  The PDR and ionized
phases, however, have their maximal $\epsilon_{\rm [C\textsc{ii}]}$ at $\log
\Z\gs0$ and $\Pe\sim10^4\,{\rm cm^{-3}\,K}$, i.e.\ at $\sim 100\times$ lower
pressures than the molecular phase, and can maintain a significant \cii
efficiency ($\gs 20\%$) even at $\Pe/k_{\rm B}\sim 10^2\,{\rm cm^{-3}\,K}$
provided $\log \Zn\gs 0$. This is consistent with our finding in Section
\ref{subsection:CII-profiles}, that the molecular phase dominates the \cii
emission in the central, high-pressure regions, while the PDR gas dominates
further out in the galaxy where the ISM pressure is less extreme.  In all three
panels, G6 and G7 lie in regions of high $\epsilon_{\rm [C\textsc{ii}]}$ ($\gs
50\%$ for the molecular and PDR gas). Thus,
the reason why these two galaxies have the highest total \cii luminosities is in
part due to the fact that they have the highest molecular gas masses (by nearly
an order of magnitude, see Fig.\ \ref{m_cii}), and in part due to the bulk of
their cloud population (GMCs or ionized clouds) having sufficiently high
metallicities (on average $\gs 5\times$ higher than G2) and experiencing external
pressures ($\sim 100 \times$ higher on average than the remaining galaxies)
which drives their \cii efficiencies up.

\section{Comparing with other \cii\ simulations}
A direct quantitative comparison with other \cii\ emission simulation studies in
the literature \citep{nagamine06,vallini13,vallini15,popping14b,munoz14} is complicated by
the fact that there is not always overlap in the masses and/or redshifts of the
galaxies simulated by the aforementioned studies and our model galaxies.
Furthermore, there are fundamental differences in the simulation approach, with
some adopting semi-analytical models \citep{popping14b,munoz14} and others
adopting SPH simulations \citep{nagamine06,vallini13,vallini15}. Also, differences in the
numerical resolution of both types of simulations, and in the specifics of the
sub-grid physics implemented, can lead to diverging results and make comparisons
difficult.

The simulations presented here combine cosmological SPH galaxy simulations with
a sub-grid treatment of a multi-phase ISM that is locally heated by FUV
radiation and CRs in a manner that depends on the local SFR 
density within the galaxies.  We have applied our method to a high
resolution ($\delta m_{\rm SPH}\simeq 1.7\times 10^4\,{\rm \msun}$) cosmological
SPH simulation of $z=2$ star-forming galaxies (i.e., baryonic mass resolution
$\delta m_{\rm SPH}\simeq 1.7\times 10^4\,{\rm \msun}$ and gravitational
softening length $\simeq 0.6\,h^{-1}\,{\rm kpc}$, see Section \ref{sph}).  A
novel feature of our simulations is the inclusion of molecular, neutral and
ionized gas as contributors to the \cii emission.  Another unique feature of our
model is the inclusion of CRs as a route to produce C$^+$ deep inside the GMCs
where UV photons cannot penetrate (see Section \ref{cii_em}). 

Our simulations probably come closest -- both in terms of methodology and
galaxies simulated -- to those of \cite{nagamine06} who 
employed cosmological simulations with \gadgettwo (the precursor for \gadget used here) to predict
total \cii luminosities
from dark matter halos associated with $z\simeq 3$ LBGs 
with $M_{\rm *} \sim 10^{10}\,{\rm \msun}$ and ${\rm SFR}
\gs 30\,{\rm \msun\,yr^{-1}}$ \citep[see also][]{nagamine04}.  Their simulations
employed gas mass resolutions in the range $\delta m_{\rm SPH}\simeq 3\times
10^{5-8}\,{\rm \msun}$ and gravitational softening lengths of typically
$>1h^{-1}\,{\rm kpc}$.  In their simulations the ISM consist of a CNM 
($T\sim 80\,{\rm K}$ and $n\sim 10\,{\rm cm^{-3}}$) and a warm neutral
medium ($T\sim 8000\,{\rm K}$ and $n\sim 0.1\,{\rm cm^{-3}}$) in
pressure-equilibrium, and the assumption is made that the \cii emission only
originates from the former phase. The thermal balance calculation of the ISM
includes heating by grain photo-electric effect, CRs, X-rays, and
photo-ionization of C\,{\sc i} -- all of which are assumed to scale on the local
SFR surface density.  Their simulations predict $L_{\rm [C{\sc II}]} \sim (0.3 -
1)\times 10^8\,{\rm \lsun}$ for their brightest LBGs (${\rm SFR} \gs 30\,{\rm
\msun\,yr^{-1}}$) which is slightly below the prediction of our integrated ${\rm
\cii-SFR}$ relation ($L_{\rm [C{\sc II}]} \sim 6\times 10^8\,{\rm \lsun}$; Fig.\
\ref{cii_sfr}).

Employing a semi-analytical model of galaxy formation, \cite{popping14b} made
predictions of the integrated \cii luminosity from galaxies at $z=2$ with
$M_{\rm *}\sim 10^8-10^{12}\,{\rm \msun}$ and ${\rm SFR \sim 0.1-100\,{\rm
\msun\,yr^{-1}}}$ \citep[see also][]{popping14a}. The galaxies are assumed to
have exponential disk gas density profiles, with randomly placed
`over-densities' mimicking GMCs (constituting $\sim 1\%$ of the total volume).
The gas is embedded in a background UV radiation field of 1\,Habing, with local
variations in the radiation field set to scale with the local SFR surface
density. The C$^+$ abundance is set to scale with the abundance of carbon in the cold gas. 
The excitation of \cii\ occurs via collisions with $e^-$, H, and H$_2$, and the
line emission is calculated with a 3D radiative transfer code that takes into
account the kinematics and optical depth effects of the gas.  For galaxies with
SFRs similar to our model galaxies ($\sim 5 - 50\,{\rm \msun\,yr^{-1}}$) their
simulations predict ensemble-median \cii\ luminosities of $\ls (1-6)\times
10^7\,{\rm \lsun}$.\footnote{Since \citet{popping14b} plots $L_{\rm [C{\sc
II}]}$ against $L_{\rm IR}$ and not SFR, we have converted our SFRs to $L_{\rm
IR}$ in order to crudely estimate their predicted \cii\ luminosities (see their
Fig.\ 11). For the ${\rm SFR}\rightarrow L_{\rm IR}$ conversion we have used
\citep{bell03}.  Not all the star formation will be obscured and so the IR
luminosities, and thereby the \cii\ luminosities given here will be upper
limits.} This is lower than the \cii\ luminosities predicted by our simulations
and also somewhat on the low side of the observed $z\sim 0$ $L_{\rm [C{\sc
II}]}-{\rm SFR}$ relation (Fig.\ \ref{cii_sfr}). Allowing for the $+$2$\sigma$
deviation from the median of the \citet{popping14b} models results in $L_{\rm
[C{\sc II}]}\ls (2-30)\times 10^7\,{\rm \lsun}$ for ${\rm SFR \sim 5
- 50\,{\rm \msun\,yr^{-1}}}$, which matches the observed $L_{\rm [C{\sc
  II}]}-{\rm SFR}$ relation.

The remaining \cii simulation studies in the literature focus on $z \gs 6$
galaxies \citep{vallini13,vallini15,munoz14}.  \citet{vallini13} uses a
GADGET-2 cosmological SPH simulation with a mass resolution $\delta m_{\rm SPH}
= 1.32\times 10^5\,{\rm \msun}$ and gravitational softening length $\sim
2h^{-1}\,{\rm kpc}$.  They adopt a two-phased ISM model (CNM$+$WNM), and
heating and cooling mechanisms similar to that of \citet{nagamine06}, in order
to predict the \cii\ emission from a $z=6.6$ Lyman alpha emitter (LAE) with
${\rm SFR \simeq 10\,{\rm \msun\,yr^{-1}}}$.  They investigated two cases of
fixed metallicity: $\Z=1$ and $\Z=0.02$.  In their simulations, the CNM ($T
\sim 250\,{\rm K}$ and $n\sim 50\,{\rm cm^{-3}}$) is found to be responsible
for $\sim 95\%$ of the total \cii emission, with the WNM phase ($T\sim
500\,{\rm K}$ and $n\sim 1\,{\rm cm^{-3}}$) contributing the remaining 5\%.
\cite{vallini15} presents an update to their 2013 model, in which the same
$z=6.6$ SPH simulation as in \cite{vallini13} is considered but now with the
implementation of a density-dependent prescription for the metallicity of the
gas, and the inclusion of \cii contributions from PDRs (in addition to their
previous two-phased CNM$+$WNM ISM model), and accounting for the effect of the
CMB on the \cii emission.  As a result of these updates, it is found that the
\cii emission is now dominated by the PDRs, with $<10\%$ coming from the CNM.
This is qualitatively consistent with the results from our simulations at $z=2$
where the PDR gas dominates the total \cii emission at least at the low SFR end
($\ls 10\,\sfru$).  However, our simulations do not incorporate the CNM to
the same extent as that of \cite{vallini15}, as the lowest densities found in
the neutral gas in our simulations is of order $\sim100\,{\rm cm^{-3}}$, i.e.,
above typical CNM densities of $\sim 20-50\,{\rm cm^{-3}}$.  It is therefore
reassuring that \cite{vallini15} find the CNM contribution to the total \cii
emission to be benign.  Assuming that $\Sigma_{\rm SFR} \propto \Sigma_{\rm H_2}$ 
and $\Sigma_{\rm H_2} \propto \CIIsd$, \cite{vallini15} scale the \cii
luminosity of their fiducial LAE model (${\rm SFR} = 10\,{\rm \msun\,yr^{-1}}$)
in order to generate a $L_{\rm [CII]} - {\rm SFR}$ relation.

\citet{munoz14} make analytical predictions of the \cii luminosities for a
range of galaxy types at $z\gs 6$ (e.g., Lyman-alpha emitters, starburst
galaxies and quasars, spanning a range in SFR from tens of ${\rm
\msun\,yr^{-1}}$ to several thousand) as part of their efforts to develop an
analytical framework for disk galaxy formation and evolution at these early
epochs.  In their study, the \cii emitting gas is assumed to come from
photo-dissociation regions only. Throughout their models, the metallicity is
kept fixed at solar.  By tuning their models, i.e., either increasing the star
formation efficiency at high redshifts or lowering the depletion of metals onto
dust grains, they arrive at the \cii$-$SFR relation: $L_{\rm [CII]}/\lsun =
5\times 10^8 \left ( {\rm SFR}/100\,{\rm \msun\,yr^{-1}}\right )^{0.9}$. While
this is essentially a linear relation, it struggles to match the expected \cii
luminosities based on observations due to the somewhat lower normalization.

\section{Conclusion}
We have developed \sigame to include simulations of the \cii\ emission from
star-forming galaxies. The code employs a multi-phased ISM consisting of UV-
and CR-heated clouds of molecular and PDR gas as well as diffuse regions
of ionized gas, and traces the \cii emission from each of these three phases.
\sigame was applied to SPH simulations of seven star-forming galaxies at $z =
2$ with stellar masses in the range $\sim (0.4 - 6.6)\times 10^{10}\,{\rm
\msun}$ and SFRs $\sim 5-60\,{\rm \msun\,yr^{-1}}$ in order to
make predictions of the \cii\ line emission from MS galaxies during
the peak of the cosmic star formation history.

A key result of our simulations is that the total \cii\ emission budget from our
galaxies is dominated by the molecular gas phase ($\gs 70\%$) in the central
regions ($R\ls 1\,{\rm kpc}$) where the bulk of the star formation occurs and is
most intense, and by PDR regions further out ($R\gs 1-2\,{\rm kpc}$) where the
molecular \cii emission has dropped by at least an order of magnitude compared
to their central values.  The PDR gas phase,
while rarely able to produce \cii emission as intense as the molecular gas in
the central regions, is nonetheless able to maintain significant levels of \cii
emission from $R\sim 2\,{\rm kpc}$ all the way out to $\sim 8\,{\rm kpc}$ from
the center. The net effect of this is that on global scales the PDR gas
can produce between 8\% and 67\% of the total \cii luminosity with the
molecular gas responsible for the remaining emission. We see a trend in which
galaxies with higher SFRs also have a higher fraction of their
total \cii luminosity coming from the molecular phase.  Our simulations
consistently show that the ionized gas contribution to the \cii\ luminosity is
negligible ($\ls 3\%$), despite the fact that this phase dominates the ISM
mass budget (see Fig.\ \ref{m_cii}). Therefore, the ionized gas phase is an
inefficient \cii line emitter in our simulations.

The integrated \cii\ luminosities of our simulated galaxies strongly correlate
with their SFRs, and in a manner that agrees well (both in terms of slope and
overall normalization) with the observed $L_{\rm [CII]} - {\rm SFR}$ relations
for normal star-forming galaxies at both low and high redshifts.  We have also
examined the relationship between the $1\,{\rm kpc}$-averaged surface densities
of $L_{\rm [C{\sc II}]}$ and ${\rm SFR}$ across our simulated galaxies. The
resulting $\CIIsd-\SFRsd$ relation spans six orders of magnitude in \SFRsd
($\sim 10^{-5}-10\,{\rm \msun\,yr^{-1}\,kpc^{-2}}$), extending beyond the
observed ranges at both the low and high end of the relation.  In the
\SFRsd-range where a direct comparison with observations can be made ($\sim
0.001-1\,{\rm \msun\,yr^{-1}\,kpc^{-2}}$) we find excellent agreement with our
simulations.

Our simulations suggest that the correlation between \cii and SFR -- both the
integrated and the resolved versions -- is determined by the combined
\cii-contribution from the molecular and PDR phases (the ionized gas
makes a negligible contribution), with the former exhibiting the steepest slope
and dominating the \cii emission at the high-SFR-end.  We argue that this is
due to the fact that the \cii luminosity scales with the amount of molecular
gas present in our simulated galaxies.  A similar luminosity-mass scaling is
not seen for the other phases.  Our work therefore suggest that the observed
$\cii-{\rm SFR}$ relation is a combination of the line predominantly tracing
the molecular gas (i.e., the star formation 'fuel') at high SFR levels/surface
densities, while at low SFRs/surface densities the line is tracing PDR
gas being exposed to a weaker, interstellar UV-field.  As a consequence, we
hypothesize that galaxies with large mid-plane pressures and large molecular
gas fractions will display a steeper ${\rm \cii - SFR}$ relationship than
galaxies where a larger fraction of the ISM is atomic/ionized gas. In the
future we will extend this study to a larger sample of model galaxies, in
particular with a larger spread in SFRs and metallicities.

\acknowledgments
\section*{Acknowledgements}
We thank Inti Pelupessy for useful discussions and help with the ionization
fractions.  Thanks also goes to Sangeeta Malhotra and Kristian Finlator for
stimulating discussions and important suggestions.  This work has benefitted
greatly from email correspondence with Paul Goldsmith regarding the \cii\
cooling rate.  We are also particularly grateful to Maria Kapala (and the SLIM
team), Ilse de Looze and Rodrigo Herrera-Camus for providing their resolved
data. Thanks also goes to Georgios Magdis for providing us with his compilation
of high-$z$ \cii detections. Finally, we would like to thank the anonymous
referee for an insightful and constructive referee report which helped improved the
paper.  Simulation analysis was done using SPHGR \citep{SPHGR} and
pyGadgetReader \citep{pygr}.  K. P. O. and S. T. gratefully acknowledge the support from
the Lundbeck foundation.  
S. T. acknowledges support from the ERC Consolidator Grant funding scheme 
(project ConTExt, grant number 648179). 
T. R. G. acknowledges support from a STFC Advanced
Fellowship.  The Dark Cosmology Centre is funded by the Danish National Research
Foundation.  D. N. was supported by the US National Science Foundation via grants
AST-1009452, AST-1442650, NASA program AR-13906.001, and a Cottrell College
Science Award funded by the Research Corporation for Science Advancement.
Support for Program number HST AR-13906.001 was provided by NASA through a grant
from the Space Telescope Science Institute, which is operated by the Association
of Universities for Research in Astronomy, Incorporated, under NASA contract
NAS5-26555.

\clearpage

\bibliographystyle{apj}
\bibliography{bibs}

\begin{thebibliography}{118}
\expandafter\ifx\csname natexlab\endcsname\relax\def\natexlab#1{#1}\fi

\bibitem[{{Acero} {et~al.}(2009){Acero}, {Aharonian}, {Akhperjanian},
  {et~al.}}]{acero09}
{Acero}, F., {Aharonian}, F., {Akhperjanian}, A.~G., {et~al.} 2009, Science,
  326, 1080

\bibitem[{{Bakes} \& {Tielens}(1994)}]{bakes94}
{Bakes}, E.~L.~O. \& {Tielens}, A.~G.~G.~M. 1994, \apj, 427, 822

\bibitem[{{Barinovs} {et~al.}(2005){Barinovs}, {van Hemert}, {Krems}, \&
  {Dalgarno}}]{barinovs05}
{Barinovs}, {\u G}., {van Hemert}, M.~C., {Krems}, R., \& {Dalgarno}, A. 2005,
  \apj, 620, 537

\bibitem[{{Bell}(2003)}]{bell03}
{Bell}, E.~F. 2003, \apj, 586, 794

\bibitem[{{Bisbas} {et~al.}(2015){Bisbas}, {Papadopoulos}, \&
  {Viti}}]{bisbas15}
{Bisbas}, T.~G., {Papadopoulos}, P.~P., \& {Viti}, S. 2015, \apj, 803, 37

\bibitem[{{Blitz} {et~al.}(2007){Blitz}, {Fukui}, {Kawamura}, {Leroy},
  {Mizuno}, \& {Rosolowsky}}]{blitz07}
{Blitz}, L., {Fukui}, Y., {Kawamura}, A., {Leroy}, A., {Mizuno}, N., \&
  {Rosolowsky}, E. 2007, Protostars and Planets V, 81

\bibitem[{{Boselli} {et~al.}(2002){Boselli}, {Gavazzi}, {Lequeux}, \&
  {Pierini}}]{boselli02}
{Boselli}, A., {Gavazzi}, G., {Lequeux}, J., \& {Pierini}, D. 2002, \aap, 385,
  454

\bibitem[{{Bovy} {et~al.}(2012){Bovy}, {Rix}, \& {Hogg}}]{bovy12}
{Bovy}, J., {Rix}, H.-W., \& {Hogg}, D.~W. 2012, \apj, 751, 131

\bibitem[{{Brauher} {et~al.}(2008){Brauher}, {Dale}, \& {Helou}}]{brauher08}
{Brauher}, J.~R., {Dale}, D.~A., \& {Helou}, G. 2008, \apjs, 178, 280

\bibitem[{{Calzetti} {et~al.}(2007){Calzetti}, {Kennicutt}, {Engelbracht},
  {et~al.}}]{calzetti07}
{Calzetti}, D., {Kennicutt}, R.~C., {Engelbracht}, C.~W., {et~al.} 2007, \apj,
  666, 870

\bibitem[{{Capak} {et~al.}(2015){Capak}, {Carilli}, {Jones}, {Casey},
  {Riechers}, {Sheth}, {Carollo}, {Ilbert}, {Karim}, {Lefevre}, {Lilly},
  {Scoville}, {Smolcic}, \& {Yan}}]{capak15}
{Capak}, P.~L., {Carilli}, C., {Jones}, G., {Casey}, C.~M., {Riechers}, D.,
  {Sheth}, K., {Carollo}, C.~M., {Ilbert}, O., {Karim}, A., {Lefevre}, O.,
  {Lilly}, S., {Scoville}, N., {Smolcic}, V., \& {Yan}, L. 2015, \nat, 522, 455

\bibitem[{{Carilli} {et~al.}(2013){Carilli}, {Riechers}, {Walter}, {Maiolino},
  {Wagg}, {Lentati}, {McMahon}, \& {Wolfe}}]{carilli13}
{Carilli}, C.~L., {Riechers}, D., {Walter}, F., {Maiolino}, R., {Wagg}, J.,
  {Lentati}, L., {McMahon}, R., \& {Wolfe}, A. 2013, \apj, 763, 120

\bibitem[{{Casey} {et~al.}(2014){Casey}, {Narayanan}, \& {Cooray}}]{casey14}
{Casey}, C.~M., {Narayanan}, D., \& {Cooray}, A. 2014, \physrep, 541, 45

\bibitem[{{Chomiuk} \& {Povich}(2011)}]{chomiuk11}
{Chomiuk}, L. \& {Povich}, M.~S. 2011, \aj, 142, 197

\bibitem[{{Crawford} {et~al.}(1985){Crawford}, {Genzel}, {Townes}, \&
  {Watson}}]{crawford85}
{Crawford}, M.~K., {Genzel}, R., {Townes}, C.~H., \& {Watson}, D.~M. 1985,
  \apj, 291, 755

\bibitem[{{Daddi} {et~al.}(2010){Daddi}, {Elbaz}, {Walter},
  {et~al.}}]{daddi10a}
{Daddi}, E., {Elbaz}, D., {Walter}, F., {et~al.} 2010, \apjl, 714, L118

\bibitem[{{Dav{\'e}} {et~al.}(2013){Dav{\'e}}, {Katz}, {Oppenheimer},
  {Kollmeier}, \& {Weinberg}}]{Dave13}
{Dav{\'e}}, R., {Katz}, N., {Oppenheimer}, B.~D., {Kollmeier}, J.~A., \&
  {Weinberg}, D.~H. 2013, \mnras, 434, 2645

\bibitem[{{De Breuck} {et~al.}(2014){De Breuck}, {Williams}, {Swinbank},
  {et~al.}}]{debreuck14}
{De Breuck}, C., {Williams}, R.~J., {Swinbank}, M., {et~al.} 2014, \aap, 565,
  A59

\bibitem[{{de Looze} {et~al.}(2011){de Looze}, {Baes}, {Bendo}, {Cortese}, \&
  {Fritz}}]{delooze11}
{de Looze}, I., {Baes}, M., {Bendo}, G.~J., {Cortese}, L., \& {Fritz}, J. 2011,
  \mnras, 416, 2712

\bibitem[{{De Looze} {et~al.}(2014){De Looze}, {Cormier}, {Lebouteiller},
  {et~al.}}]{delooze14}
{De Looze}, I., {Cormier}, D., {Lebouteiller}, V., {et~al.} 2014, \aap, 568,
  A62

\bibitem[{{D{\'{\i}}az-Santos} {et~al.}(2013){D{\'{\i}}az-Santos}, {Armus},
  {Charmandaris}, {et~al.}}]{diaz-santos13}
{D{\'{\i}}az-Santos}, T., {Armus}, L., {Charmandaris}, V., {et~al.} 2013, \apj,
  774, 68

\bibitem[{{Draine}(2011)}]{draine11}
{Draine}, B.~T. 2011, {Physics of the Interstellar and Intergalactic Medium}

\bibitem[{{Durier} \& {Dalla Vecchia}(2012)}]{Durier12}
{Durier}, F. \& {Dalla Vecchia}, C. 2012, \mnras, 419, 465

\bibitem[{{Elmegreen}(1989)}]{elmegreen89b}
{Elmegreen}, B.~G. 1989, ApJ, 344, 306

\bibitem[{{Farrah} {et~al.}(2013){Farrah}, {Lebouteiller}, {Spoon},
  {et~al.}}]{farrah13}
{Farrah}, D., {Lebouteiller}, V., {Spoon}, H.~W.~W., {et~al.} 2013, \apj, 776,
  38

\bibitem[{{Ferland} {et~al.}(2013){Ferland}, {Porter}, {van Hoof},
  {et~al.}}]{ferland13}
{Ferland}, G.~J., {Porter}, R.~L., {van Hoof}, P.~A.~M., {et~al.} 2013, RxMAA,
  49, 137

\bibitem[{{Ford} {et~al.}(2015){Ford}, {Werk}, {Dave}, {et~al.}}]{Ford15}
{Ford}, A.~B., {Werk}, J.~K., {Dave}, R., {et~al.} 2015, ArXiv e-prints

\bibitem[{{Gallerani} {et~al.}(2012){Gallerani}, {Neri}, {Maiolino},
  {et~al.}}]{gallerani12}
{Gallerani}, S., {Neri}, R., {Maiolino}, R., {et~al.} 2012, \aap, 543, A114

\bibitem[{{Genzel} {et~al.}(2010){Genzel}, {Tacconi}, {Gracia-Carpio},
  {et~al.}}]{genzel10}
{Genzel}, R., {Tacconi}, L.~J., {Gracia-Carpio}, J., {et~al.} 2010, \mnras,
  407, 2091

\bibitem[{{Goldsmith} {et~al.}(2012){Goldsmith}, {Langer}, {Pineda}, \&
  {Velusamy}}]{goldsmith12}
{Goldsmith}, P.~F., {Langer}, W.~D., {Pineda}, J.~L., \& {Velusamy}, T. 2012,
  \apjs, 203, 13

\bibitem[{{Gullberg} {et~al.}(2015){Gullberg}, {De Breuck}, {Vieira},
  {et~al.}}]{gullberg15}
{Gullberg}, B., {De Breuck}, C., {Vieira}, J.~D., {et~al.} 2015, \mnras, 449,
  2883

\bibitem[{{Hahn} \& {Abel}(2011)}]{MUSIC}
{Hahn}, O. \& {Abel}, T. 2011, \mnras, 415, 2101

\bibitem[{{Hailey-Dunsheath} {et~al.}(2010){Hailey-Dunsheath}, {Nikola},
  {Stacey}, {Oberst}, {Parshley}, {Benford}, {Staguhn}, \&
  {Tucker}}]{hailey-dunsheath10}
{Hailey-Dunsheath}, S., {Nikola}, T., {Stacey}, G.~J., {Oberst}, T.~E.,
  {Parshley}, S.~C., {Benford}, D.~J., {Staguhn}, J.~G., \& {Tucker}, C.~E.
  2010, \apjl, 714, L162

\bibitem[{{Heiderman} {et~al.}(2010){Heiderman}, {Evans}, {Allen}, {Huard}, \&
  {Heyer}}]{heiderman10}
{Heiderman}, A., {Evans}, II, N.~J., {Allen}, L.~E., {Huard}, T., \& {Heyer},
  M. 2010, \apj, 723, 1019

\bibitem[{{Herrera-Camus} {et~al.}(2015){Herrera-Camus}, {Bolatto}, {Wolfire},
  {et~al.}}]{herrera-camus15}
{Herrera-Camus}, R., {Bolatto}, A.~D., {Wolfire}, M.~G., {et~al.} 2015, \apj,
  800, 1

\bibitem[{{Hodge} {et~al.}(1999){Hodge}, {Balsley}, {Wyder}, \&
  {Skelton}}]{hodge99}
{Hodge}, P.~W., {Balsley}, J., {Wyder}, T.~K., \& {Skelton}, B.~P. 1999, \pasp,
  111, 685

\bibitem[{{Hollenbach} \& {Tielens}(1999)}]{hollenbach99}
{Hollenbach}, D.~J. \& {Tielens}, A.~G.~G.~M. 1999, Reviews of Modern Physics,
  71, 173

\bibitem[{{Hopkins}(2013)}]{Hopkins13DISPH}
{Hopkins}, P.~F. 2013, \mnras, 428, 2840

\bibitem[{{Iono} {et~al.}(2006){Iono}, {Yun}, {Elvis}, {et~al.}}]{iono06}
{Iono}, D., {Yun}, M.~S., {Elvis}, M., {et~al.} 2006, \apjl, 645, L97

\bibitem[{{Kanekar} {et~al.}(2013){Kanekar}, {Wagg}, {Ram Chary}, \&
  {Carilli}}]{kanekar13}
{Kanekar}, N., {Wagg}, J., {Ram Chary}, R., \& {Carilli}, C.~L. 2013, \apjl,
  771, L20

\bibitem[{{Kapala} {et~al.}(2015){Kapala}, {Sandstrom}, {Groves},
  {et~al.}}]{kapala15}
{Kapala}, M.~J., {Sandstrom}, K., {Groves}, B., {et~al.} 2015, \apj, 798, 24

\bibitem[{{Katz} {et~al.}(1996){Katz}, {Weinberg}, \& {Hernquist}}]{Katz96}
{Katz}, N., {Weinberg}, D.~H., \& {Hernquist}, L. 1996, \apjs, 105, 19

\bibitem[{{Keenan} {et~al.}(1986){Keenan}, {Lennon}, {Johnson}, \&
  {Kingston}}]{keenan86}
{Keenan}, F.~P., {Lennon}, D.~J., {Johnson}, C.~T., \& {Kingston}, A.~E. 1986,
  \mnras, 220, 571

\bibitem[{{Kennicutt}(1998)}]{kennicutt98}
{Kennicutt}, Jr., R.~C. 1998, \araa, 36, 189

\bibitem[{{Krumholz} {et~al.}(2008){Krumholz}, {McKee}, \&
  {Tumlinson}}]{Krumholz08}
{Krumholz}, M.~R., {McKee}, C.~F., \& {Tumlinson}, J. 2008, \apj, 689, 865

\bibitem[{{Krumholz} {et~al.}(2009){Krumholz}, {McKee}, \&
  {Tumlinson}}]{Krumholz09}
---. 2009, \apj, 693, 216

\bibitem[{{Krumholz} \& {Tan}(2007)}]{Krumholz07}
{Krumholz}, M.~R. \& {Tan}, J.~C. 2007, \apj, 654, 304

\bibitem[{{Kurtz} {et~al.}(1983){Kurtz}, {Smyers}, {Russell}, {Harwit}, \&
  {Melnick}}]{kurtz83}
{Kurtz}, N.~T., {Smyers}, S.~D., {Russell}, R.~W., {Harwit}, M., \& {Melnick},
  G. 1983, \apj, 264, 538

\bibitem[{{Lada} {et~al.}(2010){Lada}, {Lombardi}, \& {Alves}}]{Lada10}
{Lada}, C.~J., {Lombardi}, M., \& {Alves}, J.~F. 2010, \apj, 724, 687

\bibitem[{{Langer} \& {Pineda}(2015)}]{langer15}
{Langer}, W.~D. \& {Pineda}, J.~L. 2015, \aap, 580, A5

\bibitem[{{Larson}(1981)}]{larson81}
{Larson}, R.~B. 1981, \mnras, 194, 809

\bibitem[{{Leroy} {et~al.}(2012){Leroy}, {Bigiel}, {de Blok},
  {et~al.}}]{leroy12}
{Leroy}, A.~K., {Bigiel}, F., {de Blok}, W.~J.~G., {et~al.} 2012, \aj, 144, 3

\bibitem[{{Luhman} {et~al.}(1998){Luhman}, {Satyapal}, {Fischer}, {Wolfire},
  {Cox}, {Lord}, {Smith}, {Stacey}, \& {Unger}}]{luhman98}
{Luhman}, M.~L., {Satyapal}, S., {Fischer}, J., {Wolfire}, M.~G., {Cox}, P.,
  {Lord}, S.~D., {Smith}, H.~A., {Stacey}, G.~J., \& {Unger}, S.~J. 1998,
  \apjl, 504, L11

\bibitem[{{Luhman} {et~al.}(2003){Luhman}, {Satyapal}, {Fischer}, {Wolfire},
  {Sturm}, {Dudley}, {Lutz}, \& {Genzel}}]{luhman03}
{Luhman}, M.~L., {Satyapal}, S., {Fischer}, J., {Wolfire}, M.~G., {Sturm}, E.,
  {Dudley}, C.~C., {Lutz}, D., \& {Genzel}, R. 2003, \apj, 594, 758

\bibitem[{{Madden} {et~al.}(1992){Madden}, {Genzel}, {Herrmann}, {Poglitsch},
  {Geis}, {Townes}, \& {Stacey}}]{madden92}
{Madden}, S.~C., {Genzel}, R., {Herrmann}, F., {Poglitsch}, A., {Geis}, N.,
  {Townes}, C.~H., \& {Stacey}, G.~J. 1992, in Bulletin of the American
  Astronomical Society, Vol.~24, American Astronomical Society Meeting
  Abstracts, 1268

\bibitem[{{Magdis} {et~al.}(2014){Magdis}, {Rigopoulou}, {Hopwood},
  {et~al.}}]{magdis14}
{Magdis}, G.~E., {Rigopoulou}, D., {Hopwood}, R., {et~al.} 2014, \apj, 796, 63

\bibitem[{{Maiolino} {et~al.}(2015){Maiolino}, {Carniani}, {Fontana},
  {Vallini}, {Pentericci}, {Ferrara}, {Vanzella}, {Grazian}, {Gallerani},
  {Castellano}, {Cristiani}, {Brammer}, {Santini}, {Wagg}, \&
  {Williams}}]{maiolino15}
{Maiolino}, R., {Carniani}, S., {Fontana}, A., {Vallini}, L., {Pentericci}, L.,
  {Ferrara}, A., {Vanzella}, E., {Grazian}, A., {Gallerani}, S., {Castellano},
  M., {Cristiani}, S., {Brammer}, G., {Santini}, P., {Wagg}, J., \& {Williams},
  R. 2015, \mnras, 452, 54

\bibitem[{{Maiolino} {et~al.}(2009){Maiolino}, {Caselli}, {Nagao}, {Walmsley},
  {De Breuck}, \& {Meneghetti}}]{maiolino09}
{Maiolino}, R., {Caselli}, P., {Nagao}, T., {Walmsley}, M., {De Breuck}, C., \&
  {Meneghetti}, M. 2009, \aap, 500, L1

\bibitem[{{Maiolino} {et~al.}(2005){Maiolino}, {Cox}, {Caselli},
  {et~al.}}]{maiolino05}
{Maiolino}, R., {Cox}, P., {Caselli}, P., {et~al.} 2005, \aap, 440, L51

\bibitem[{{Malhotra} {et~al.}(1997){Malhotra}, {Helou}, {Stacey},
  {et~al.}}]{malhotra97}
{Malhotra}, S., {Helou}, G., {Stacey}, G., {et~al.} 1997, \apjl, 491, L27

\bibitem[{{Malhotra} {et~al.}(2001){Malhotra}, {Kaufman}, {Hollenbach},
  {et~al.}}]{malhotra01}
{Malhotra}, S., {Kaufman}, M.~J., {Hollenbach}, D., {et~al.} 2001, \apj, 561,
  766

\bibitem[{{McKee} \& {Krumholz}(2010)}]{mckee10}
{McKee}, C.~F. \& {Krumholz}, M.~R. 2010, \apj, 709, 308

\bibitem[{{Meijerink} {et~al.}(2007){Meijerink}, {Spaans}, \&
  {Israel}}]{meijerink07}
{Meijerink}, R., {Spaans}, M., \& {Israel}, F.~P. 2007, \aap, 461, 793

\bibitem[{{Mezger} {et~al.}(1982){Mezger}, {Mathis}, \& {Panagia}}]{mezger82}
{Mezger}, P.~G., {Mathis}, J.~S., \& {Panagia}, N. 1982, \aap, 105, 372

\bibitem[{{Mu{\~n}oz} \& {Furlanetto}(2014)}]{munoz14}
{Mu{\~n}oz}, J.~A. \& {Furlanetto}, S.~R. 2014, \mnras, 438, 2483

\bibitem[{{Nagamine} {et~al.}(2004){Nagamine}, {Springel}, {Hernquist}, \&
  {Machacek}}]{nagamine04}
{Nagamine}, K., {Springel}, V., {Hernquist}, L., \& {Machacek}, M. 2004,
  \mnras, 350, 385

\bibitem[{{Nagamine} {et~al.}(2006){Nagamine}, {Wolfe}, \&
  {Hernquist}}]{nagamine06}
{Nagamine}, K., {Wolfe}, A.~M., \& {Hernquist}, L. 2006, \apj, 647, 60

\bibitem[{{Narayanan} {et~al.}(2008{\natexlab{a}}){Narayanan}, {Cox}, {Kelly},
  {et~al.}}]{narayanan08}
{Narayanan}, D., {Cox}, T.~J., {Kelly}, B., {et~al.} 2008{\natexlab{a}}, \apjs,
  176, 331

\bibitem[{{Narayanan} {et~al.}(2008{\natexlab{b}}){Narayanan}, {Cox},
  {Shirley}, {Dav{\'e}}, {Hernquist}, \& {Walker}}]{narayanan08a}
{Narayanan}, D., {Cox}, T.~J., {Shirley}, Y., {Dav{\'e}}, R., {Hernquist}, L.,
  \& {Walker}, C.~K. 2008{\natexlab{b}}, \apj, 684, 996

\bibitem[{{Narayanan} {et~al.}(2012){Narayanan}, {Krumholz}, {Ostriker}, \&
  {Hernquist}}]{narayanan12a}
{Narayanan}, D., {Krumholz}, M.~R., {Ostriker}, E.~C., \& {Hernquist}, L. 2012,
  \mnras, 421, 3127

\bibitem[{{Narayanan} {et~al.}(2015){Narayanan}, {Turk}, {Feldmann},
  {Robitaille}, {Hopkins}, {Thompson}, {Hayward}, {Ball},
  {Faucher-Gigu{\`e}re}, \& {Kere{\v s}}}]{narayanan15}
{Narayanan}, D., {Turk}, M., {Feldmann}, R., {Robitaille}, T., {Hopkins}, P.,
  {Thompson}, R., {Hayward}, C., {Ball}, D., {Faucher-Gigu{\`e}re}, C.-A., \&
  {Kere{\v s}}, D. 2015, \nat, 525, 496

\bibitem[{{Oey} \& {Clarke}(1997)}]{oey97}
{Oey}, M.~S. \& {Clarke}, C.~J. 1997, \mnras, 289, 570

\bibitem[{{Olsen} {et~al.}(2015){Olsen}, {Greve}, {Brinch}, {Sommer-Larsen},
  {Rasmussen}, {Toft}, \& {Zirm}}]{olsen15}
{Olsen}, K.~P., {Greve}, T.~R., {Brinch}, C., {Sommer-Larsen}, J., {Rasmussen},
  J., {Toft}, S., \& {Zirm}, A. 2015, ArXiv e-prints

\bibitem[{{Oppenheimer} \& {Dav{\'e}}(2008)}]{Oppenheimer08}
{Oppenheimer}, B.~D. \& {Dav{\'e}}, R. 2008, \mnras, 387, 577

\bibitem[{{Ouchi} {et~al.}(2013){Ouchi}, {Ellis}, {Ono}, {et~al.}}]{ouchi13}
{Ouchi}, M., {Ellis}, R., {Ono}, Y., {et~al.} 2013, \apj, 778, 102

\bibitem[{{Papadopoulos} {et~al.}(2014){Papadopoulos}, {Zhang}, {Xilouris},
  {et~al.}}]{papa14}
{Papadopoulos}, P.~P., {Zhang}, Z.-Y., {Xilouris}, E.~M., {et~al.} 2014, ApJ,
  788, 153

\bibitem[{{Pelupessy} \& {Papadopoulos}(2009)}]{pelupessy09}
{Pelupessy}, F.~I. \& {Papadopoulos}, P.~P. 2009, \apj, 707, 954

\bibitem[{{Pelupessy} {et~al.}(2006){Pelupessy}, {Papadopoulos}, \& {van der
  Werf}}]{pelupessy06}
{Pelupessy}, F.~I., {Papadopoulos}, P.~P., \& {van der Werf}, P. 2006, \apj,
  645, 1024

\bibitem[{{Pineda} {et~al.}(2014){Pineda}, {Langer}, \& {Goldsmith}}]{pineda14}
{Pineda}, J.~L., {Langer}, W.~D., \& {Goldsmith}, P.~F. 2014, \aap, 570, A121

\bibitem[{{Planck Collaboration} {et~al.}(2014){Planck Collaboration}, {Ade},
  {Aghanim}, {Armitage-Caplan}, {Arnaud}, {Ashdown}, {Atrio-Barandela},
  {Aumont}, {Baccigalupi}, {Banday}, \& et~al.}]{Planck14}
{Planck Collaboration}, {Ade}, P.~A.~R., {Aghanim}, N., {Armitage-Caplan}, C.,
  {Arnaud}, M., {Ashdown}, M., {Atrio-Barandela}, F., {Aumont}, J.,
  {Baccigalupi}, C., {Banday}, A.~J., \& et~al. 2014, \aap, 571, A16

\bibitem[{{Popping} {et~al.}(2014{\natexlab{a}}){Popping},
  {P{\'e}rez-Beaupuits}, {Spaans}, {Trager}, \& {Somerville}}]{popping14b}
{Popping}, G., {P{\'e}rez-Beaupuits}, J.~P., {Spaans}, M., {Trager}, S.~C., \&
  {Somerville}, R.~S. 2014{\natexlab{a}}, \mnras, 444, 1301

\bibitem[{{Popping} {et~al.}(2014{\natexlab{b}}){Popping}, {Somerville}, \&
  {Trager}}]{popping14a}
{Popping}, G., {Somerville}, R.~S., \& {Trager}, S.~C. 2014{\natexlab{b}},
  \mnras, 442, 2398

\bibitem[{{Robertson} \& {Kravtsov}(2008)}]{Robertson08}
{Robertson}, B.~E. \& {Kravtsov}, A.~V. 2008, \apj, 680, 1083

\bibitem[{{R{\"o}llig} {et~al.}(2006){R{\"o}llig}, {Ossenkopf}, {Jeyakumar},
  {Stutzki}, \& {Sternberg}}]{rollig06}
{R{\"o}llig}, M., {Ossenkopf}, V., {Jeyakumar}, S., {Stutzki}, J., \&
  {Sternberg}, A. 2006, \aap, 451, 917

\bibitem[{{Russell} {et~al.}(1980){Russell}, {Melnick}, {Gull}, \&
  {Harwit}}]{russell80}
{Russell}, R.~W., {Melnick}, G., {Gull}, G.~E., \& {Harwit}, M. 1980, \apjl,
  240, L99

\bibitem[{{Saitoh} \& {Makino}(2009)}]{Saitoh09}
{Saitoh}, T.~R. \& {Makino}, J. 2009, \apjl, 697, L99

\bibitem[{{Saitoh} \& {Makino}(2013)}]{Saitoh13}
---. 2013, \apj, 768, 44

\bibitem[{{Schaerer} {et~al.}(2015){Schaerer}, {Boone}, {Zamojski}, {Staguhn},
  {Dessauges-Zavadsky}, {Finkelstein}, \& {Combes}}]{schaerer15}
{Schaerer}, D., {Boone}, F., {Zamojski}, M., {Staguhn}, J.,
  {Dessauges-Zavadsky}, M., {Finkelstein}, S., \& {Combes}, F. 2015, \aap, 574,
  A19

\bibitem[{{Schaye} \& {Dalla Vecchia}(2008)}]{Schaye08}
{Schaye}, J. \& {Dalla Vecchia}, C. 2008, \mnras, 383, 1210

\bibitem[{{Schmidt}(1959)}]{Schmidt59}
{Schmidt}, M. 1959, \apj, 129, 243

\bibitem[{{Sch{\"o}ier} {et~al.}(2005){Sch{\"o}ier}, {van der Tak}, {van
  Dishoeck}, \& {Black}}]{schoier05}
{Sch{\"o}ier}, F.~L., {van der Tak}, F.~F.~S., {van Dishoeck}, E.~F., \&
  {Black}, J.~H. 2005, A\&A, 432, 369

\bibitem[{{Seon} {et~al.}(2011){Seon}, {Edelstein}, {Korpela},
  {et~al.}}]{seon11}
{Seon}, K.-I., {Edelstein}, J., {Korpela}, E., {et~al.} 2011, \apjs, 196, 15

\bibitem[{{Speagle} {et~al.}(2014){Speagle}, {Steinhardt}, {Capak}, \&
  {Silverman}}]{speagle14}
{Speagle}, J.~S., {Steinhardt}, C.~L., {Capak}, P.~L., \& {Silverman}, J.~D.
  2014, \apjs, 214, 15

\bibitem[{{Spergel} {et~al.}(2003){Spergel}, {Verde}, {Peiris},
  {et~al.}}]{spergel2003}
{Spergel}, D.~N., {Verde}, L., {Peiris}, H.~V., {et~al.} 2003, ApJS, 148, 175

\bibitem[{{Springel}(2005)}]{Springel05}
{Springel}, V. 2005, \mnras, 364, 1105

\bibitem[{{Springel} \& {Hernquist}(2003)}]{Springel03}
{Springel}, V. \& {Hernquist}, L. 2003, \mnras, 339, 289

\bibitem[{{Stacey} {et~al.}(1991){Stacey}, {Geis}, {Genzel}, {Lugten},
  {Poglitsch}, {Sternberg}, \& {Townes}}]{stacey91}
{Stacey}, G.~J., {Geis}, N., {Genzel}, R., {Lugten}, J.~B., {Poglitsch}, A.,
  {Sternberg}, A., \& {Townes}, C.~H. 1991, \apj, 373, 423

\bibitem[{{Stacey} {et~al.}(2010){Stacey}, {Hailey-Dunsheath}, {Ferkinhoff},
  {Nikola}, {Parshley}, {Benford}, {Staguhn}, \& {Fiolet}}]{stacey10}
{Stacey}, G.~J., {Hailey-Dunsheath}, S., {Ferkinhoff}, C., {Nikola}, T.,
  {Parshley}, S.~C., {Benford}, D.~J., {Staguhn}, J.~G., \& {Fiolet}, N. 2010,
  \apj, 724, 957

\bibitem[{{Stacey} {et~al.}(1983){Stacey}, {Smyers}, {Kurtz}, \&
  {Harwit}}]{stacey83}
{Stacey}, G.~J., {Smyers}, S.~D., {Kurtz}, N.~T., \& {Harwit}, M. 1983, \apjl,
  268, L99

\bibitem[{{Stahler} \& {Palla}(2005)}]{stahler05}
{Stahler}, S.~W. \& {Palla}, F. 2005, {The Formation of Stars} (Wiley)

\bibitem[{{Swinbank} {et~al.}(2011){Swinbank}, {Papadopoulos}, {Cox}, {Krips},
  {Ivison}, {Smail}, {Thomson}, {Neri}, {Richard}, \& {Ebeling}}]{swinbank11}
{Swinbank}, A.~M., {Papadopoulos}, P.~P., {Cox}, P., {Krips}, M., {Ivison},
  R.~J., {Smail}, I., {Thomson}, A.~P., {Neri}, R., {Richard}, J., \&
  {Ebeling}, H. 2011, ApJ, 742, 11

\bibitem[{{Thompson}(2014)}]{pygr}
{Thompson}, R. 2014, {pyGadgetReader: GADGET snapshot reader for python},
  Astrophysics Source Code Library

\bibitem[{{Thompson}(2015)}]{SPHGR}
---. 2015, {SPHGR: Smoothed-Particle Hydrodynamics Galaxy Reduction},
  Astrophysics Source Code Library

\bibitem[{{Thompson} {et~al.}(2015){Thompson}, {Dav{\'e}}, {Huang}, \&
  {Katz}}]{thompson15}
{Thompson}, R., {Dav{\'e}}, R., {Huang}, S., \& {Katz}, N. 2015, ArXiv e-prints

\bibitem[{{Thompson} {et~al.}(2014){Thompson}, {Nagamine}, {Jaacks}, \&
  {Choi}}]{thompson14}
{Thompson}, R., {Nagamine}, K., {Jaacks}, J., \& {Choi}, J.-H. 2014, \apj, 780,
  145

\bibitem[{{Vallini} {et~al.}(2013){Vallini}, {Gallerani}, {Ferrara}, \&
  {Baek}}]{vallini13}
{Vallini}, L., {Gallerani}, S., {Ferrara}, A., \& {Baek}, S. 2013, \mnras, 433,
  1567

\bibitem[{{Vallini} {et~al.}(2015){Vallini}, {Gallerani}, {Ferrara},
  {Pallottini}, \& {Yue}}]{vallini15}
{Vallini}, L., {Gallerani}, S., {Ferrara}, A., {Pallottini}, A., \& {Yue}, B.
  2015, \apj, 813, 36

\bibitem[{{Venemans} {et~al.}(2012){Venemans}, {McMahon}, {Walter},
  {et~al.}}]{venemans12}
{Venemans}, B.~P., {McMahon}, R.~G., {Walter}, F., {et~al.} 2012, \apjl, 751,
  L25

\bibitem[{{Wagg} {et~al.}(2010){Wagg}, {Carilli}, {Wilner}, {Cox}, {De Breuck},
  {Menten}, {Riechers}, \& {Walter}}]{wagg10}
{Wagg}, J., {Carilli}, C.~L., {Wilner}, D.~J., {Cox}, P., {De Breuck}, C.,
  {Menten}, K., {Riechers}, D.~A., \& {Walter}, F. 2010, \aap, 519, L1

\bibitem[{{Walter} {et~al.}(2012){Walter}, {Decarli}, {Carilli},
  {et~al.}}]{walter12}
{Walter}, F., {Decarli}, R., {Carilli}, C., {et~al.} 2012, \apj, 752, 93

\bibitem[{{Walter} {et~al.}(2009){Walter}, {Riechers}, {Cox}, {Neri},
  {Carilli}, {Bertoldi}, {Weiss}, \& {Maiolino}}]{walter09}
{Walter}, F., {Riechers}, D., {Cox}, P., {Neri}, R., {Carilli}, C., {Bertoldi},
  F., {Weiss}, A., \& {Maiolino}, R. 2009, \nat, 457, 699

\bibitem[{{Wang} {et~al.}(2013){Wang}, {Wagg}, {Carilli}, {et~al.}}]{wang13}
{Wang}, R., {Wagg}, J., {Carilli}, C.~L., {et~al.} 2013, \apj, 773, 44

\bibitem[{{Webber}(1998)}]{webber98}
{Webber}, W.~R. 1998, \apj, 506, 329

\bibitem[{{Whitaker} {et~al.}(2011){Whitaker}, {Labb{\'e}}, {van Dokkum},
  {et~al.}}]{whitaker11}
{Whitaker}, K.~E., {Labb{\'e}}, I., {van Dokkum}, P.~G., {et~al.} 2011, \apj,
  735, 86

\bibitem[{{Wiersma} {et~al.}(2009){Wiersma}, {Schaye}, {Theuns}, {Dalla
  Vecchia}, \& {Tornatore}}]{wiersma09}
{Wiersma}, R.~P.~C., {Schaye}, J., {Theuns}, T., {Dalla Vecchia}, C., \&
  {Tornatore}, L. 2009, \mnras, 399, 574

\bibitem[{{Willott} {et~al.}(2013){Willott}, {Omont}, \&
  {Bergeron}}]{willott13}
{Willott}, C.~J., {Omont}, A., \& {Bergeron}, J. 2013, \apj, 770, 13

\bibitem[{{Wilson} \& {Bell}(2002)}]{wilson02}
{Wilson}, N.~J. \& {Bell}, K.~L. 2002, \mnras, 337, 1027

\bibitem[{{Wolfire} {et~al.}(2010){Wolfire}, {Hollenbach}, \&
  {McKee}}]{wolfire10}
{Wolfire}, M.~G., {Hollenbach}, D., \& {McKee}, C.~F. 2010, \apj, 716, 1191

\end{thebibliography}

\clearpage

\appendix
\section{\cii\ excitation and emission} \label{apD}
\begin{table}[htbp]
\centering
\caption{Provides a summary of the various heating and
cooling mechanisms adopted in the molecular and atomic gas regions of our GMC models,
along with the physical quantities on which they depend.} 
\begin{threeparttable}
\begin{tabular*}{\columnwidth}{p{2cm}p{6cm}p{4cm}p{4cm}} 
\hline
\hline\\ 
\multicolumn{4}{c}{Cooling and Heating Rates in GMC Models} 	\\ [0.1cm]
Process			&												& Parameters			&	Reference		\\
\toprule 
\Hpe 			& Photo-electric heating 						& \g0, \Tk, \ne, \nH	&  	\cite{bakes94}		\\
\Hcrhi 			& Cosmic ray heating in atomic gas				& \cri, \nhi, \xe		&	\cite{draine11}		\\ 
\Hcrh2 			& Cosmic ray heating in molecular gas			& \cri, \nh2			&	\cite{stahler05}		\\ 
\Ch2 			& H2 line cooling								& \nh2, \Tk				&	\cite{papa14} \\
\Coi 			& \oi ~line cooling								& \nH, \Tk				&	\cite{rollig06} \\
\Ccii	 		& \cii\ line cooling in ionized gas				& \g0, \Tk, \nH, \Xc, $\sigma_v$ 	&	\cite{goldsmith12} \\
\bottomrule
\end{tabular*}
{\bf Note.} The most important
references are also given, and we also refer to \cite{olsen15} for a
detailed description.
\label{lookup}
\end{threeparttable}
\end{table}

For a two-level system such as \cii embedded in a radiation field with energy
density $U$ at the transition frequency, the general rate equation governing the
population levels can be written as: 
\begin{align}
	\frac{n_u}{n_l}	=	\frac{B_{lu}U+C_{lu}}{A_{ul}+B_{ul}U+C_{ul}} 
					=	\frac{gB_{ul}U+gKC_{ul}}{A_{ul}+B_{ul}U+C_{ul}}, 
\end{align}
where $A_{ul}$ ($=2.3\e{-6}$\,\ps) is the spontaneous emission rate, and
$B_{ul}$ and $B_{lu}$ are the stimulated emission and absorption rate
coefficients, respectively \citep{goldsmith12}. Here we have invoked detailed
balance, i.e.\ $B_{lu}U=gB_{ul}U$ and $C_{lu}=gKC_{ul}$, where $g$ ($=g_u/g_l$)
is the ratio of the statistical weights and $K = e^{-h\nu/k_{\rm B}T_{\rm k}} =
e^{-91.25\,{\rm K}/T_{\rm k}}$.  We can then write the fraction of C$^+$ in the
upper level, $f_u$, as:
\begin{align}
	f_u				&=	\frac{n_u}{n_l+n_u} =	\frac{1}{\frac{n_l}{n_u}+1}  \nonumber\\
					&=	\frac{gB_{ul}U+gKC_{ul}}{A_{ul}+(1+g)B_{ul}U+C_{ul}+gKC_{ul}}  \nonumber\\	
					&=	\frac{gK+gB_{ul}U/C_{ul}}{1+gK+A_{ul}/C_{ul}+(1+g)B_{ul}U/C_{ul}}.
					\label{eq:f_u}
\end{align}
In general, $U = (1-\beta)U(T_{\rm ex}) + \beta U(T_{\rm bg})$, where $U(T_{\rm
bg})$ is the energy density from a background field (e.g.\ CMB or radiation from
dust), and $\beta$ is the escape probability fraction at the \cii frequency
\citep[see also][]{goldsmith12}.  We shall ignore any background, however, in which case we
can write the stimulated downward rate as:
\begin{align}
	B_{ul}U 		= 	\frac{(1-\beta)A_{ul}}{e^{-91.25\,{\rm K}/T_{\rm ex}}-1},
	\label{eq:BulU}
\end{align}
and the excitation temperature as:
\begin{align}
	e^{-91.25\,{\rm K}/T^{\rm ex}}	=	K\left( 1+\frac{\beta A_{ul}}{C_{ul}} \right),
	\label{eq:Tex}
\end{align}
see \cite{goldsmith12}. For the escape probability we assume a spherical
geometry with a radial velocity gradient proportional to radius, such that \
$\beta = (1 - \exp(-\tau))/\tau$.  In calculating the integral in eq.\
\ref{eq:CII-integral} we split the integral up into 100 radial bins of thickness
$\Delta R$ (see Section \ref{cii_em}), and we approximate the optical depth in
each such bin with that of a homogeneous static slab of gas of thickness $\Delta
R$ \citep{draine11}:
\begin{align}
	\tau	=	\frac{g_u}{g_l}\frac{A_{ul}c^3}{4(2\pi)^{3/2}\nu^3\sigma_v}n_l\Delta R
				\left( 1-\frac{n_ug_l}{n_lg_u} \right), 
				\label{eq:tau}
\end{align}
where $\sigma_v$ is the gas velocity dispersion, which is calculated according
to eq.\,\ref{sigma_v_Pe} for the PDR and molecular gas regions in the GMCs, and
is set to the local velocity dispersion in the SPH simulation in the case of the
ionized clouds.  We use eqs.\ \ref{eq:f_u}, \ref{eq:BulU}, \ref{eq:Tex}, and
\ref{eq:tau} to iteratively solve for consistent values of $f_u$ and $\beta$ in
each $\Delta R$ bin. This is done by first assuming optically thin emission ($\beta=1$) 
in order to get an initial estimate of $f_u$ (eq.\
\ref{eq:f_u}), which is subsequently used to calculate $\tau$ and $\beta$ (eq.\
\ref{eq:tau}) and from that $T_{\rm ex}$ and $B_{ul}U$ (eqs.\ \ref{eq:Tex} and
\ref{eq:BulU}), etc.  Once consistent values for $f_u$ and $\tau$ have been
reached, we calculate the total cooling rate according to eq.\,\ref{eq:ccii}.
This cooling rate is used to determine the thermal balance at various points
within the GMCs as well as the \cii emission from the ionized clouds as
described in Section \ref{cii_em}.  We emphasize that our methods assumes that
the \cii emission from the different $\Delta R$ bins within a cloud is
radiatively de-coupled, and that the total \cii emission from a cloud is
therefore the sum of the emission from each bin.

\bigskip

\begin{figure}[htbp] 
\centering
\includegraphics[width=0.8\columnwidth]{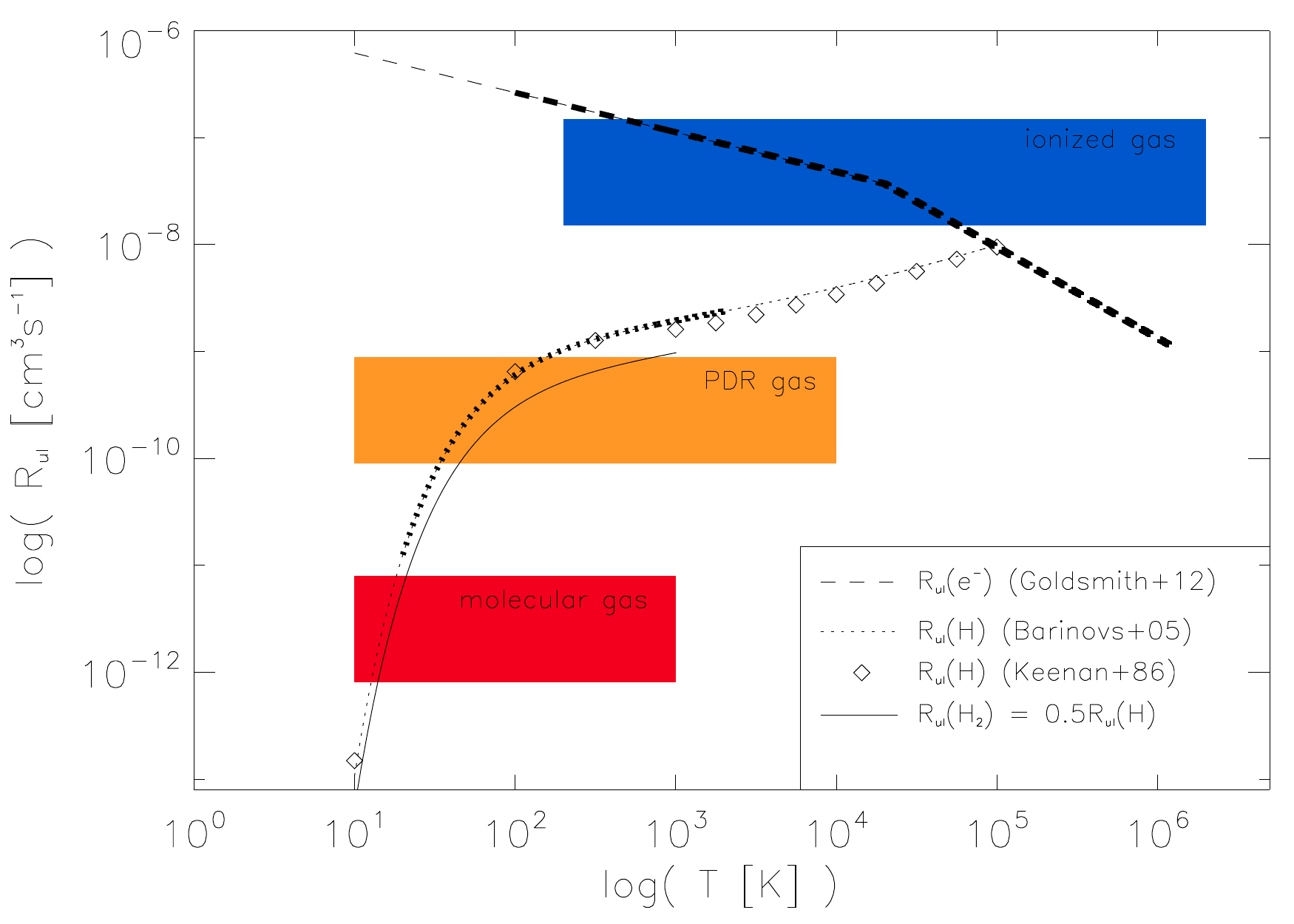}
\caption{\cii deexcitation coefficients ($R_{ul}$) as a function of temperature
for collisions with $e^-$ (dashed line), H\,{\sc i} (dotted line), and H$_2$
(solid line and diamonds). The curves are thicker over the temperature ranges where 
they are formally applicable. The blue, orange, and red rectangles indicate the
temperature range encountered in the ionized, atomic, and molecular phase in our
simulated galaxies, respectively (see sections \ref{split12} and \ref{split2}).}
\label{figure:rate-coefficients-vs-T}
\end{figure}

As explained in Section \ref{cii_em} \cii is collisionally (de)excited by H$_2$
in the molecular phase, by $e^-$ and H{\sc i} in the PDR region, and by $e^-$ in
the ionized gas. For a single collision partner the collision rates are equal to
the density ($n$) of the collision partner times the rate coefficients, i.e., 
$C_{ul} = n R_{ul}$ and $C_{lu} = n R_{lu}$. In case of two collision partners
we have $C_{ul} = n_1 R_{ul,1} + n_2 R_{ul,2}$ and $C_{lu} = n_1 R_{lu,1} + n_2
R_{lu,2}$. Fig.\ \ref{figure:rate-coefficients-vs-T} shows the \cii\ deexcitation
rate coefficients that we have adopted in our work for collisions with $e^-$
H\,{\sc i}, and H$_2$ as a function of temperature.  For collisions with
electrons, we adopt the following expression for the deexcitation rate
coefficient as a function of electron temperature ($T_e$):
\begin{align}
	R_{ul}(e^-)		=	8.7\e{-8}(T_{e}/2000\,{\rm K})^{-0.37}\,{\rm cm^3\,s^{-1}},
\label{equation:rate_coefficient_e}
\end{align}
which is applicable for temperatures from $\simeq 100$\,K to $20{,}000$\,K
\citep{goldsmith12}. At temperatures $>20{,}000$\,K we set $R_{ul}(e^-) \sim
T_{e}^{-0.85}$ \citep{wilson02,langer15}. For the PDR gas, the electron
density is calculated using CLOUDY and the electron temperature is set to the
kinetic temperature of the gas (calculated according to eq.\ \ref{Tk_rH2}). For
the ionized gas we assume $n_{e}=n_{\rm H\textsc{ii}}$, i.e., $x_{e}=1$, and
the electron temperature gas is set to the SPH gas kinetic temperature (see
Section \ref{split2}).  For the deexcitation rate coefficient for collisions
with H\,{\sc i} we use the analytical expression provided by \cite{barinovs05}
and shown as the dotted curve in Fig.\ \ref{figure:rate-coefficients-vs-T}:
\begin{equation}
R_{ul}({\rm H\,\textsc{i}}) = 7.938\times 10^{11} \exp (-91.2/T_{\rm k}) \left ( 16 + 0.344\sqrt{T_{\rm k}} - 47.7/T_{\rm k}\right )\, {\rm cm^3\,s^{-1}}.
\label{equation:rate_coefficient_H}
\end{equation}
While \cite{barinovs05} cites an application range for the above expression of
$15\,{\rm K} < T_{\rm k} < 2000\,{\rm K}$, a comparison with $R_{ul}({\rm
H\,\textsc{i}})$-values found by \cite{keenan86} over the temperature range
$10\,{\rm K} < T_{\rm k} < 100,000\,{\rm K}$ shows that eq.\
\ref{equation:rate_coefficient_H} provides an excellent match over this larger
temperature range (Fig.\ \ref{figure:rate-coefficients-vs-T}).  For collisions
with H$_2$ we follow \cite{goldsmith12} and assume that the collision rate
coefficients are approximately half those for collisions with H over the
relevant temperature range for the molecular gas, i.e.\ $R_{ul}({\rm H}_2) =
0.5\times R_{ul}({\rm H})$.

\end{document}